\documentclass[12pt, nofootinbib]{article}

\usepackage{amssymb}
\usepackage{amsmath,bm}
\usepackage{amssymb}
\usepackage{graphicx}
\usepackage{amsfonts}         
\usepackage{fancybox}

\usepackage[usenames,dvipsnames]{xcolor}

\usepackage[pagebackref,  
bookmarks={false}, pdfauthor={Rolando Ramirez Camasca, John McGreevy}, pdftitle={Yay, physics!}]{hyperref}
\hypersetup{colorlinks=true, 
linkcolor=BrickRed, 
citecolor=Violet, 
filecolor=OliveGreen, 
urlcolor=RoyalBlue, 
filebordercolor={.8 .8 1}, 
urlbordercolor={.8 .8 0}}
\usepackage{soul}
\setstcolor{Red}

\definecolor{darkgreen}{rgb}{0,0.4,0}
\definecolor{darkred}{rgb}{0.4,0,0}
\definecolor{darkblue}{rgb}{0,0,0.4}
\definecolor{lightblue}{rgb}{.6,.6,0.9}

\definecolor{uglybrown}{rgb}{0.8,  0.7,  0.5}

\definecolor{palatinatepurple}{rgb}{0.41, 0.16, 0.38}
\definecolor{celebrationcolor}{rgb}{0.75,  0.0,  0.9}

\definecolor{shadecolor}{rgb}{0.90,0.90,0.90}
\definecolor{DVcolor}{rgb}{0.95,  0.5,  0.2}
\definecolor{lightbluemuons}{rgb}{0.0,.65,1.0}

\definecolor{chartreuse}{rgb}{0.70, 1.00, 0.00}

\usepackage{tikz}

\tikzset{
    vector/.style={decorate, decoration={snake}, draw},
    fermion/.style={postaction={decorate},
        decoration={markings,mark=at position .55 with {\arrow{>}}}},
    fermionbar/.style={draw, postaction={decorate},
        decoration={markings,mark=at position .55 with {\arrow{<}}}},
    fermionnoarrow/.style={},
    gluon/.style={decorate,
        decoration={coil,amplitude=4pt, segment length=5pt}},
    scalar/.style={dashed, postaction={decorate},
        decoration={markings,mark=at position .55 with {\arrow{>}}}},
    scalarbar/.style={dashed, postaction={decorate},
        decoration={markings,mark=at position .55 with {\arrow{<}}}},
    scalarnoarrow/.style={dashed,draw},
	vectorscalar/.style={loosely dotted,draw=black, postaction={decorate}},
}

\usepackage{enumitem}

\usepackage{slashed}

\usepackage[font=small,labelfont=bf]{caption}
\DeclareCaptionFont{tiny}{\tiny}
\usepackage{epstopdf}
\usepackage{wasysym}

\usepackage[yyyymmdd,hhmmss]{datetime}   
\usepackage{framed}  
 \usepackage{mdframed}
 \usepackage{wrapfig}
 \usepackage{yfonts}  

\def\ketbra#1#2{ | #1 \rangle\hskip-2pt\langle #2|}

\newmdenv[%
        backgroundcolor=lightgray,
    linecolor=black,
    outerlinewidth=2pt,
]{boxedandshaded}

\def\parfig#1#2{
\parbox{#1\textwidth}
{\includegraphics[width=#1\textwidth]{#2}}
}

\numberwithin{equation}{section}

\renewcommand{\theequation}{\arabic{section}.\arabic{equation}}

\def\nd{{ \vphantom{\dagger}}}

\newcommand{\vev}[1]{\langle #1 \rangle}

\newlength{\extraspace}
\newlength{\extraspaces}
\def\be{\begin{equation}}
\def\ee{\end{equation}}

\newcommand{\bea}{\begin{eqnarray}}
\newcommand{\eea}{\end{eqnarray}}

\def\eps{\epsilon}
\def\half{{1\over 2}}

\def\tr{{\rm tr}}

\def\bra#1{\left\langle#1\right|}
\def\ket#1{\left|#1\right\rangle}

\def\vev#1{\left\langle{#1}\right\rangle}

\def\CO{{\cal O}}

\def\II{\relax{I\kern-.10em I}}

\def\IZ{\mathbb{Z}}
\def\IB{\relax{\rm I\kern-.18em B}}

\def\ID{\relax{\rm I\kern-.18em D}}
\def\IE{\relax{\rm I\kern-.18em E}}
\def\IF{\relax{\rm I\kern-.18em F}}
\def\IG{\relax\hbox{$\inbar\kern-.3em{\rm G}$}}
\def\IGa{\relax\hbox{${\rm I}\kern-.18em\Gamma$}}
\def\IH{\relax{\rm I\kern-.18em H}}
\def\II{\relax{\rm I\kern-.18em I}}
\def\IK{\relax{\rm I\kern-.18em K}}

\def\inbar{\,\vrule height1.5ex width.4pt depth0pt}

\def\IR{\mathbb{R}}

\def\simgt{\hskip0.05in\relax{ 
\raise3.0pt\hbox{ $>$
{\lower5.0pt\hbox{\kern-1.05em $\sim$}} }} \hskip0.05in}

\def\lp10{\ell_p^{10}}
\def\lp11{\ell_p^{11}}
\def\R11{R_{11}}

\def\frac#1#2{{#1 \over #2}}

\def\up{\uparrow}
\def\down{\downarrow}

\def\Ione{\hbox{$1\hskip -1.2pt\vrule depth 0pt height 1.53ex width 0.7pt
                  \vrule depth 0pt height 0.3pt width 0.12em$}}

\newdimen\tableauside\tableauside=1.0ex
\newdimen\tableaurule\tableaurule=0.4pt
\newdimen\tableaustep
\def\phantomhrule#1{\hbox{\vbox to0pt{\hrule height\tableaurule width#1\vss}}}
\def\phantomvrule#1{\vbox{\hbox to0pt{\vrule width\tableaurule height#1\hss}}}
\def\sqr{\vbox{%
  \phantomhrule\tableaustep
  \hbox{\phantomvrule\tableaustep\kern\tableaustep\phantomvrule\tableaustep}%
  \hbox{\vbox{\phantomhrule\tableauside}\kern-\tableaurule}}}
\def\squares#1{\hbox{\count0=#1\noindent\loop\sqr
  \advance\count0 by-1 \ifnum\count0>0\repeat}}
\def\tableau#1{\vcenter{\offinterlineskip
  \tableaustep=\tableauside\advance\tableaustep by-\tableaurule
  \kern\normallineskip\hbox
    {\kern\normallineskip\vbox
      {\gettableau#1 0 }%
     \kern\normallineskip\kern\tableaurule}%
  \kern\normallineskip\kern\tableaurule}}
\def\gettableau#1 {\ifnum#1=0\let\next=\null\else
  \squares{#1}\let\next=\gettableau\fi\next}

\def\({\left(}
\def\){\right)}

\def\ii{{\bf i}}

\def\lsim{\mathrel{\mathstrut\smash{\ooalign{\raise2.5pt\hbox{$<$}\cr\lower2.5pt\hbox{$\sim$}}}}}
\def\gsim{\mathrel{\mathstrut\smash{\ooalign{\raise2.5pt\hbox{$>$}\cr\lower2.5pt\hbox{$\sim$}}}}}

\def\overleftrightarrow#1{\vbox{\ialign{##\crcr
     $\leftrightarrow$\crcr\noalign{\kern-0pt\nointerlineskip}
     $\hfil\displaystyle{#1}\hfil$\crcr}}}
     
     \def\overleftarrow#1{\vbox{\ialign{##\crcr
     $\leftarrow$\crcr\noalign{\kern-0pt\nointerlineskip}
     $\hfil\displaystyle{#1}\hfil$\crcr}}}

\def\eg{{\it e.g.}}
\def\ie{{\it i.e.}}

\def\gSO{\textsf{SO}}

\def\gSU{\textsf{SU}}
\def\gU{\textsf{U}}

\newif{\ifeq}           
\eqtrue                 
\newcounter{lecturecounter}

\def\baselinestretch{1.1}
\renewcommand{\title}[1]{\vbox{\center\LARGE{#1}}\vspace{5mm}}
\renewcommand{\author}[1]{\vbox{\center#1}\vspace{5mm}}
\newcommand{\address}[1]{\vbox{\center\em#1}}

\renewcommand{\date}[1]{\vbox{\center#1}}

\usepackage{geometry}
\usepackage{pdflscape}

\usepackage{amsmath} 
\usepackage{mathtools}

\usepackage{amsthm}
\usepackage{thmtools}

\usepackage{diagbox}
\usepackage{boldline}

\usepackage{bbold}

\makeatother

\theoremstyle{plain}

\theoremstyle{definition}

\theoremstyle{plain}

\theoremstyle{plain}

\theoremstyle{plain}

\theoremstyle{plain}

\definecolor{RRCgreen}{RGB}{50,100,50}
\definecolor{JMblue}{RGB}{25,25,125}
\definecolor{l-blue}{RGB}{32, 168, 232}
\definecolor{d-blue}{RGB}{10, 68, 168}
\definecolor{l-purple}{RGB}{201, 121, 237}
\definecolor{d-purple}{RGB}{111, 6, 158}
\definecolor{my-orange}{RGB}{237, 140, 12}
\definecolor{my-red}{RGB}{194, 29, 29}

\def\Pf{{\rm Pf}}

\begin{document}

\title{Dimer piling problems and interacting field theory}

\author{Rolando Ramirez Camasca and John McGreevy}

\address{Department of Physics, University of California San Diego, La Jolla, CA 92093, USA}

\begin{abstract}
The dimer tiling problem asks in how many ways can the edges of a graph be covered by dimers so that each site is covered once.  
In the special case of a planar graph, this problem has a solution in terms of a free fermionic field theory.
We rediscover and explore an expression for the number of coverings of an arbitrary graph by arbitrary objects in terms of an interacting fermionic field theory first proposed by Samuel.
Generalizations of the dimer tiling problem, which we call `dimer piling problems,' demand that each site be covered $N$ times by indistinguishable dimers.
Our field theory provides a solution of these problems in the large-$N$ limit.
We give a similar path integral representation for certain lattice coloring problems.

\end{abstract}

\vfill

\today

\vfill\eject


\setcounter{tocdepth}{2}    

\renewcommand{\baselinestretch}{0.75}\normalsize
\tableofcontents
\renewcommand{\baselinestretch}{1.1}\normalsize

\vfill\eject

\section{Introduction}

The dimer tiling problem for a graph $\Gamma$ asks in how many ways can we place dimers on the edges of $\Gamma$ 
such that every vertex is covered by a single dimer.  
In the special case of a {\it planar} graph, this (fully-packed) dimer tiling problem is exactly solvable
\cite{kasteleyn1961statistics, temperley1961dimer, fisher1961statistical, kasteleyn1963dimer,
fisher1963statistical, heilmann1972theory, fendley2002classical}.  One representation of the solution is as an integral over grassmann variables at the sites of the graph, 
with a gaussian measure determined by a signed adjacency matrix of the graph.  The signs are chosen so that there is $\pi$ flux through each plaquette of the graph, and this weights each dimer configuration with a $+$ sign.  Such a weighting is not possible on a non-planar graph. 

The dimer tiling problem admits many generalizations.  For example, we could allow monomers, sites that are not covered by dimers.  
More generally, one can introduce the following partition function describing coverings by monomers and dimers even on non-planar graphs: 
\be Z(\Gamma,x) = \sum_\text{dimer configurations on $\Gamma$} x^{\# \text{of monomers}} \ee 
where the sum runs over dimer configurations that cover each site of $\Gamma$ at most once; uncovered sites are called monomers and are weighted with monomer fugacity $x$.  
Even on a planar graph, the monomer-dimer problem is not exactly solvable.   

Many other generalizations of the dimer problem have been studied.  We could weight the dimers of different orientation differently.
We could replace the dimers by objects of a different shape, such as linear trimers \cite{ghosh2007random, dhar2021entropy} or wedges, 
or sphinxes \cite{Huber2023}.

\begin{figure}
    \centering
    \parfig{.3}{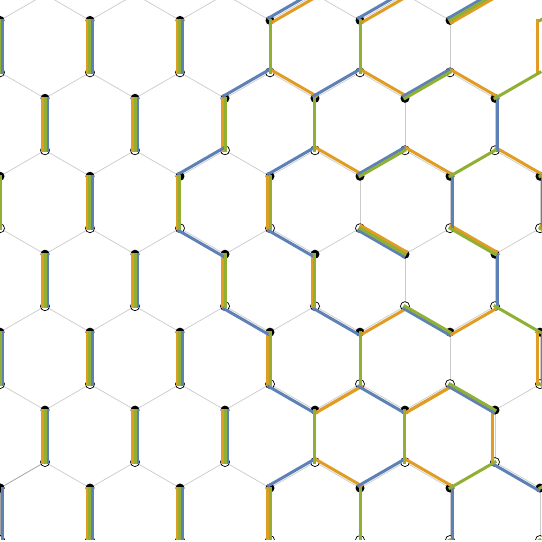}
    \caption{A dimer piling configuration with $N=3$ on a patch of the honeycomb lattice.  An $N$-dimer piling can be regarded (not canonically) as $N$ independent dimer coverings of the lattice, and here we choose a particular coloring of the dimer piling to make this explicit.}
    \label{fig:sample-piling}
\end{figure}

A family of generalizations on which we will focus is where the number of dimers incident on each site is $N$ instead of just $1$.  
We call such generalizations {\it dimer piling problems}.
A sample configuration is shown in Fig.~\ref{fig:sample-piling}; a dimer piling configuration is really just $N$ independent dimer coverings of the graph, but the weights with which we will count them do not factorize.
Our main contribution is an integral representation for three large-$N$ generalizations (for certain weights of the configurations) that turn out to be solvable by standard field theory methods.  

One motivation for our study is as a controlled test of vector-like large-$N$ limits so common in condensed matter physics and high-energy physics.  Along the way we will encounter a number of artifacts of the large-$N$ expansion, as well as an example of its breakdown, through the appearance of terms that go like $\log N$.  
Similar phenomena are known to happen in large-$N$ quantum systems dual to black holes,  
\eg~\cite{Sen:2012kpz, Sen:2012dw,H:2023qko}.

The $N=1$ version of this integral representation was studied previously in \cite{samuel1980use, samuel1980-2,samuel1980-3}.
Even at finite $N$, it can be used as a starting point for an auxiliary-field Monte Carlo evaluation of these partition functions.  

Other graph coloring problems can also be studied along the same lines. In particular, we show that the problem of counting $N$-colorings of the 
links of a graph, so that no two links sharing a vertex are of the same color, can be expressed as an integral over fermionic variables. We find two such integral representations: one that works for planar lattices only, and the other for non-planar graphs. 
We also find an interesting integral representation for the sum of such colorings weighted by a sign-valued invariant.
Each of these integrals are amenable to field theory techniques, which we use to give estimates.

In \S\ref{sec:integral-rep} we introduce the non-gaussian grassmann integral representation of the dimer model on a general graph.
In \S\ref{sec:dimer-piling-problems} we generalize these integrals to $N>1$ colors of dimers, and describe the combinatorial problems defined by these integrals with $N>1$.
In \S\ref{sec:large-N-analysis} we study the large-$N$ behavior of these integrals using both saddle point methods and Feynman diagrams.  
In \S\ref{sec:graph-coloring} we apply the same methods to graph-coloring problems.
In \S\ref{sec:QDM}, we construct a family of quantum systems, generalizing quantum dimer models, some of whose groundstate correlation functions are computed by dimer piling models.  
This was our original motivation for revisiting the study of dimer models.
In \S\ref{sec:discussion} we discuss many related questions.

\section{Integral representations of dimer problems}
\label{sec:integral-rep}

We begin by describing our route to discovering Samuel's integral representation of the general dimer problem.  
Consider the following integral, associated with a graph $\Gamma$:
\be 
Z_1(\Gamma) \equiv \int D(\tilde \eta,  \eta)~e^{ \sum_{\vev{ij}} A_{ij} \eta_i \tilde \eta_i \eta_j \tilde \eta_j } . \label{eq:dimer-integral-N=1}
\ee
In this expression, $(\eta_i, \tilde \eta_i)$ are two grassmann variables on each site of the graph $\Gamma$.  
\be A_{ij} = \begin{cases} 1, & \text{if $ij$ is an edge of $\Gamma$} \\ 0, & \text{else} \end{cases}\ee
is an unsigned adjacency matrix of $\Gamma$.  
$\vev{ij}$ denotes an edge of $\Gamma$ connecting sites $i$ and $j$.
In this paper, we do not use the Einstein summation convention for the site indices.
The integration measure is 
\be D(\tilde \eta, \eta) \equiv \prod_i \( d \tilde \eta_i d \eta_i \) ~.\ee
We assume $\Gamma$ has an even number of sites, since otherwise $Z_1 = 0$.

The partition function~\eqref{eq:dimer-integral-N=1} counts the number of perfect matchings of the graph $\Gamma$. This is the number of ways of placing dimers on the edges of $\Gamma$ such that each lattice site is paired with only one dimer. 
We can see this by doing the integral over the $\tilde \eta$s: 
\be Z_1(\Gamma) = \int D\eta \Pf_{ij} \(  A_{ij} \eta_i  \eta_j \) ~.\ee
The argument of the Pfaffian is an antisymmetric $2n \times 2n$ matrix in the $ij$ indices, where $2n$ is the number of sites of the graph $\Gamma$.
Using the definition of the Pfaffian of a $2n \times 2n$ antisymmetric matrix as 
\be \Pf_{ij}  \( M_{ij} \) =  { C_n} \sum_{\sigma \in S_{2n}} (-1)^\sigma M_{\sigma_1 \sigma_2} M_{\sigma_3 \sigma_4}  \cdots M_{\sigma_{2n-1} \sigma_{2n}}  \ee
(where $(-1)^\sigma$ is the sign of the permutation, and $C_n \equiv { 1\over 2^n n!}$)
we can write $Z_1$ as 
\begin{align} Z_1(\Gamma) & =  {C_n}  \int D\eta  \sum_{\sigma \in S_{2n}} (-1)^\sigma \prod_{a=1}^n \( A_{\sigma_{2a-1} \sigma_{2a}} \eta_{\sigma_{2a-1}}   \eta_{\sigma_{2a}}  \) 
\\ 
& = {C_n}  \sum_{\sigma \in S_{2n}} \prod_{a=1}^n \( A_{\sigma_{2a-1} \sigma_{2a}} \)  (-1)^{\sigma} 
\underbrace{ \int D\eta \prod_{a=1}^n \( \eta_{\sigma_{2a-1}}   \eta_{\sigma_{2a}}  \) }_{=(-1)^\sigma} 
\\ & = {1\over 2^n n!}  \sum_{\sigma \in S_{2n}} \prod_{a=1}^n \( A_{\sigma_{2a-1} \sigma_{2a}}  \)
 ~.\end{align}
This is manifestly the number of dimer coverings of the graph $\Gamma$.
(The factor of ${1\over 2^n n!}$ cancels the overcounting from swapping $ \sigma_{2a-1} \leftrightarrow \sigma_{2a}$ and swapping the labels on the $n$ dimers.)
Thus
\begin{equation}\label{eq:one-dimer}
    Z_1(\Gamma) = \int D(\tilde \eta,  \eta)~e^{ \sum_{\vev{ij}} A_{ij} \eta_i \tilde \eta_i \eta_j \tilde \eta_j } = \text{Hf}(A), 
\end{equation}
the number of perfect matchings of a graph $\Gamma$ is Hf$(A)$, the Haffnian of its adjacency matrix. 
Note that the Haffnian is not a linear algebra object, in that it does not behave well under basis transformations.

Interestingly, Kasteleyn showed that, if one considers planar lattices, the dimer tiling problem reduces to evaluating the pfaffian of the so called Kasteleyn matrix \cite{kasteleyn1961statistics}. This is the signed adjacency matrix $\tilde A$, whose edges are oriented so that every face of the graph $\Gamma$ has an odd number of clockwise oriented edges:
\be \tilde A_{ij} = \begin{cases} 1, & \text{if $ij$ is an edge of $\Gamma$ and $i$ points to $j$} \\ -1, & \text{if $ij$ is an edge of $\Gamma$ and $j$ points to $i$} \\ 0, & \text{else.} \end{cases}\ee
We can summarize these findings as 
\be \text{Dimer coverings} = {\rm Hf}(A) \buildrel{\text{planar}}\over{=} {\rm Pf}(\tilde A), \ee
where the last equality is only true if the graph is planar and one can find the Kasteleyn matrix. 
It is interesting to note that the Kasteleyn matrix can be regarded as a discretization of the 2d Dirac operator.  
The pfaffian is a linear algebra object.

Quite generally, we could also include monomers in our partition function. If we wish to include a fugacity for monomers in the integral, we can modify it to 
\be Z_1(\Gamma,x ) \equiv \int D(\tilde \eta , \eta)~e^{\sum_i x_i \eta_i \tilde \eta_i + \sum_{\vev{ij}} A_{ij} \eta_i \tilde \eta_i \eta_j \tilde \eta_j } ~.\ee
Pulling down a power of $x_i$ from the action removes the site $i$ from the dimer coverings.

{\bf Schwinger-Dyson equation.} The following line of thought gives an independent proof of the correctness of the non-gaussian grassmann integral representation for the monomer-dimer problem.  
Consider the following Schwinger-Dyson equation:
\begin{align} 0 &= \int D(\tilde \eta , \eta) ~{ \partial \over \partial \eta_i } \left( \eta_i e^{ \sum_i x_{\,} \eta_i \tilde \eta_i + \sum_{\vev{ij}} A_{ij} \eta_i \tilde \eta_i \eta_j \tilde \eta_j} \right) 
             \\ &= Z_1(\Gamma,x) - x Z_1(\Gamma - i,x) - \sum_{j} A_{ij} Z_1(\Gamma- i - j,x) . \label{eq:SD}\end{align}
Here $Z(\Gamma - ...,x)$ means the partition function with the indicated ... removed from the graph.  
Each term can be interpreted as follows:
The first term is all possible configurations of the graph $\Gamma$. The second term is all configurations with a monomer at site $i$.  The third term is all configurations where a dimer ends at site $i$. 

In fact, equation \eqref{eq:SD} is the main tool used by Heilmann and Lieb \cite{heilmann1972theory} to demonstrate the absence of a phase transition in this model at finite monomer fugacity. They show that the zeros of the partition function, as a function of the complex fugacity, all lie on the imaginary axis. And thus, the only value where a phase transition can occur is at $x=0$.

{\bf Parton construction.} The integrand of $Z_1(\Gamma,x)$ has a local invariance under transformations of $\eta, \tilde \eta$ that preserve the product $\eta \tilde \eta$.  It is tempting to interpret the integral as arising from a parton decomposition (see \eg~\cite{lee2006doping, Lee:2010fy, McGreevy:2016myw} for reviews) of the following formal integral.  

Define real `even grassmann variables' $\zeta_i$, satisfying the algebra
\begin{equation}
\label{eq:even-grassmann}
    \zeta_i \zeta_j = \zeta_j \zeta_i, \quad \zeta_i^2 = 0.
\end{equation}
We {\it define} integrals over such variables by the same condition as for odd grassmann integrals:
\be \int d\zeta \zeta = 1, ~~~\int d\zeta 1 = 0 ~.\ee
Functions of $\zeta$ are  defined by Taylor expansion, \eg~$ e^\zeta = 1 + \zeta$.
With these definitions, the monomer-dimer tiling partition sum can be expressed as 
\be Z_1(\Gamma, x) = \int \prod_{i=1}^n  d\zeta~ e^{\sum_{\vev{ij}} \zeta_i A_{ij} \zeta_j + \sum_i x_i \zeta_i }
\ee

Now we apply the parton construction to these $\zeta$ variables, by writing them (at each site)
as a product of two ordinary grassmann variables
\be \zeta_i = \eta_i \tilde \eta_i~.\ee  
One can check that the grassmann algebra implies the algebra \eqref{eq:even-grassmann}.  
As with any parton construction, this decomposition comes with a gauge redundancy: any transformation of $\eta, \tilde \eta$ that preserves $ \zeta = \eta \tilde \eta$ is a redundancy of this description that does not act on the physical degrees of freedom. Regarding $\tilde \eta$ as the conjugate of $\eta$, i.e. $\tilde \eta \equiv \eta^\dagger$, we can see that there is a large gauge redundancy. Each site of the graph $\Gamma$ has an on-site $\gU(1)$, given by 
\be \eta_i \to e^{i \theta_i} \eta_i, ~~~ \tilde \eta_i \to e^{-i \theta_i } \tilde \eta_i .  \label{eq:gauge_redundancy}
\ee
This gauge redundancy will become crucial later on, as we will check in section~\S\ref{sec:large-N-analysis}.
However, unlike in other parton constructions, we will not encounter any propagating gauge fields in our analysis.  This is because of the ultra-local nature of the quadratic terms.

We can rewrite the integral over grassmanns as a bosonic integral by a Hubbard-Stratonovich transformation.
First, introduce a complex scalar on each link, by 
\be 
e^{ A_{ij} \eta_i \tilde \eta_i \eta_j \tilde \eta_j} 
= \int  d\phi_{ij}^\nd d\phi^\star_{ij}
e^{ -  \phi_{ij}^\nd\phi_{ij}^\star + \sqrt{A_{ij}} \( \eta_i \tilde \eta_i \phi_{ij} + \eta_j^\alpha\tilde \eta_j^\alpha \phi_{ij}^\star  \)} ~.\ee
Then the integral over the grassmanns is gaussian and we are left with
\be Z_1 = \int \prod_{\vev{ij}} d\phi_{ij} d\phi^\star_{ij} e^{ - S[\phi]}, ~~
S[\phi] = \sum_{\vev{ij}} \phi_{ij}^\nd\phi^\star_{ij} 
- \sum_i \log \( x + \sum_{ \vev{i|j} }\sqrt{A_{ij}} \( \phi_{ij}^\nd + \phi_{ij}^\star\) \) ~, \ee
where $\vev{i|j}$ indicates a sum over neighbors $j$ of the fixed site $i$.  
This is a strange-looking field theory, where the bosonic variables on different links are coupled only by the determinant term.  

This integral is, however, difficult to analyze.  One may attempt to use a saddle-point approximation \cite{samuel1980-3}, however no small parameter controls this approximation.  
In the following section, we introduce a modification of the problem which does have such a small parameter.  

\section{Dimer piling problems} \label{sec:dimer-piling-problems}

The grassmann integral representation for the general dimer problem is a strongly interacting field theory.  A time-honored approach to such problems is to modify the problem to conjure a small parameter in which to expand.
With this in mind, let's contemplate the extension to $N$ colors. That is, instead of a single pair of grassmann variables at each site of the graph $\Gamma$, we introduce variables $(\eta^\alpha_i, \tilde \eta^\alpha_i)_{\alpha=1..N, i\in\Gamma}$.  We will call the index $\alpha$ a color index.  

Consider the following partition function:
\begin{equation}
    Z_N(\Gamma, x, V) \equiv \int D(\tilde \eta, \eta)^\alpha~ e^{\sum_i x \,\eta_i^\alpha \tilde \eta_i^\alpha + \frac{1}{N}V(\eta_i^\alpha, ~\tilde \eta_i^\beta, ~ \eta_j^\rho,~ \tilde \eta_j^\sigma)}, \label{eq:partition_function_all}
\end{equation}
where the integration measure is 
\be
    D(\tilde \eta, \eta)^\alpha  \equiv \prod_{i,\alpha} \( d \tilde \eta_i^\alpha d \eta_i^\alpha \) ~.
\ee
We have in the action~\eqref{eq:partition_function_all} two terms. The first term is a Gaussian term describing colorful monomers. The second term is the potential $V$, which could, in principle, describe different kinds of colorful objects, such as dimers, trimers, wedges, or sphinxes. In this paper, we will focus on quartic interactions, which represent colorful dimers. Note that a factor of $1/N$ is placed in front of the potential. As we will see later on, this factor is designed to make the free energy extensive in $N$.  

There are four distinct maximally-symmetric possibilities for the potential $V$, which we will label from $A$ to $D$. All of these possibilities can be organized by symmetry, as showcased in table~\ref{table:colorings}. However, we could also require that when the number of colors $N$ is set to $1$,  the model reduces to the usual monomer-dimer problem on $\Gamma$. This constrains the number of possibilities to three, $A$ through $C$. 
If we further demand that the weights of the associated combinatorial problem are positive on an arbitrary graph, we are left with models A and B.

\begin{table}[t!]
    \centering
    \begin{tabular}{ |p{4cm}|p{4cm}|p{4cm}|  }
    \hline
 \multicolumn{2}{|c|}{Colorings} & {\centering \quad \quad \,\, Invariance} \\
 \hline
 \centering Type A  &    $ \quad  V_A = A_{ij\,} \eta^\alpha_i \tilde \eta_i^\alpha \eta_j^\beta \tilde \eta_j^\beta $ & 
 \quad \quad \,\,\,\, $\gU(N,\mathbb{C})^{n_s}$ \\[0.05cm]
 \centering Type B  &   $\quad  V_B = A_{ij\,} \eta^\alpha_i \tilde \eta^\beta_i \eta^\alpha_j \tilde \eta^\beta_j $ & 
 \quad \,\, $\mathsf{O}(N, \mathbb{C}) \times \gU(1)^{n_s}$ \\[0.05cm]
 \centering Type C  & $\quad    V_C = A_{ij\,} \eta^\alpha_i \tilde{\eta}^\beta_i \eta^\beta_j  \tilde{\eta}^\alpha_j$ & 
 \quad \,\, $\gU(N,\mathbb{C}) \times \gU(1)^{n_s}$
 \\[0.05cm]
 \centering Type D  & $\quad    V_D = A_{ij\,} \eta^\alpha_i \eta^\beta_i \tilde \eta^\beta_j  \tilde{\eta}^\alpha_j$ & \quad \quad \,\,\,\,\, $\gU(N,\mathbb{C})$ \\[0.12cm]
 \hline
\end{tabular}
\caption{\label{table:colorings} The possible colorings of the dimer model.  Models $A$ through $C$ reduce to the dimer tiling problem when $N \to 1$.  We've written the invariance group of the model on a generic graph.  On special graphs there can be an enhancement (for example, model B on a bipartite lattice has the same symmetry as model C).}
\end{table}

We pause to make a brief comment on the `naturalness' of our models, in the context of model A for definiteness.  
No symmetry forbids the addition of a term of the form $ (\eta_i^\alpha \tilde \eta_i^\alpha)^2 \neq 0$.  
Comparing to the combinatorial description, adding such a term corresponds to a modification of the graph to have self-loops, \ie~links connecting a site $i$ to itself.
And no symmetry forbids the addition of sextic terms made from products of 
$\eta_i^\alpha \tilde \eta_i^\alpha$.  This corresponds to a fugacity for trimers, as we discuss further below.

\subsection{Feynman rules} 

A clear way to see the distinctions between the models we've introduced (and to understand the sense in which this exhausts the possibilities) is to consider the Feynman diagram expansion about large monomer fugacity, $x\gg 1$. 
That is, we 
expand around the gaussian problem 
\be \label{eq:free-partition-function}
Z_0 = \int D(\tilde\eta, \eta)^\alpha~ e^{ \sum_i x \eta^\alpha_i \tilde \eta^\alpha_i}
= x^{Nn_s}
\ee
where $n_s$ is the number of sites of the graph.    
We define $ \vev{\cdot}_0 
\equiv {1\over Z_0} \int D(\tilde \eta, \eta)^\alpha~ e^{ x \eta \tilde \eta} \cdot$.

For the large-$N$ analysis below, it will be convenient to employ an 't Hooft double-line notation \cite{t1974Hooftplanar} where one line represents the position index $i$ and another (in red) the color index.  The lines are oriented to distinguish $\eta$ from $\tilde \eta$.  
Including the $e^{ x \eta \tilde \eta} $ term in the measure, 
the propagator is ultralocal:
\be \label{eq:propagator} \vev{\eta_i^\alpha \tilde \eta_j^\beta}_0 = {1\over x } \delta_{ij} \delta^{\alpha\beta} = \parfig{.2}{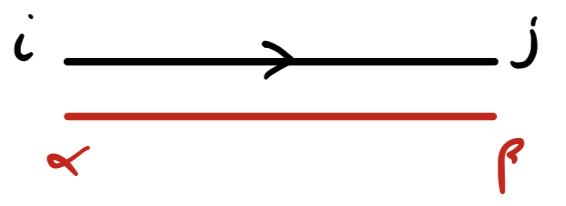} ~~.\ee
The interaction vertices for the various models in Table~\ref{table:colorings} are then: 
\be\label{eq:vertices}
A: \parfig{.19}{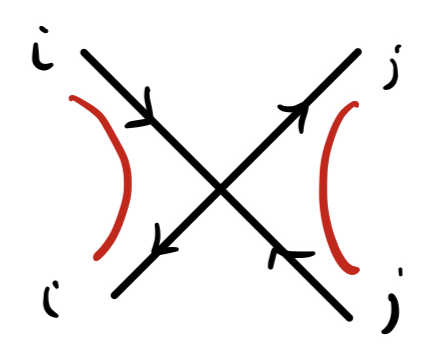}
~~~~
B: \parfig{.19}{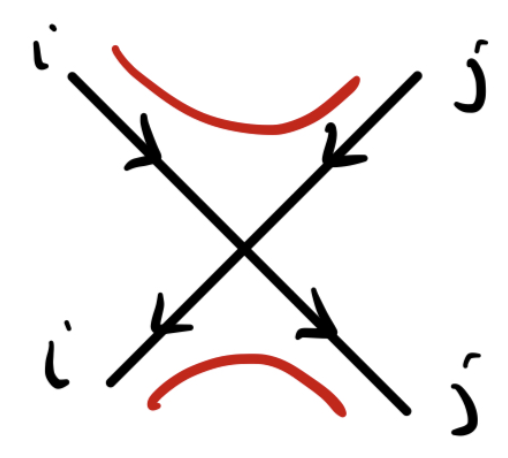}
~~~~
C: \parfig{.19}{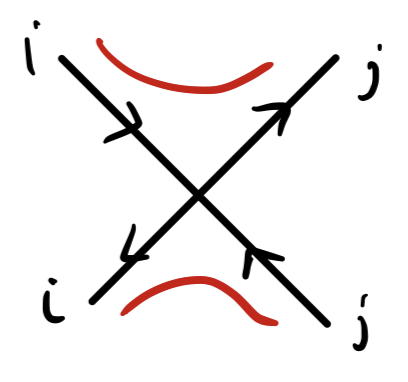}
~~~~
D: \parfig{.19}{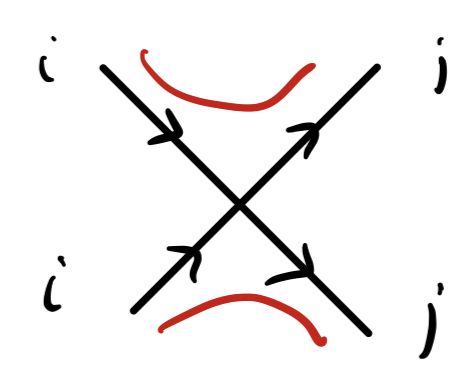}
~~~~ = {A_{ij}\over 2N}
\ee
(with color indices suppressed).
We will use these Feynman rules in \S\ref{subsec:large-N-diagrams}
to evaluate the dominant contributions at large $N$.

\subsection{Combinatorial problems from grassmann integrals} \label{subsec:combinatorial-integral}

Define ``the model-A dimer $N$-piling problem on a graph $\Gamma$" by whatever is counted by the integral defined above, with $2N$ grassmanns at each site of $\Gamma$ and  with $V=V_A$:
\be Z^A_N(\Gamma) = \int D(\tilde \eta, \eta)^{\alpha} \, e^{S_0} \ee
with 
\be S_0 = {1 \over N} \sum_{\vev{ij}} A_{ij} \eta_i^\alpha \tilde \eta_i^\alpha \eta_j^\beta \tilde \eta_j^\beta  .\ee

Note that because of the couplings between the colors $\alpha \neq \beta$, this is not just a decoupled stack of $N$ dimer problems: $Z^A_N \neq (Z_1)^N$.
The right-hand side here counts multi-dimer tilings of $\Gamma$ with each site covered by exactly $N$ dimers.  
For example, for $N=2$, any collection of closed loops is allowed, including loops that backtrack across a link.  Such a collection of closed loops can be regarded as (the symmetric difference of) an arbitrary pair of dimer tilings (as defined in e.g.~\cite{moore2011nature}).  But we must specify the measure with which such configurations are counted.

To see what combinatorial problem we have defined, observe that the only nonzero contributions to $Z^A_N$ arise from integrals of the form
\be \int \prod_{\alpha} d\tilde\eta^\alpha_i  d\eta^\alpha_i  ( \sum_\alpha \eta_i^\alpha \tilde \eta_i^\alpha)^N = N!  \label{eq:single-site} \ee
for each site $i$.  
\begin{align}
Z^A_N(\Gamma) 
& = \int D(\tilde \eta, \eta)^\alpha \prod_{\ell} e^{ S_\ell }
\\ & = \int D(\tilde \eta, \eta)^\alpha \prod_{\ell}   \( \sum_{n_\ell=0}^\infty {1\over  n_\ell !} S_\ell^{n_\ell} \) 
\label{eq:expand-link}
\end{align}
Here the product is over (undirected) links of $\Gamma$ and $S_{\vev{ij}} \equiv {1 \over N} A_{ij} \eta_i^\alpha \tilde \eta_i^\alpha \eta_j^\beta \tilde \eta_j^\beta$.

In order to get a nonzero contribution to the integral over the grassmanns at site $i$ in \eqref{eq:expand-link}, 
the $n_\ell$ must satisfy 
\be \sum_{\vev{i|j} } n_{ij} = N~, \label{eq:dimer-piling-condition}\ee 
where the notation $\sum_{\vev{i|j}}$ means sum over $j$ with $i$ fixed.
This constraint can be regarded as a Gauss law for the $\gU(1)^n$ redundancy of model A.  (A similar constraint will appear in models B and C, which also have such a $\gU(1)^n$ gauge redundancy.)
Below, this will interpreted as a local number conservation of a class of ``particles".

Using \eqref{eq:single-site} at each site, the result is 
\begin{align}
Z^A_{N}(\Gamma)  =  (N!)^{n_s} \sum_{ \text{dimer pilings} \{ n_\ell \}  }  
 \prod_\ell {1 \over N^{n_\ell } ( n_\ell )! } 
\end{align}
where $n_s$ is the number of sites of the graph $\Gamma$, and the sum over dimer pilings indicates 
a non-negative integer for each link, satisfying \eqref{eq:dimer-piling-condition}.
In a nonzero contribution, we interpret $n_\ell$ as the number of dimers on the link $\ell$.  
We can interpret the factor of ${1\over (n_\ell)!}$ as indicating that the dimers are {\it indistinguishable}.

In the same spirit, we can define ``the model-$B$ dimer $N$-piling problem on a graph $\Gamma$"as the integral with $V=V_B$:
\be Z^B_N(\Gamma) = \int D(\tilde \eta, \eta)^{\alpha} \, e^{S_0} \label{eq:piling_B} \ee
with 
\be S_0 = {1 \over N} \sum_{\vev{ij}} A_{ij} \eta_i^\alpha \tilde \eta_i^\beta \eta_j^\alpha \tilde \eta_j^\beta \label{eq:color-B-interaction} ,\ee 
where we now instead contract the color indices $\eta^\alpha_i$ with $\eta_j^\alpha$, and $\tilde \eta_i^\beta$ with $\tilde \eta_j^\beta$.

It is worth emphasizing the effect of this choice of interaction: it decouples the $\eta$ from the $\tilde \eta$ such that each variable can be colored \emph{independently}. That is, for every piling $\mathcal{C}$ coming from expanding~\eqref{eq:piling_B}, the integral factorizes and the $\eta$ and $\tilde \eta$ variables are colorwise independent. 

For example, consider tiling the two-site open chain. The only term contributing to the partition comes from filling its link $\vev{ij}$ with $N$ dimers:
\begin{align}
Z^B_N(\Gamma) 
& = \int D(\tilde \eta, \eta)^\alpha ~\frac{1}{N!} \( \frac{1}{N} \sum_{\alpha, \beta} \eta_i^\alpha \tilde \eta_i^\beta \eta_j^\alpha \tilde \eta_j^\beta \)^N
\\ & = \int D(\tilde \eta, \eta)^\alpha ~\frac{(-1)^N}{N! N^N}  \Big( \sum_{\alpha}\eta_i^\alpha \eta_j^\alpha \Big)^N \Big( \sum_\beta \tilde \eta_i^\beta  \tilde \eta_j^\beta \Big)^N  = \frac{(N!)^2}{N^N~N!}
\label{eq:expand-link-B}
\end{align}
We get~$N!$ from each integral, one from piling the $\eta$ and the other from the $\tilde \eta$.

How can we interpret the resulting factor in~\eqref{eq:expand-link-B}? We see that pilings $\mathcal{C}$ of model $B$ define the same piling structure in the $\eta$'s and the $\tilde \eta$'s. In other words, it defines copies $\mathcal{C}_{\eta}$ and $\mathcal{C}_{\tilde \eta}$ living on the variables $\eta$ and $\tilde \eta$ respectively. At each of these copies, we still need to satisfy that, at each site, there are $N$ different colorful grassmanns.   
As a consequence, the total factor one gets is the number of ways of arranging colorful $\eta$-dimers in $\mathcal{C}_\eta$ times the number of ways of arranging colorful $\tilde \eta$-dimers in $\mathcal{C}_{\tilde \eta}$, satisfying such constraint. Thus this equals the square of the number of ways of coloring the piling $\mathcal{C}$.

These facts, combined with the condition~\eqref{eq:dimer-piling-condition}, lead to
\begin{equation}
    Z^B_N(\Gamma) = \sum_{ \text{dimer pilings }  \mathcal{C} =  \{n_\ell\}  }  \prod_\ell {1 \over N^{n_\ell } ( n_\ell )! } \bigg( \substack{ \text{\# ways of coloring} \\ \text{ dimer piling $\mathcal{C}$} } \bigg)^2, \label{eq:partition-dimer-fuction-B}
\end{equation}
where now it gets a factor depending on the number of ways of distributing the colorful dimers, which is squared as we are coloring $\eta$'s and $\tilde \eta$'s independently.

We also see in~$\eqref{eq:expand-link-B}$ that negative signs appear from permuting the grassmann variables. This could imply that different dimer pilings come with relative sign factors, leading to a sum of alternating sign terms in the partition function. This is not the case for model $B$: all terms in the expansion do come with the same positive sign, as is manifest in \eqref{eq:partition-dimer-fuction-B}.  
In the appendix~\ref{appendix:combinatorial-problems}, we give a graphical method to understand how the signs from the grassmann variables conspire to cancel out in model B, unlike in model C.

Lastly, we define the ``the model-C dimer $N$-piling problem on a graph $\Gamma$" as the integral with $V=V_C$:
\be Z^C_N(\Gamma) = \int D(\tilde \eta, \eta)^{\alpha} \, e^{S_0} \label{eq:piling_C} \ee
with 
\be S_0 = {1 \over N} \sum_{\vev{ij}} A_{ij} \eta_i^\alpha \tilde \eta_i^\beta \eta_j^\beta \tilde \eta_j^\alpha \label{eq:color-C-interaction} ,\ee 
which defines the combinatorial problem over the two copies $\mathcal{C}_\eta$ and $\mathcal{C}_{\tilde \eta}$ of a piling $\mathcal{C}$, just as in model $B$, but contracts the color indices $\eta_i^\alpha$ with $\tilde \eta_j^\alpha$, and $\tilde \eta_i^\beta$ with $\eta_j^\beta$.

Model $C$ is indeed very much related to model $B$. As a matter of fact, they are the same on a bipartite graph. This can be seen by substiting in~\eqref{eq:piling_B} 
the following transformation on one of the two sublattices: $(\eta_i^\alpha, \tilde \eta_i^\alpha) \to (\tilde \eta_i^\alpha, - \eta_i^\alpha)$, which leads to~\eqref{eq:piling_C}. 
Therefore, on a bipartite lattice, the invariance group of model B is enlarged to that of $C$\footnote{The explicit expression for the $\gU(N)$ symmetry of model B on a bipartite graph is
\begin{align}
    \eta_A^\alpha  &\to U^{\alpha \beta} \eta_A^\beta, \quad \quad  \quad \, \eta_B^\alpha \to (U^\dagger)^{\alpha \beta} \eta_B^\beta, \nonumber \\
    \tilde \eta_A^\alpha  &\to (U^\dagger)^{\alpha \beta} \tilde \eta_A^\beta, \quad \quad \tilde \eta_B^\alpha \to U^{\alpha \beta} \tilde \eta_B^\beta, \nonumber
\end{align}
where the subscript $A/B$ refers to the two sublattices.
}.
Under this condition, the partition function $Z^C_N$ is just~\eqref{eq:partition-dimer-fuction-B}.

However, on non-bipartite graphs, these two models are not the same: the argument forbidding negative sign contributions fails for model $C$, and negative terms do appear. This in turn means that we need to find out how do type-$C$ dimers color the two copies of the piling $\mathcal{C}$.  This is not an easy task using integrals, but straightforward using the graphical methods of Appendix~\ref{appendix:combinatorial-problems}. For that reason, we present the partition function of model $C$ here and leave the explanation in the appendix.
\be
    Z^C_N(\Gamma) = \sum_{ \text{dimer pilings } \mathcal{C}=\{ n_\ell \}  }  \prod_\ell {(-1)^{\mathcal{C}} \over N^{n_\ell } ( n_\ell )! } \bigg( \substack{ \text{\# ways of coloring $\mathcal{C}_\eta$ and $\mathcal{C}_{\tilde \eta}$} \\ \text{ with type-$C$ dimers }} \bigg). \label{eq:partition-dimer-fuction-C}
\ee

One can add a monomer fugacity in any model, the same way as for $N=1$, by perturbing the action by an on-site quadratic term
\be S[\eta] = \sum_{i,\alpha} x_i \eta_i^\alpha \tilde \eta_i^\alpha + S_0~. 
\ee 
The way this monomer term interplays with the dimer term differs when considering model $A$ vs $B/C$. For model $A$, at the cost of a factor of $x_i$, this allows the site $i$ to be covered by one fewer dimer. But for the other models, this will remove one site of the same color $\alpha$ at each $\mathcal{C}_\eta$ and $\mathcal{C}_{\tilde \eta}$.

\section{Large-$N$ analysis} \label{sec:large-N-analysis}

In this section, we will make use of large-$N$ field theory methods to analyze the models defined above. We begin by presenting the partition function for the different types of coloring, and introduce an auxiliary field $\phi_{ij}$ for each of them. We shall denote this field $\phi_{ij}$ as a link field since it lives on the links $\vev{ij}$ of the graph. An effective action of these link fields is obtained after integrating out the original degrees of freedom, and at large $N$ is amenable to saddle point techniques. We study the resulting mean-field equations and find the dominant saddle configurations. We consider fluctuations around the saddle to get an effective partition function and its free energy density.  
In the case of model B on a bipartite lattice, this fluctuation analysis will lead us to the surprise that in a certain regime the system crystallizes, that is, the dominant saddle point is not uniform.  

In the subsequent subsection, we will also study the partition function using a diagrammatic expansion, and compare the result with the saddle point calculation. We identify the leading connected diagrams for each coloring, and resum them to find the corresponding free energy density.

{\bf Model A.} Consider the first generalization to $N$ colors, which we called coloring type $A$. 
Let us perform a Hubbard-Stratonovich decoupling of the quartic term by introducing auxiliary link variables $\phi_{ij}$:
\begin{equation}\label{eq:model-A-HS}
    e^{\frac{1}{N}\sum_{\vev{ij}} A_{ij} \eta^\alpha_i \tilde\eta^\alpha_i \eta_j^\beta \tilde \eta_j^\beta } = \int D(\phi,\phi^*) ~e^{\sum_{\vev{ij}} -N \phi_{ij} \phi_{ij}^* + \sqrt{A_{ij}} (\eta^\alpha_i \tilde \eta_i^\alpha \phi_{ij} + \eta^\alpha_j \tilde \eta_j^\alpha \phi^*_{ij})  }.
\end{equation}
\\
After integrating out the $\eta, \tilde \eta$ variables, we can rewrite the partition function as a path integral over the field $\phi_{ij}$,
\begin{equation}
    Z^A_N(\Gamma, x) = \int D(\phi,\phi^*)~e^{-N S_{A}[\phi,\phi^*]},
\end{equation}
where the effective action $S_{A}[\phi,\phi^*]$ is 
\begin{equation}
    S_{A}[\phi,\phi^*] = \sum_{\vev{ij}} \phi_{ij} \phi_{ij}^* - \sum_i \ln \left( x+ \sum_{\vev{i|j}} \sqrt{A_{ij}} (\phi_{ij} +\phi_{ji}^*) \right). \label{eq:effective-action-A}
\end{equation}
By solving its equation of motion in \eqref{eq:model-A-HS}, we can relate $\phi$ to an expectation value of monomers at a site. This is because, at the saddle point, $\phi_{ij}^{\text{saddle}} = \frac{1}{N}\sum_{\alpha} \( \eta_i^\alpha \tilde \eta_i^\alpha + \eta_j^\alpha \tilde \eta_j^\alpha \) $.
Model A has a time-reversal symmetry under which $ \eta \leftrightarrow \tilde \eta, \phi_{ij} \to \phi_{ij}^\star = \phi_{ji}$ 
which shows that the action is real and all configurations contribute with positive weight.

{\bf Model B.} Next, consider the second generalization, which we named coloring type $B$. 
We introduce an auxiliary Hubbard-Stratonovich link field $\phi_{ij}$ to decouple the type-$B$ interaction:
\begin{equation}
    e^{\frac{1}{N} \sum_{\vev{ij}} A_{ij} \eta_i^\alpha \tilde \eta_i^\beta \eta_j^\alpha \tilde \eta_j^\beta } = \int D(\phi, \phi^*)~e^{\sum_{\vev{ij}} -N \phi_{ij}\phi_{ij}^* + \sqrt{A_{ij}}(\eta_i^\alpha \eta_j^\alpha \phi_{ij}^* - \tilde \eta_i^\alpha \tilde \eta_j^\alpha \phi_{ij})  }. \label{eq:HS_B}
\end{equation}
Doing the integral over the $\eta, \tilde \eta$ fields, one gets the type-$B$ partition function as a integral over the complex field $\phi_{ij}$,
\begin{equation}
    Z^B_{N}(\Gamma,x) = \int D(\phi,\phi^*)~e^{-NS_B[\phi,\phi^*]},
\end{equation}
where the effective action $S_B[\phi,\phi^*]$ is 
\begin{equation}
    S_B[\phi,\phi^*] = \sum_{\vev{ij}} \phi_{ij} \phi_{ij}^* - \frac{1}{2} \tr \ln \begin{pmatrix}
        -B(\phi^*) & -x\mathbb{1} \\ 
        x\mathbb{1} & B(\phi)
    \end{pmatrix}, \label{eq:model-B-action}
\end{equation}
and the block matrix $B$ is 
\begin{equation}
    B(\phi) = \sum_{\vev{ij}} \phi_{ij}(\ket{i}\hspace{-0.1cm}\bra{j}- \ket{j}\hspace{-0.1cm}\bra{i}). \label{eq:block-matrix-B}
\end{equation}
The HS variable of model $B$ at the saddle point of~$\eqref{eq:HS_B}$ is related to the dimer density at a link according to $\phi_{ij}^{\text{saddle}} = \frac{1}{N} \sum_{\alpha} \eta_i^\alpha \eta_j^\alpha$. Importantly, this implies that $\phi$ transforms under the gauge redundancy in~\eqref{eq:gauge_redundancy} as $\phi_{ij} \to e^{i(\theta_i+\theta_j)} \phi_{ij}$, making $\phi_{ij}$ a gauge non-invariant variable. This decoupling channel for model B is analogous to the BCS channel in the mean-field analysis of a superconductor.  We note that other possible channels, such as 
$ \phi_i^{\alpha\beta} \equiv \eta_i^\alpha \tilde \eta_i^\beta $, where the color indices on the order parameter are not contracted, will produce saddle points that are not competitive at large $N$ with the one described above.

Model B also has a time-reversal symmetry under which $ \eta \leftrightarrow \tilde \eta, \phi_{ij} \to \phi_{ij}^\star$ 
which shows that the action is real and all configurations contribute with positive weight.

{\bf Model C.} Lastly, we implement the decoupling for model $C$. The Hubbard-Stratonovich transformation that factors the quartic interaction is:
\begin{equation}
    e^{\frac{1}{N} \sum_{\vev{ij}} A_{ij} \eta_i^\alpha \tilde \eta_i^\beta \eta_j^\beta \tilde \eta_j^\alpha} = \int D(\phi,\phi^*)~e^{\sum_{\vev{ij}}-N \phi_{ij} \phi_{ij}^* + \sqrt{A_{ij}} (\eta_i^\alpha \tilde \eta_j^\alpha \phi_{ij}^* + \tilde \eta_i^\alpha \eta_j^\alpha \phi_{ij} ) }. \label{eq:HS_C}
\end{equation}
After integrating out the original variables, the partition function for type-$C$ becomes
\begin{equation}
    Z^C_N(\Gamma,x) = \int D(\phi, \phi^*)~e^{-N S_C[\phi,\phi^*]},
\end{equation}
where the effective action $S_C[\phi,\phi^*]$ is 
\begin{equation}
    S_C[\phi,\phi^*] = \sum_{\vev{ij}} \phi_{ij} \phi_{ij}^* -\frac{1}{2} \tr \ln \begin{pmatrix}
        0 & C(\phi,\phi^*) \\ 
        - C^T(\phi,\phi^*) & 0
    \end{pmatrix},
\end{equation}
and the block matrix $C$ is 
\begin{equation}
    C(\phi,\phi^*) = -\sum_i x \ket{i}\hspace{-0.1cm}\bra{i} +\sum_{\vev{ij}} \left( -\phi^*_{ij} \ket{i}\hspace{-0.1cm}\bra{j} + \phi_{ij} \ket{j}\hspace{-0.1cm}\bra{i} \right). 
\end{equation}
The solution of the equations of motion in~\eqref{eq:HS_C} lead to $\phi_{\text{saddle}}= \frac{1}{N} \sum_{\alpha} \eta_i^\alpha \tilde \eta_j^\alpha$. As in model $B$, we see that we can similarly interpret the field as the dimer density in the link.

Model C does {\it not} have a time-reversal symmetry and indeed the action $S_C$ is not real in general.  We will see below that (on a non-bipartite graph) some configurations of model C contribute with negative weight.

\subsection{Saddle point analysis} \label{subsec:saddle-point-analysis}

In the large-$N$ limit, the theory of $(\eta,\tilde \eta)$, \eqref{eq:partition_function_all}, has a  semiclassical description in terms of the  $(\phi,\phi^*)$ fields introduced above. Thus, in the limit $N \to \infty$, the partition function is determined by the configuration that makes the action stationary:
\begin{equation}
    \left.\frac{\delta S}{\delta \phi_{ij}} \right|_{\phi_0, \phi_0^*} = \left.\frac{\delta S}{\delta \phi_{ij}^*}\right|_{\phi_0,\phi_0^*} = 0 \label{eq:first-variation}
\end{equation} 
and has the smallest action.

With the mean field determined, we proceed to expand the functional in fluctuations around $(\phi_0)_{ij}$. We will denote such fluctuations as $\tilde \phi_{ij} = \phi_{ij} - (\phi_0)_{ij}$, and $\tilde \phi_{ij}^* = \phi^*_{ij} - (\phi_0^*)_{ij}$. The expansion of the action around the mean field is thus:
\begin{equation}
    S[\phi, \phi^*]  = S_0 +S^{(1)}[\tilde \phi, \tilde \phi^*]+ S^{(2)}[\tilde \phi, \tilde \phi^*] +\ldots,
\end{equation}
where the first term $S_0 \equiv S[\phi_0, \phi_0^*]$ is an overall constant multiplying the partition function $e^{-N S_0} = Z_0$. The second term $S^{(1)}$, linear in the fields, is equal to zero as we are expanding around the saddle. The third term $S^{(2)}$, quadratic in the fields, is the second-order fluctuation and corresponds to one-loop corrections:
\begin{equation}
    S^{(2)}[\tilde \phi, \tilde \phi^*] = \frac{1}{2} \left.\frac{\delta^2 S}{\delta \phi_{ij} \delta \phi_{nm}}\right|_{\phi_0, \phi_0^*} \hspace{-0.4cm}\tilde \phi_{ij} \tilde \phi_{nm} + \frac{1}{2}\left.\frac{\delta^2 S}{\delta \phi_{ij}^* \delta \phi_{nm}^*}\right|_{\phi_0, \phi_0^*} \hspace{-0.4cm}\tilde \phi^*_{ij} \tilde \phi^*_{nm} + \left.\frac{\delta^2 S}{\delta \phi_{ij} \delta \phi^*_{nm}}\right|_{\phi_0, \phi_0^*} \hspace{-0.4cm}\tilde \phi_{ij} \tilde \phi^*_{nm}. \label{eq:second-order-variation}
\end{equation}

The effective partition function after integrating over these Gaussian fluctuations is then
\begin{equation}
    Z_{\text{eff}} = e^{-N S_0} \int D(\tilde \phi, \tilde \phi^*)~e^{-N S^{(2)}[\tilde \phi, \tilde \phi^*]} = 
    e^{ - N S_0 - N^{0} S_1}, \label{eq:effect-partition}
\end{equation}
which defines the free energy density $f = -\lim_{n_s\to \infty}\frac{1}{n_s} \ln Z_{\text{eff}}$. 
Notice that $Z_\text{eff}$ omits the terms in the free energy that go like negative powers of $N$ coming from higher-loop contributions.
At large $N$, $Z_{\text{eff}}$ counts the weighted monomer-dimer configurations described in~\S\ref{appendix:combinatorial-problems}. 
However, at finite $N$, $Z_\text{eff}$ is not necessarily an integer.

{\bf Model A.} Taking the functional derivative of~\eqref{eq:effective-action-A}, we get 
\begin{equation}
    \frac{\delta S_A}{\delta \phi_{ij}} = \phi_{ij}^* - \frac{\sqrt{A_{ij}}}{x+ \sum\limits_{\vev{ik}}\sqrt{A_{ik}}(\phi_{ik}+\phi^*_{ki}) }, \,\,\,\, \frac{\delta S_A}{\delta \phi^*_{ij}} = \phi_{ij} - \frac{\sqrt{A_{ji}}}{x+ \sum\limits_{\vev{jk}}\sqrt{A_{jk}}(\phi_{jk}+\phi^*_{kj}) }. \label{eq:first-variation-type-A}
\end{equation}

Equation~\eqref{eq:first-variation-type-A} is a system of equations, which can be solved in principle, that highly depends on the graph we are considering. 
For simplicity, we restrict our attention to translation-invariant lattices, for which a spatially uniform solution is naturally the first guess to solve these equations.  
For model A, we shall see that indeed this configuration is free-energetically favorable and provides a stable extremum. 

Using the homogeneous configuration as an ansatz in~\eqref{eq:first-variation}, we see that it is real $(\phi_0 = \phi_0^*)$ and equal to 
\begin{equation}
    \phi_0 = \frac{-x \pm \sqrt{x^2 + 4z}}{2z}, \label{eq:meanfield_type_A}
\end{equation}
where $z$ is the coordination number.

We see that, in \eqref{eq:meanfield_type_A}, two solutions exist. These two mean fields will dominate in different regimes. The positive root solution dominates when $x>0$, while the negative root dominates when $x<0$. 

Let's give some examples. We start with the simplest one, the $1D$ periodic chain. We can find its exact solution using transfer matrix methods and use it to compare it with our saddle point approach. 
We do this in Appendix~\ref{appendix:transfer-matrix}.
As we already know the mean field, we will continue with the second-order fluctuations. And since we are dealing with translation invariant graphs, we can simplify~\eqref{eq:second-order-variation} by going to Fourier space, 
\begin{equation}
    S^{(2)}_{A}[\tilde \phi, \tilde \phi^*] = \frac{1}{2} \sum_{p \in BZ}
    \begin{pmatrix}
        \tilde \phi_p & \tilde \phi_{-p}^*
    \end{pmatrix}
    \begin{pmatrix}
        1+e^{ip}\phi_0^2 & \phi_0^2 \\ 
        \phi_0^2 & 1+e^{-ip}\phi_0^2
    \end{pmatrix}
    \begin{pmatrix}
        \tilde \phi^*_p \\ \tilde \phi_{-p}
    \end{pmatrix}.
\end{equation}
and compute the free energy density 
\be
    f_{1D} =  N(\phi_0^2 - \ln\left|x+2\phi_0\right|) - \frac{1}{2} \ln\bigg(\frac{2}{ 1+\sqrt{1-4\phi_0^4}}\bigg). \label{eq:1D-free-energy-density}
\ee
\begin{figure}[h!]
    \centering
    \includegraphics[width=\textwidth]{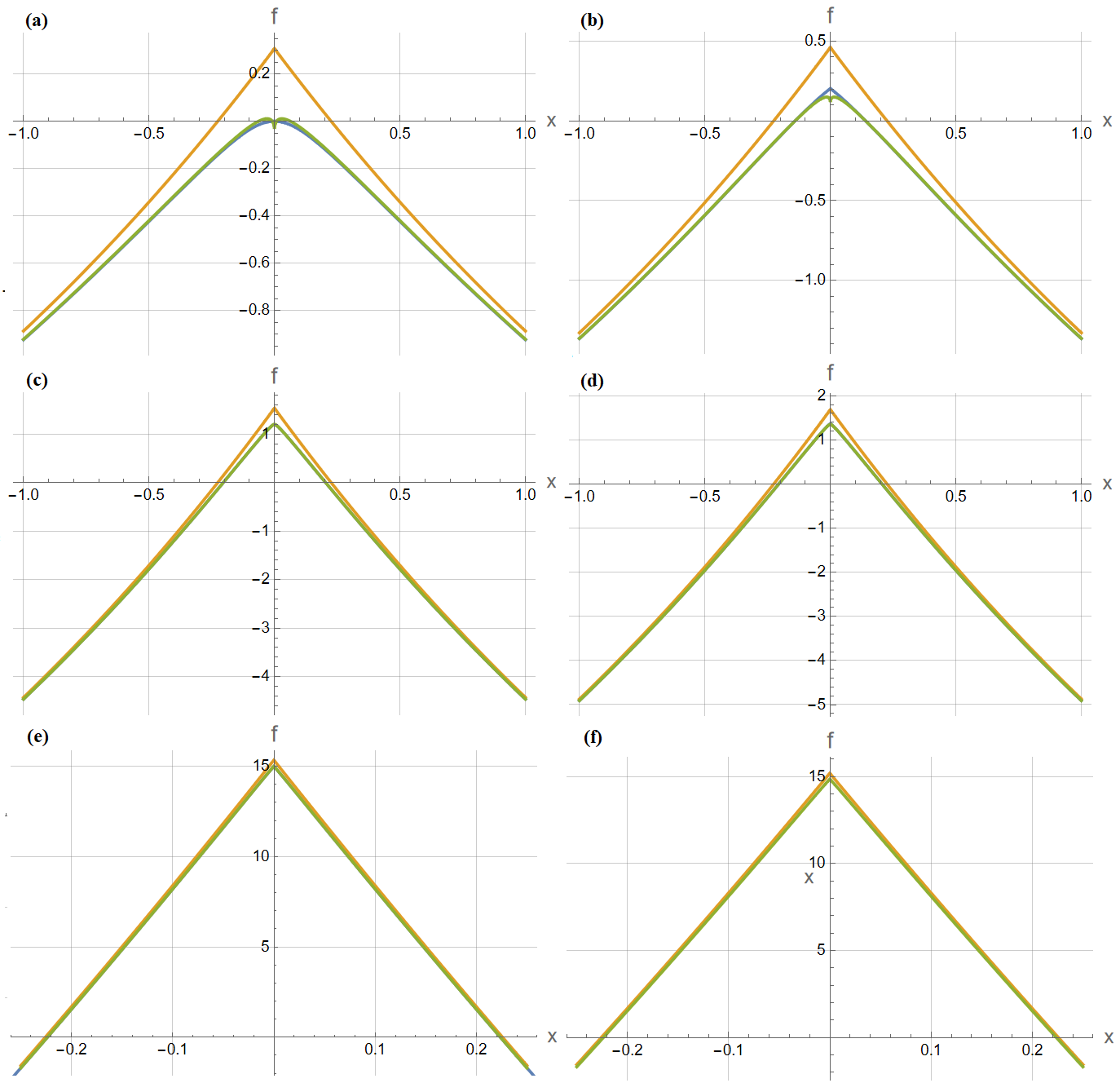}
    \caption{$1D$, type-$A$ free energy density $f$ for the exact solution (blue), mean field (orange) and mean field including the one-loop correction (green) as a function of the coupling $x$ for different number parity: evens (a) $N=2$, (c) $N=10$, (e) $N=100$, and odds (b) $N=3$, (d) $N=11$, (f) $N=99$. Even at small values of $N$, the effective free energy density~\eqref{eq:1D-free-energy-density} represents an extremely well-approximation of the exact solution. }
    \label{fig:coloringA}
\end{figure}
Plots of the effective free energy density compared to the exact solution obtained by transfer matrix methods, for different number of colors $N$, are presented in Figure~\ref{fig:coloringA}. As expected, exact solution (blue) and mean field plus one-loop corrections (green) become indistinguishable as we consider larger values of $N$. As first observed in~\cite{samuel1980-3}, it also gives pretty good results for $N=1$. Notice the sharp difference between odd/even number of colors. Series expansion around $x=0$ shows that for even colors the leading correction is $x^2$ while for odd colors it is $\left|x\right|$.

\begin{figure}[t!]
    \centering
    \includegraphics[width=\textwidth]{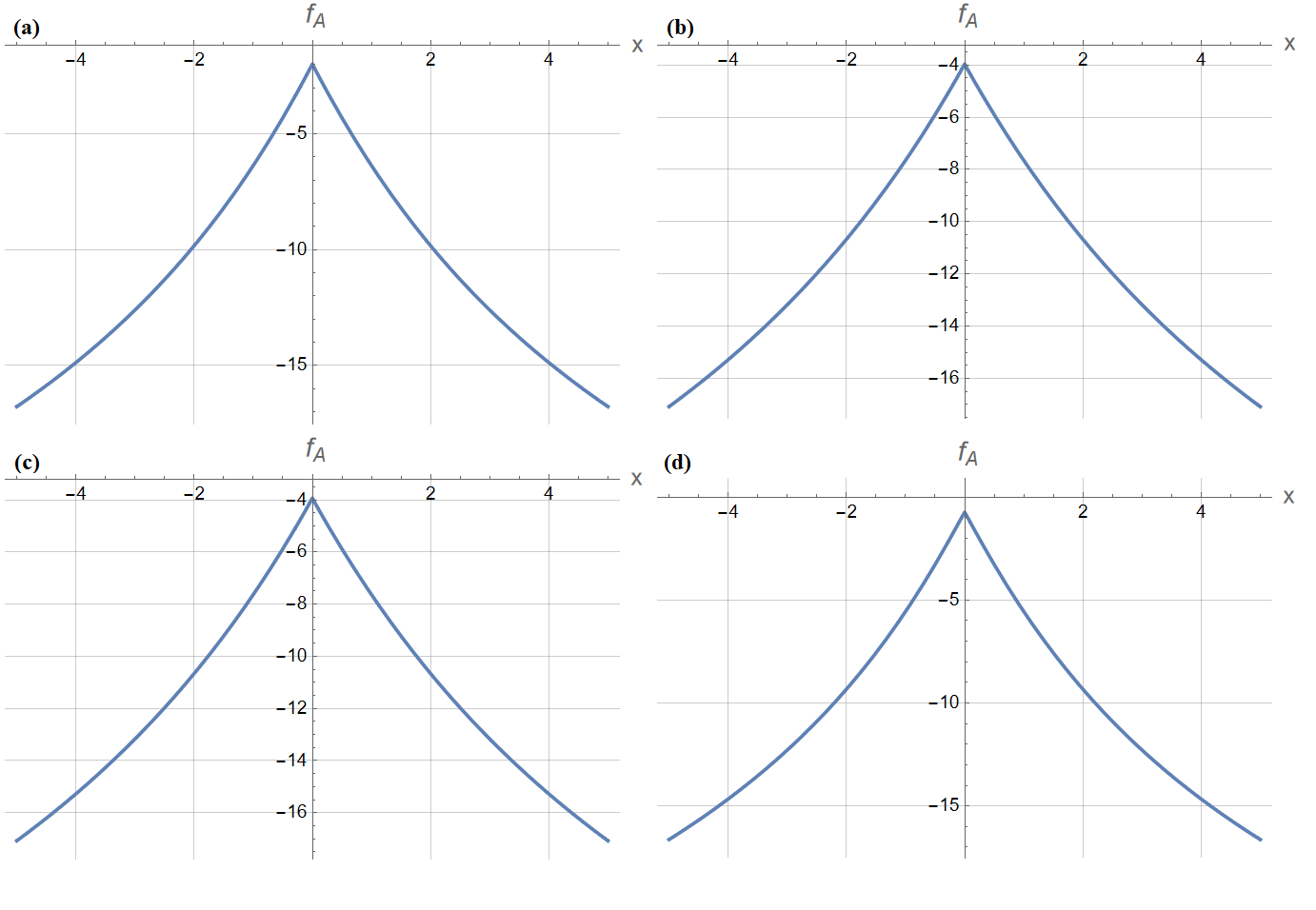}
    \caption{Type-A free energy density $f_A$ for $N=10$ and different lattices. We consider Bravais lattices ((a) to (c)), which follow~\eqref{eq:free-energy-density-A}: (a) square lattice, (b) cubic lattice, (c) triangular lattice; but also (d) honeycomb lattice. As discussed, no phase transition occurs at finite $x$.  }
    \label{fig:free-energy-density-lattices}
\end{figure}

We can do the same procedure for other translation invariant graphs. 
On any lattice, we can diagonalize the Gaussian fluctuations, find its spectrum, and get the effective free energy density. In particular, for a Bravais lattice with coordination number $z$, we find that the free energy density is:
\begin{equation}
    f_A = N \bigg( \frac{1}{2}z \phi_0^2 - \ln\left|x+z \phi_0 \right| \bigg) +  \frac{{\nu_0}}{2} \int_{BZ} \frac{d^dp}{(2\pi)^d} \ln\Big(1+2\phi_0^2 \sum_{i}\cos(\vec p \cdot \hat e_i) \Big), \label{eq:free-energy-density-A}
\end{equation}
where $\nu_0$ is the unit cell volume, and $\hat e_i$ are the primitive vectors. 
In Appendix~\ref{appendix:saddle-point-details}, we give the details of this calculation along with the generalization to the case with separate dimer fugacities for each link direction.

We note that when the monomer fugacity $x \neq 0$, the argument inside the logarithm is positive and thus, the uniform saddle point is stable. The free energy density is analytic in $x$ away from $x=0$ and no phase transition occurs at finite monomer fugacity. Only in the pure dimer limit $(x=0)$ does the argument of the logarithm vanish, and the free energy is singular, in agreement with results of Heilmann and Lieb when $N=1$. 
For any $N$, the deformation to include dimer fugacities for different dimer directions also resolves the singularity (see Appendix~\ref{appendix:saddle-point-details}), as it does for $N=1$
\cite{kasteleyn1963dimer,fisher1963statistical,fendley2002classical}.

{\bf Model B.} Let's proceed with model $B$ and begin by identifying the mean-field configurations of $\phi$ for which the action is minimized. 

\begin{figure}[t]
    \centering
    \includegraphics[width=\textwidth]{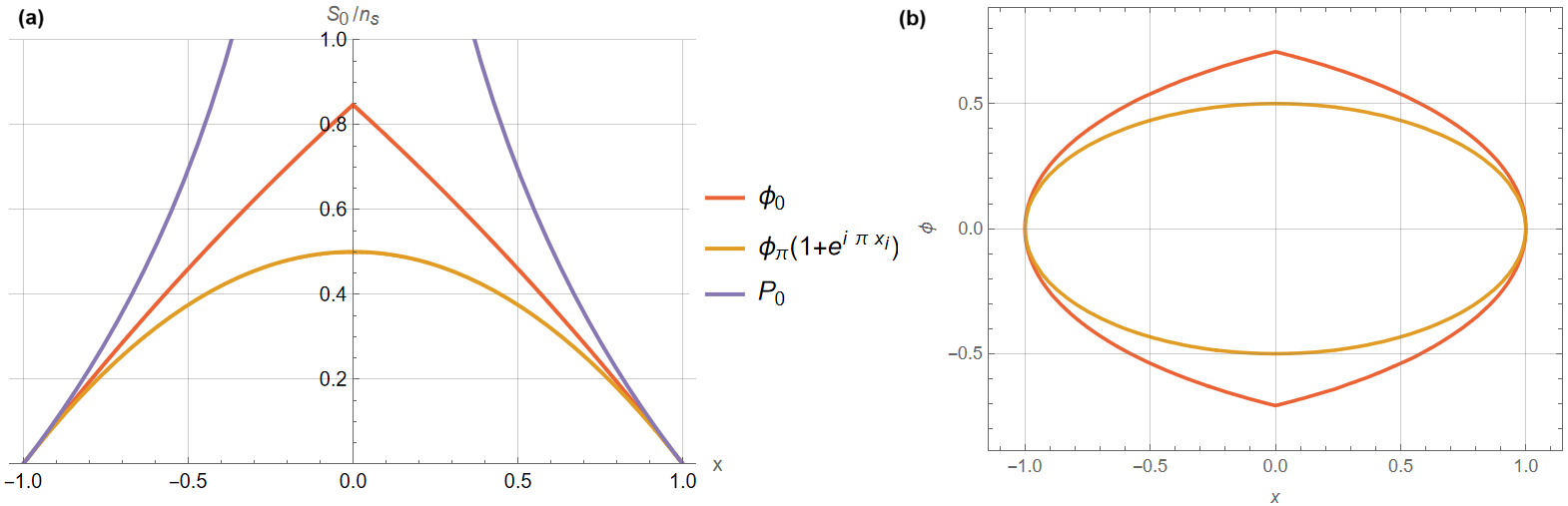}
    \caption{(a)~Leading order term $S_0$ over number of sites $n_s$ in~\eqref{eq:effect-partition} of the $1D$ periodic chain for the homogeneous solution $(\phi_0)_{ij}=\phi_0$ (red) and the textured solution $(\phi_0)_{ij} = \phi_\pi (1+e^{i \pi x_i})$ (orange). We also plot the leading contribution obtained from decoupling in the  alternative channel using the solution $(P_0)_{ij}^{\alpha \beta} = P_0\, \delta_{\alpha \beta}$, as described in Appendix~\ref{appendix:saddle-point-details}.  We see that the alternating mean field is energetically favorable for all values of $x$. (b) Gap equation solutions for the homogeneous $\phi_0$ and alternating mean field $\phi_\pi$ for the 1D periodic chain. }
    \label{fig:action-solutions}
\end{figure}

The first guess would be to try a homogeneous solution, just like in the previous model. However, the homogeneous solution to the mean-field equations turns out to be unstable and is not the dominant saddle point.
Rather, a textured configuration that breaks translation symmetry will be energetically superior. This can be seen in figure~\ref{fig:action-solutions} for the $1D$ periodic chain, where the textured configuration $(\phi_0)_{ij} = \phi_\pi (1+e^{i \pi x_i})$ provides a stable extremum, as we will see. We shall call this configuration the alternating mean field.
We also explore other decoupling channels to compare with this saddle, and find that they are subleading; a detailed analysis of such ansatze is presented in Appendix~\ref{appendix:saddle-point-details}.

Let's continue exploring the periodic chain. The first variation of the action~\eqref{eq:model-B-action} evaluated at a such a configuration leads to the gap equation,  
\begin{equation}
    \left.\frac{\delta S_B}{\delta \phi_{ij}}\right|_{\phi_{\pi},\phi_{\pi}^*} = 2 n_s \phi_{\pi} - \frac{4 \phi_{\pi} }{x^2 + 4\phi_{\pi}^2} \sum_p  \sin^2(p) =0 . \label{eq:gap-equation-B-alt}
\end{equation}
Solving~\eqref{eq:gap-equation-B-alt}, we get that either $\phi_\pi =0$ or $\phi_\pi^2 = \frac{1}{4}(1-x^2)$. We note that the non-trivial solution only exists on the interval $\left| x \right| <1$, while outside this interval $\phi_\pi=0$ is the only solution, as seen in figure~\ref{fig:action-solutions}.

Interestingly, the conclusion is that a phase transition occurs at finite monomer fugacity: there exists a critical value $x_c = \pm 1$ for which $\phi$ starts to condense. We can understand this picture by remembering that $\phi$ at the saddle is proportional to the number of dimers at a link of the graph. At small values of $x$, it is favorable to fully cover the graph with mostly dimers. As the monomer fugacity starts increasing in value, the number of sites covered by monomers increases, reducing the number of dimers per link, until it eventually reaches the critical point $x=1$, beyond which the monomers are free and the system has no motivation to include any dimers.

We proceed next to calculate the second-order fluctuations~\eqref{eq:second-order-variation} around the alternating mean field
for $\left|x\right|<x_c$.  
This configuration $\phi_\pi$ preserves translations by two lattice sites.  Hence, we double the unit cell size and work in Fourier space to get
\begin{equation}
    S^{(2)}_B[\tilde \phi, \tilde \phi^*] = \sum\limits_{p \in (0, \frac{\pi}{2})} \begin{pmatrix}
        \tilde \phi_p & \tilde \phi_{p+\pi} & \tilde \phi_{-p}^* & \tilde \phi^*_{-p+\pi}
    \end{pmatrix}
    \begin{pmatrix}
        1-x^2 & ~0 & ~2\phi_\pi^2 & ~2\phi_\pi^2 \\ 
        0 & 1-x^2 & ~2\phi_\pi^2 & ~2\phi_\pi^2 \\ 
        2\phi_\pi^2 & ~2\phi_\pi^2 & 1-x^2 & ~0 \\ 
        2\phi_\pi^2 & ~2\phi_\pi^2 & ~0 & 1-x^2  
    \end{pmatrix}
    \begin{pmatrix}
        \tilde \phi_p^* \\ 
        \tilde \phi_{p+\pi}^* \\ 
        \tilde \phi_{-p} \\ 
        \tilde \phi_{-p+\pi}
    \end{pmatrix}. \label{eq:singular-matrix-1D-chain-B}
\end{equation}
Yet, looking at its spectrum, this matrix is singular: it has zeromodes with zero eigenvalues for every value of $p$. This is a manifestation of the gauge redundancy $\phi_{ij} \to e^{i (\theta_i + \theta_j)}\phi_{ij}$ of this model. Under the gauge transformation, the action~\eqref{eq:model-B-action} is invariant\footnote{ The action $S_B$ transforms as 
\be
    S_B \to S_B' = \sum_{\vev{ij}} \left| \phi_{ij}\right|^2 - \frac{1}{2} \tr \ln \Bigg(\begin{smallmatrix}
        -\sum\limits_{\vev{ij}} e^{-i(\theta_i + \theta_j)} \phi_{ij}^* (\ket{i}\hspace{-0.05cm}\bra{j}- \ket{j}\hspace{-0.05cm}\bra{i}) & -x\mathbb{1} \\ 
        x\mathbb{1} & \sum\limits_{\vev{ij}} e^{i(\theta_i + \theta_j)} \phi_{ij} (\ket{i}\hspace{-0.05cm}\bra{j}- \ket{j}\hspace{-0.05cm}\bra{i}) 
    \end{smallmatrix}
    \Bigg),
\ee
which using the invariance property of the trace $\tr(A) = \tr(U^\dagger A U)$ under the unitary transformation 
\be
U_\gamma = \begin{pmatrix}
    0 & \sum_j e^{-i \theta_j} \ket{j}\bra{j} \\ \sum_j e^{i \theta_j} \ket{j}\bra{j} & 0 
\end{pmatrix}
\ee
shows the action invariance $S_B' = S_B$.}.

To lift the gauge redundancy, we do gauge fixing, details of which are presented in appendix~\ref{appendix:saddle-point-details}. We present here the result for the free energy density
\be    f_{1D} = \frac{N}{2}(1-x^2) - \frac{1}{4} \ln(8\pi N) + \frac{1}{2} \ln(1-x^2), \quad \left|x\right|<1. \label{eq:first-free-energy-density-1D-B} 
\ee
Remarkably, we get a contribution proportional to $\ln N$. This is a product of the gauge fixing calculation and comes from the measure of the integral as shown in \S\ref{appendix:saddle-point-details}. 
Such violations of the naive expansion in powers of $1/N$ are poorly understood, and it is useful to have more examples.  
Such logarithmic terms are known to arise in systems dual to black holes, as in \eg~\cite{Sen:2012kpz, Sen:2012dw,H:2023qko}, where their microscopic origin \cite{H:2023qko} also involves zeromodes of the path integral.  

The last term in \eqref{eq:first-free-energy-density-1D-B} is actually singular at the transition point $x_c=\pm 1$; we will see below that this is a large-$N$ artifact.

\begin{figure}[h!]
    \centering
    \includegraphics[width=\textwidth]{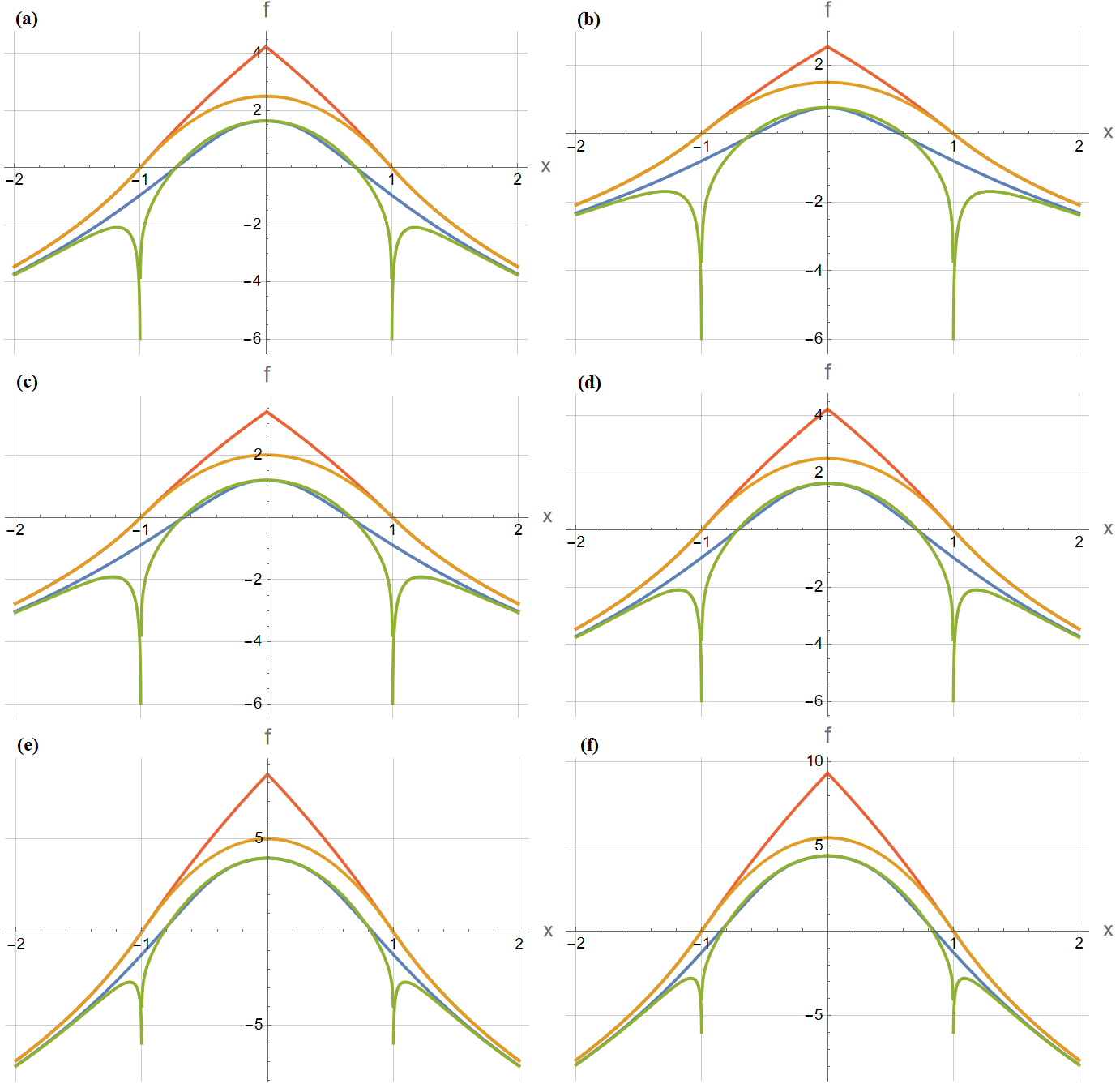}
    \caption{$1D$, type-$B$ free energy density $f$ of the exact solution (blue), alternating mean field (orange), and alternating mean field including the one-loop correction (green) as a function of the coupling $x$. Moreover, the free energy density of the homogeneous mean field (red) is also plotted. Evens (a) $N=2$, (c) $N=4$, (e) $N=10$, and odds (b) $N=3$, (d) $N=5$, (f) $N=11$. The exact solution (blue) does not present the mean-field transition at $x=1$.}
    \label{fig:1d_type_B_free_energy}
\end{figure}

What about the regime where the monomer coupling is $\left|x\right|>x_c$? We know that the solution to the mean-field equations is $\phi = 0$. Hence, we look at the fluctuations~\eqref{eq:second-order-variation} around this mean field, 
\begin{equation}
    S^{(2)}_B[\tilde \phi, \tilde \phi^*] =  \sum_p |\tilde \phi_p|^2 + \frac{1}{2x^2} \tr(B(\tilde \phi)B(\tilde \phi^*)) =  \sum_p\Big(1- \frac{1}{x^2}\Big) |\tilde \phi_p|^2,
\end{equation}
where the $B$ matrix is the block matrix~\eqref{eq:block-matrix-B}, and, in the last equality, we took the trace in Fourier space.   
The free energy is thus  
\begin{equation}
    f_{1D} = -N \ln(x) + \ln\Big(1-\frac{1}{x^2}\Big), \quad \left| x \right| >1, \label{eq:second-free-energy-density-1D-B}
\end{equation}
which shows the same singularity at the boundary $x=\pm 1$ as in~\eqref{eq:first-free-energy-density-1D-B}.

In the case of the 1d periodic chain, an exact solution of model B is available by transfer matrix methods (see Appendix~\ref{appendix:transfer-matrix}).
Figure~\ref{fig:1d_type_B_free_energy} summarizes the mean-field results~\eqref{eq:first-free-energy-density-1D-B} and~\eqref{eq:second-free-energy-density-1D-B} for model B on the $1D$ periodic chain. The free energy density is presented for different numbers of colors, and compared with the exact transfer matrix solution and the homogenous ansatz. We see that the free energy presents a mean-field transition at $x=1$ which the solution by transfer matrix does not possess {\it at any finite $N$}.  

To understand the origin of this singularity, consider the simplest possible graph on which we can put model B: two sites connected by a single link.  
The partition function is (see \eqref{eq:two-sites-coloring} for some explanation)
\be Z_{N, \text{one link}}(x) = x^{2N} \sum_{k=1}^N {1\over N^k k!} x^{-2k} 
C^N_k (k!)^2 
= x^{2N} \sum_{k=1}^N {N! \over (N-k)! N^k} x^{-2k} . 
\ee
Here $k$ is the number of dimers on the link, 
and $C^N_k \equiv {N! \over (N-k)!k!}$.
In the large-$N$ limit, this is 
\be Z_{N, \text{one link}}(x)/x^{2N} \buildrel{N\to\infty}\over{=} \sum_{k=1}^\infty x^{-2k} =  {1\over 1 - 1/x^2} ,\ee
 producing precisely the singularity we saw above.  
 Thus, this singularity is present even in the absence of any thermodynamic limit, and is purely a large-$N$ artifact.  
 It results from the failure of the limits $x\to 1$ and $N \to \infty$ to commute.  
 In the appendix, we show that an expansion in the number of dimers precisely matches the terms in this expansion on other graphs.

We can also study Model B on graphs other than the periodic chain. Again, the mean-field equations lead to a gap equation. There is certainly a homogenous solution, but it is not the only one and generally not the dominant saddle point. Textured mean fields will be energetically superior.  We do not know the correct mean field configuration for a general lattice in the interval $\left|x\right|<x_c$.

However, for $\left|x\right|>x_c$, we can actually find an analytical expression for the effective free energy density by expanding around $(\phi_0)_{ij}=0$. 
Let's discuss the case of a Bravais lattice with coordination number $z$.
The fluctuations~\eqref{eq:second-order-variation} are 
\be
S^{(2)}_B[\tilde \phi, \tilde \phi^*] =  \sum_{\vec p, \hat e_i} |\tilde \phi_p^{(\hat e_i)}|^2 + \frac{1}{2x^2} \tr(B(\tilde \phi)B(\tilde \phi^*)), 
\ee
where now we have $z \over 2$ link fields in the independent directions $\hat e_i$. Expanding the trace term, we get 
\begin{align}
S^{(2)}_B &= \sum_{\vec p, \hat e_i} |\tilde \phi_p^{(\hat e_i)}|^2 + \frac{1}{2x^2 n_s} \hspace{-0.1cm} \sum_{\vec p, \vec k, \hat e_i, \hat e_j} \hspace{-0.2cm} \tilde \phi_{\vec p}^{(\hat e_i)} \tilde \phi_{\vec p}^{*(\hat e_j)}(e^{-i(\vec p + \vec k)\cdot \hat e_i} \hspace{-0.15cm}- e^{i \vec k \cdot \hat e_i})(e^{-i \vec k \cdot \hat e_j}\hspace{-0.15cm}-e^{i(\vec p + \vec k)\cdot \hat e_j}) \label{eq:expansion-B1}  \\
&=  \sum_{\vec p, \hat e_i} |\tilde \phi_p^{(\hat e_i)}|^2  +\frac{1}{2x^2 n_s}  \sum_{\vec p, \hat e_i} (-2 n_s)|\tilde \phi_p^{(\hat e_i)}|^2   =\sum_{\vec p, \hat e_i} \left(1-\frac{1}{x^2}\right)| \tilde \phi_{\vec p}^{(\hat e_i)}  |^2, \label{eq:expansion-B2}
\end{align}
where in getting to the last line, we have summed over the Fourier modes $\vec k$ and used the fact that the only contributing factors will be when $\hat e_i = \hat e_j$.

Integrating the Gaussian fluctuations in~\eqref{eq:expansion-B2}, we find that the free energy density is 
\begin{equation}
    f_B = -N \ln(x) + \frac{z}{2}\ln\left(1-\frac{1}{x^2}\right), \quad \left| x \right| >1, \label{eq:free-energy-density-general-B}
\end{equation}
where we see that the saddle point $(\phi_0)_{ij} = 0$ is stable in the regime $|x|>1$. There is a singularity at the same value of the monomer fugacity $x_c = \pm 1$.

The large-$N$ analysis therefore shows a singularity at $x_c=1$ on any Bravais lattice.
The large-$N$ mechanism we showed above for producing such a singularity on a single link applies in the large-$N$ limit, and we conclude that this singularity is a ubiquitous large-$N$ artifact.  

{\bf Comment on Elitzur's Theorem.} As we noted above, model B enjoys local symmetries under which each site transforms independently $ \eta_i \to e^{ \ii \alpha_i} \eta_i,
\tilde \eta_i \to e^{ - \ii \alpha_i }\tilde \eta_i$, 
and $\phi_{ij} \to e^{ \ii ( \alpha_i + \alpha_j)} \phi_{ij}$.  Since $\phi_{ij}$ transforms nontrivially under these local invariances, Elitzur's Theorem forbids it from getting an expectation value.  
Recall that Elitzur's Theorem \cite{Elitzur:1975im,Batista:2004sc} says that no symmetry acting  only on a finite number of degrees of freedom can be spontaneously broken.   And indeed, because of the local invariance, the saddle point equations do not determine the phase of $\phi_{ij}$.  This variable is thus a collective coordinate of the saddle.  Integrating over this coordinate restores the local invariance, and indeed $\vev{\phi_{ij}}=0$.  However, 
$\vev{|\phi_{ij}|^2}$ does indeed have a macroscopic value that is well-approximated by the saddle-point calculation.  The same logic applies equally well to the standard mean-field analysis of BCS theory \cite{schrieffer2018theory}, where one can find the superconducting (Higgs) groundstate by extremizing an action for a non-gauge-invariant order parameter.

\subsection{Large-$N$ Feynman diagrams}
\label{subsec:large-N-diagrams}

When $x$ is large, we can study our integral by a diagrammatic expansion familiar from perturbative quantum field theory, using the Feynman rules written in \eqref{eq:propagator} and \eqref{eq:vertices}.
The leading contribution to the partition function, independent of the choice of interactions, is \eqref{eq:free-partition-function}. 
This produces a term 
\be \label{eq:bare-free-energy} f_0(x) = - N \log x 
\ee in the free energy density $f \equiv - {1\over n_s} \log Z$.   

As usual we can associate a diagram with a term arising by expanding 
the exponentiated interaction term 
$e^{ A \eta \tilde \eta \eta \tilde \eta }$.  
Disconnected diagrams exponentiate and only connected diagrams contribute to the free energy.

At large $N$, the structure of the corrections to \eqref{eq:bare-free-energy}  due to the interaction terms simplify dramatically.
To identify the dominant diagrams at large $N$, the double-line notation is useful: we wish to find diagrams with the most index loops for a given number of interaction vertices.  

{\bf Model A.} For model A the dominant diagrams are bubble chains, familiar from many vector-large-$N$ field theories, such as the $\mathsf{O}(N)$ model and the Gross-Neveu model \cite{ma1972critical,gross1974dynamical}.
Each propagator in the bubble chain is the full propagator, which at leading order in large-$N$ is a sum of cactus diagrams (shown below).
Let's first use the bare propagator, and then we will see that 
incorporating all the 1PI corrections will be easy.  
Since the propagator is ultralocal, each bubble in the chain is labeled by a single site, and neighboring bubbles must be associated with neighboring sites.
So a diagram with $k+1$ bubbles involves a walk on the graph with $k$ steps.  
On an arbitrary homogeneous graph of coordination number $z$, therefore, the diagram with $k+1$ bubbles contributes
\be \label{eq:type-A-correction}
\parfig{.35}{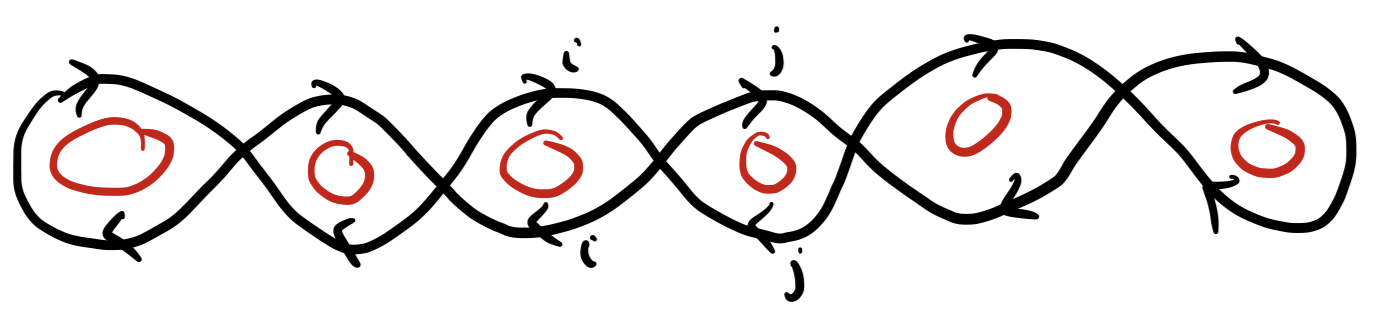} = 
\frac{1}{2} Nn_s \( -{  \lambda z \over x^2 }\)^k  \ee
where $\lambda$ is the 't Hooft coupling (set to 1 above).
(In more detail: The vertices give a factor of $ ( - \lambda / N)^n$.  The index loops give $N^{n+1}$.   
The $2n$ propagators give $ x^{-2n}$.
The choice of labels on the edges (the number of $n$-step walks on the graph) give $ n_s z^n$.  
The symmetry factor is 1/2 from a reflection in the center of the graph.)
The sum of these is 
\be f_\text{bare}(x, \lambda) = - \frac{1}{2} N \sum_{k=1} \( - { \lambda z \over x^2 }\)^k 
= - {1\over 2} N  \(- { { z \lambda \over x^2 } \over 1 - { z \lambda \over x^2 } } \).  \ee

Now let's correct the propagator. The self-energy $\Sigma$ and the full propagator $\Delta$ satisfy the Schwinger-Dyson equations
\begin{align}  \label{eq:large-N-SD-equations-model-A}
\Delta &= { 1\over x - \Sigma}, ~~~~~~~ 
\parfig{.5}{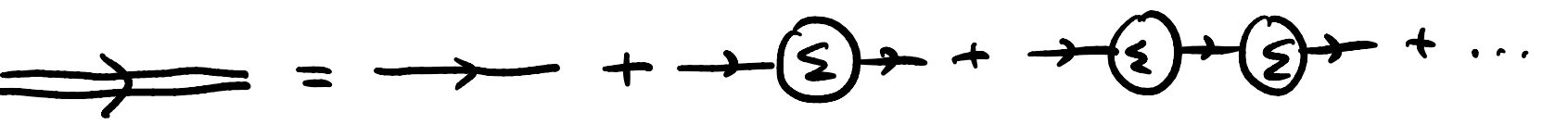}
\\ 
\nonumber
\Sigma_{ij} &= -\delta_{ij} \sum_{\ell } A_{i\ell} \Delta_{\ell\ell} + \CO(1/N)  , ~~~~~~ 
\parfig{.5}{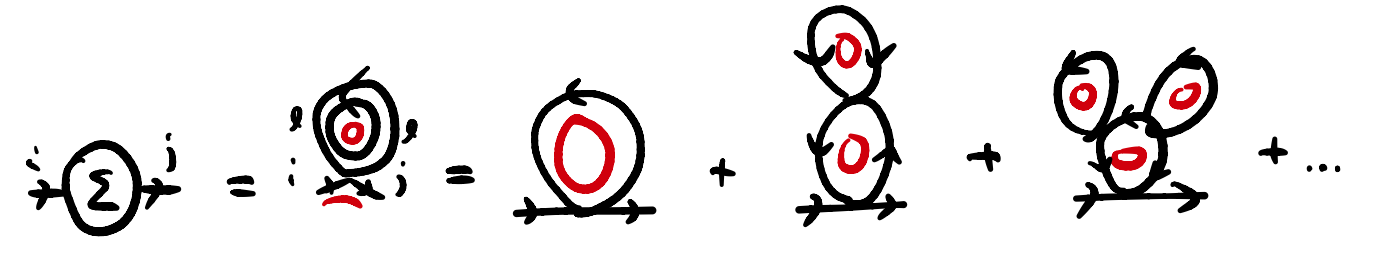}
\end{align}
where we have suppressed color and site indices when possible.  
The second equation shows that at leading order in $N$ only the diagonal entries in the self-energy are nonzero: $\Sigma_{ij} = \delta_{ij} \Sigma_j $.  
We note that the vertex appearing in the 1PI self-energy is the bare vertex; putting the full corrected vertex described below would double count diagrams.
Therefore the Schwinger-Dyson equations reduce to
\be \Sigma_i  = - \sum_\ell { A_{i\ell} \over x - \Sigma_\ell}. \ee
We do not know how to solve this equation in general.  For a translation-invariant graph (of volume $V$), the Green's function is 
\be A^{-1}_{j\ell} = {1\over V} \sum_k e^{ \ii k ( j-\ell) }D(k) , 
\ee
and therefore the SD equation is 
\be
{1\over V} \sum_k e^{ \ii k j } D(k) \tilde \Sigma(k) = - {1\over x - \Sigma_j}. 
\ee
Assuming translation invariance of the solution, $\Sigma_j  = \Sigma, \tilde \Sigma(k) = \delta_{k,0} V \Sigma$, this gives 
\be \label{eq:quadratic-equation-for-Sigma}
\Sigma(x - \Sigma) = - z  
\ee
where $ z = { 1\over D(0)} $ is the coordination number of the graph.
This is a quadratic equation for $\Sigma$ whose solution is
\be
\Sigma = \frac{1}{2} \( x \pm \sqrt{x^2 +4z}\)
\ee
Two solutions to the SD equation exist, just as in~\eqref{eq:meanfield_type_A}. To decide which of the two we shall use, we look at the one that minimizes the action (negative root for $x>0$, and positive root for $x<0$). Looking back at the form of the full propagator in \eqref{eq:large-N-SD-equations-model-A}, we see that the end result of the propagator corrections from cacti is simply to replace $x \to x - \Sigma$.

The same correction needs to be done to the vertex. At large $N$, the full vertex $\Gamma$ and the bare vertex $\Gamma^0_{ij} \equiv { A_{ij} \over 2N} $ satisfy the Bethe-Salpeter equation
\begin{align}
&\Gamma_{ij} = \Gamma^0_{ij} + \sum_{k, l} \Gamma^0_{ik}  \Delta_{kl}  \Delta_{kl}  \Gamma_{lj} ~~~ 
\\ & 
~~~ 
\parfig{.8}{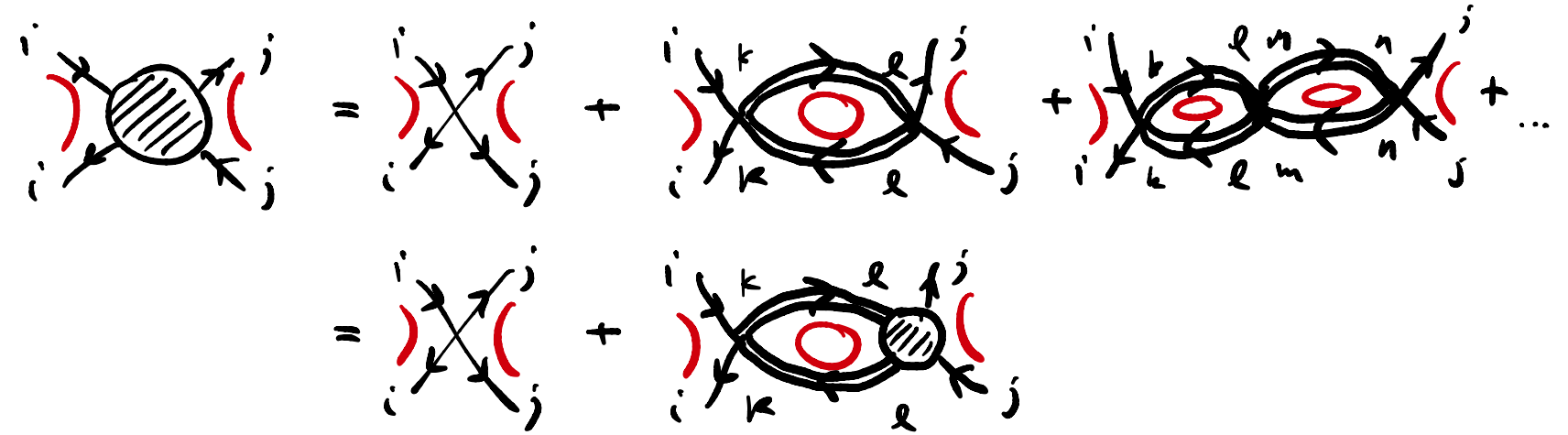}
\nonumber
\end{align}
where color indices have been suppressed. Assuming that the full vertex is independent of the color indices and using the form of the full propagator~$\Delta$, we can simplify the equation to
\be
\Gamma_{ij} = \frac{A_{ij}}{2N} + \frac{z}{(x- \Sigma)^2} \Gamma_{ij},
\ee
whose solution is 
\be
\Gamma_{ij} =  \(\frac{(x-\Sigma)^2}{ (x-\Sigma)^2 -z}\) \frac{A_{ij}}{2N} \equiv \alpha  \frac{A_{ij}}{2N} .
\ee
Making the comparison with the bare vertex, we see that the corrections amount to a rescaling by a factor of $\alpha$. This change can be taken into account by simply rescaling the 't Hooft coupling $\lambda \to \alpha \lambda$. 

All these corrections together make the free energy density to be 
\be
f_A(x) = f_0(x- \Sigma) + f_{\text{bare}}(x- \Sigma, \alpha \lambda),
\ee
which is the same as Eq.~\eqref{eq:free-energy-density-A} that we found using saddle point techniques to leading order in $1/N$.

{\bf Model B.} In model B, the dominant diagrams at large $N$ have a different structure.  
They are instead rings of bubbles, where each bubble is associated with {\it the same} pair of neighboring sites.
The contribution from the diagram with $k$ bubbles is 
\be 
\parfig{.2}{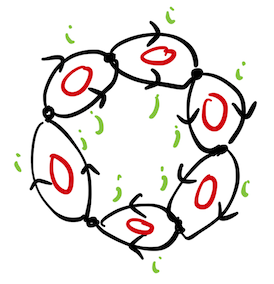} = 
 n_\text{links} {1\over k} \( { \lambda \over x^2 }\)^k  ,\ee
where $n_\text{links} = n_s z /2 $ is the number of links in the graph.
The factor of $1/k$ is a symmetry factor from cyclic permutations of the bubbles.
Notice that in this case, the correction is of order $N^0$, in contrast with the expression for type $A$ \eqref{eq:type-A-correction}.
In Model B the propagator and vertex corrections are down by powers of $N$.

The sum of these diagrams (for a homogeneous graph of valence $z$) is 
\be {n_sz \over 2} \sum_{k=1} {1\over k} \( { \lambda  \over x^2 }\)^k 
= {n_sz \over 2} \log \( 1 - { \lambda \over x } \).  \ee
Therefore
\be 
\Delta f_B 
=  {z \over 2}  \log \( 1 - { \lambda  \over x^2 }\), \ee
which is the same as we found in~\eqref{eq:free-energy-density-general-B}. The singularity for type B occurs at $x=\pm 1$ independent of the graph.

{\bf Model C.} For a graph with no self-loops ($A_{ii} = 0, \forall i$), 
models B and C are the same at leading order in large $N$.
The dominant diagrams for model $C$ are: 
\be \parfig{.2}{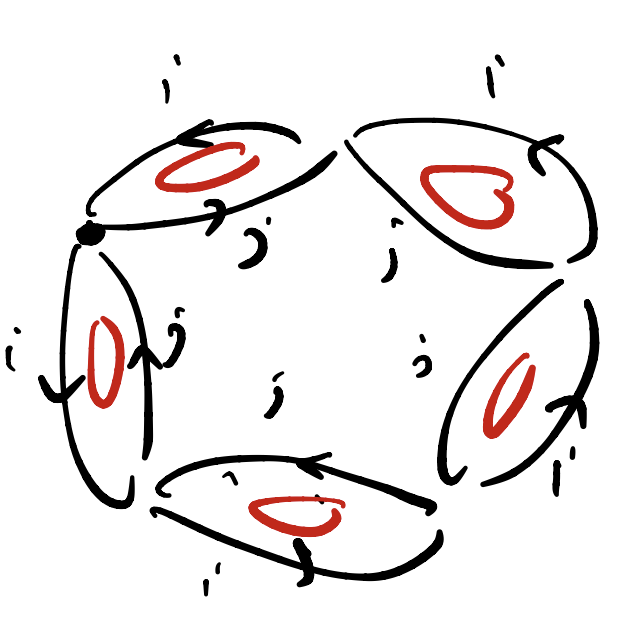}
\ee
They only differ by the orientations of the lines.

{\bf Model D.} The large-$N$ diagram story for model D is similar to models B and C, except that the labels on the edges must alternate between successive bubbles,
as in: 
\be \parfig{.2}{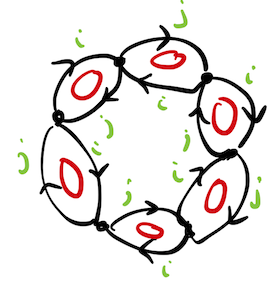}
\ee
This means that 
only rings with an even number of bubbles contribute, giving
\be 
\Delta f_D
=  {z } \log \( 1 - 
\( {\lambda \over x^2 }\)^2
\). \ee

\section{Graph coloring problems}\label{sec:graph-coloring}

Closely-related ideas may be used to find integral representations for various graph-coloring problems.  For example, we can ask how many ways are there to 3-color the links of the honeycomb lattice so that no links sharing a vertex are the same color?  This problem was solved exactly by Baxter \cite{baxter1970colorings}, 
is equivalent to the 
$T=0$ limit of the Kagome lattice classical $\gU(1)$ antiferromagnet \cite{huse1992classical}, 
is a regularization of the $\gSU(3)_1$ CFT \cite{kondev1996kac,kondev1996operator}, and has various generalizations and possible experimental realizations \cite{castelnovo2004dynamical, verpoort2018color};
quantum versions have also been studied \cite{castelnovo2005quantum,castelnovo2005quantum2}.
The honeycomb lattice 3-coloring problem is also closely related to ensembles of fully-packed loops, since if in a given coloring we regard one of the colors as nothing, and then the other two form a collection of closed loops.  

We give several integral representations for such problems.  
The first and simplest counts 3-colorings of {\it planar} 3-valent lattices, because it takes advantage of Kasteleyn's signed matrix.  
We then find a related integral that works on an arbitrary 3-valent graph, but counts colorings with a certain sign.  
In the special case of the honeycomb lattice, this modification is likely still solvable by Baxter's methods.
Finally, we combine all of these ideas to find a grassmann integral representation for the 3-coloring problem (without sign) on an {\it arbitrary} 3-valent graph.  
In each case we use field theory techniques to give an estimate of the answer.

\subsection{Planar 3-coloring problem} \label{subsec:planar-three-coloring}

Place $N_c=2$ species of grassmann variables $ \eta^{\alpha=R,G}_i$ at each site of a 3-fold coordinated graph.  
Let $ \tilde A_{ij}$ be a signed adjacency matrix of the (planar) graph with the following properties:
\begin{itemize}
\item $ \tilde A_{ij} = - \tilde A_{ji}$
\item $ \prod_{\ell \in \partial p} \tilde A_\ell = - 1$, where $p$ is any elementary plaquette.  
\end{itemize}
This is only possible for planar graphs and is equal to Kasteleyn's matrix, \ie~a choice of $\IZ_2$ gauge field configuration with $\pi$-flux through every plaquette, up to a gauge transformation. 
Then the following integral computes the number of 3-colorings of the graph:
\be Z = \int \prod_{\alpha, i} d \eta_i^\alpha \prod_{\vev{ij} }
( 1 +  \tilde A_{ij} \eta_i^\alpha \eta_j^\alpha ) = \int D\eta~ e^{ S[\eta]}.  \label{eq:3-coloring-partition}\ee
(As above, we use the summation convention for the color indices, but not for the lattice site indices.)
The idea is again that a nonzero contribution comes only from terms in the expansion of the product where each mode appears once.  
Each such term involves a choice for each link of either $1$ or $\eta_i^R \eta_j^R$ or $\eta_i^G \eta_j^G$, which we can regard as the three colors.  
A choice for a given link is incompatible with the same choice for any link sharing a vertex with it.

In order to show that the integral~$\eqref{eq:3-coloring-partition}$ does count the number of 3-colorings, we need to show that all contributions come with the same sign. A simple way to see that this is in fact true is by noting that nonzero terms $\mathcal{C}$ are given by 
\be
    Z = \int \prod_{\alpha, i} d\eta_i^\alpha~ \sum_{\mathcal{C}} \prod_{\vev{i,j}\in l_R(\mathcal{C})} \tilde A_{ij} \eta_i^R \eta_j^R \prod_{\vev{m,n}\in l_G(\mathcal{C})} \tilde A_{mn} \eta_m^G \eta_n^G, 
\ee
where $\{l_R(\mathcal{C}),l_G(\mathcal{C})\}$ are the set of all links colored $R,G$ in the 3-coloring configuration $\mathcal{C}$. Since in this expression the grassmanns come in pairs we can permute them and integrate each color independently. 
This is the same form in which contributions from a given dimer covering appear in Kasteleyn's integral representation (though here not all of them appear).  From Kasteleyn's argument we know that each term comes with the same sign, and hence all colorings contribute with the same sign.


We can represent the integrand of Eq.~\ref{eq:3-coloring-partition} in terms of the action
\be\label{eq:3-coloring-action}
S[\eta]  = 
\sum_{\vev{ij}} \log ( 1 + \tilde A_{ij} \eta_i^\alpha \eta_j^\alpha ). \ee
The nice thing about $N_c=2$ is that $ \( \eta_i^\alpha \eta_j^\alpha\)^3 = 0 $,
which means that the Taylor expansion of the logarithm terminates at the quartic terms:
\be S[\eta]  = 
\sum_{\vev{ij}} \( \tilde A_{ij} \eta_i^\alpha \eta_j^\alpha - \half 
A_{ij} \eta_i^\alpha \eta_j^\alpha \eta_i^\beta \eta_j^\beta \) \label{eq:first-action-3-coloring}  \ee
where we used the fact that $\tilde A_{ij} \tilde A_{ij} = A_{ij}$ is the unsigned, symmetric adjacency matrix.

We observe that the model specified by \eqref{eq:3-coloring-action} on a bipartite lattice has a manifest $\gSU(2)$ symmetry that acts by 
\be \eta_A^\alpha \to U^\alpha_\beta \eta_A^\beta, \quad  
\eta_B^\alpha \to (U^\dagger)^\alpha_\beta \eta_B^\beta, \ee
for $A,B$ sites in the $A$ and $B$ sublattices.  
This is thus an exact symmetry of the 3-coloring partition function on a bipartite lattice.
This is a subgroup of the $\gSU(3)$ symmetry that emerges in the continuum limit as the model approaches the $\gSU(3)_1 $ WZW model fixed point.

This model is also invariant under time-reversal. That is, there exists an antiunitary operator $\mathcal{T}$ which is a symmetry of the action, satisfying $\mathcal{T}^2=-1$ and $\mathcal{T}^{-1} i \mathcal{T} = -i$. This operator acts on the fields as:
\be
    \eta_A^\alpha \to \mathcal{T}^{-1} \eta_A^\alpha \mathcal{T} = \eta_B^\alpha, \quad \eta_B^\alpha \to \mathcal{T}^{-1} \eta_B^\alpha \mathcal{T} = -\eta_A^\alpha.
\ee
Note that this is only possible when the model is defined on a bipartite lattice.

{\bf Interpolation between 3-coloring and dimer models.} 
The quartic interaction has a simple mechanical interpretation.  In its absence, the model would factorize into $Z_1^2$, two copies of the dimer model.  Call the two kinds of dimers red $R$ and green $G$. $Z_1^2$ includes configurations where a single link has both a red and a green dimer, associated with the second-order term term in the expansion of the exponential $e^{ \tilde A_{ij} \eta_i^\alpha \eta_j^\alpha }$.  
The role of the quartic term is precisely to cancel these contributions.

In fact, we can interpolate between these models by deforming~\eqref{eq:first-action-3-coloring} as:
\be S[\eta] = \sum_{\vev{ij}} \( \tilde A_{ij} \eta_i^\alpha \eta_j^\alpha 
- { u^2\over 2 } A_{ij} \eta_i^\alpha \eta_j^\alpha \eta_i^\beta \eta_j^\beta\). \label{eq:deformed-first-action}
\ee
By varying the coefficient $u^2$, this integral can interpolate between two copies of the dimer model $(u=0)$, the honeycomb 3-coloring model $(u=1)$, and a single dimer model $(u \to \infty)$\footnote{To make explicit the connection to the single dimer model at $u\to\infty$, let us do the following change of variables $\eta_i^R \to \eta_i$ and $\eta^G_i \to \tilde \eta_i$ to get:
\be
    S[\eta, \tilde \eta] = \sum_{\vev{ij}} \( \tilde A_{ij} (\eta_i \eta_j + \tilde \eta_i \tilde \eta_j) +u^2 A_{ij} \eta_i \tilde \eta_i \eta_j \tilde \eta_j \). 
\ee
We can move the $u^2$-coupling to the Gaussian term by rescaling $\eta \to  \frac{\eta}{\sqrt{u}}, \tilde \eta \to \frac{ \tilde \eta}{\sqrt{u}}$: 
\be
    S[\eta, \tilde \eta] = \sum_{\vev{ij}} \(\frac{1}{u}\tilde A_{ij} (\eta_i \eta_j + \tilde \eta_i \tilde \eta_j) + A_{ij} \eta_i \tilde \eta_i \eta_j \tilde \eta_j \).
\ee
Hence, in the limit $u \to \infty$, only the quartic term survives and we recognize the action from \eqref{eq:one-dimer}.}.

Following the ideas discussed previously, we study the action $S[\eta]$. That is, decouple the quartic term in~\eqref{eq:deformed-first-action} using an auxiliary link field $\phi_{ij}$ and do mean-field theory. 
We note that 
this mean field theory becomes exact in a limit where $N\to \infty$, but only when $N+1=z$, the coordination number of the lattice, does the model count colorings.
The HS transformation that decouples the quartic interaction is
\be
e^{-{u^2\over 2} \sum_{\vev{ij}} A_{ij} \eta_i^\alpha \eta_j^\alpha \eta_i^\beta \eta_j^\beta} = \int D \phi~e^{\sum_{\vev{ij}} -\frac{1}{2} \phi_{ij}^2 +i u \tilde A_{ij} \phi_{ij} \eta_i^\alpha \eta_j^\alpha }, \label{eq:HS-planar-3-coloring}
\ee
where the measure is $D\phi \equiv \prod_{\vev{ij}} \frac{d\phi_{ij}}{\sqrt{2\pi}}$.
Solving the equation of motion, we see that $\phi$ at the saddle is proportional to the number of dimers via $\phi_{ij}^{\text{saddle}} = i u\sum_{\alpha} \tilde A_{ij} \eta_i^\alpha \eta_j^\alpha$. This field is symmetric under exchanging site indices $\phi_{ij} = \phi_{ji}$ and, most importantly, it is odd under time reversal $\phi_{ij} \to -\phi_{ij}$.

After integrating out the $\eta$ variables, we can rewrite the partition function as an integral over the link variables $\phi$ whose action $S[\phi]$ is given by 
\be
S[\phi] = \sum_{\vev{ij}} \frac{1}{2} \phi_{ij}^2 - \tr \ln\left( \tilde A_{ij} (1+iu \phi_{ij})\right). \label{eq:action-deformed-u-model}
\ee
We can study this link-field representation using saddle point techniques. We proceed to do mean-field theory for the honeycomb lattice.  Assuming the mean field is homogeneous $\phi_0$, the action for such configuration is: 
\be
    S[\phi_0] = \frac{3}{4}n_s \phi_0^2  -\frac{1}{2} n_s \nu_0  \int_{BZ} \frac{d^2p}{(2\pi)^2} \ln((1+iu\phi_0)(3+2\cos(p_1) - 2\cos(p_2) -2 \cos(p_1 - p_2)), \label{eq:action-tree-level-planar-3-colorings}
\ee
where $n_s$ is the number of sites, $\nu_0$ is the unit cell volume, and $p_i \equiv \vec p \cdot \hat e_i$ where $\hat e_i$ are the primitive lattice vectors. Its first variation with respect to this mean field is 
\be 
    \left.\frac{\delta S}{ \delta \phi_{ij}} \right|_{\phi_0}  = \frac{3}{2} n_s \phi_0 - \frac{n_s}{2}\frac{iu}{(1+i u\phi_0)}=0, \label{eq:mean-field-equations-deformed-u-model}
\ee
where $n_s$ is the number of sites, and whose solutions are $\phi_0^{\pm} = \pm \frac{\sqrt{12 u^2 - 9}}{6u} + \frac{1}{2u} i$. Note that although the location of the saddle point does not depend on the lattice, the free energy does. 
\begin{figure}[h!]
    \centering
    \includegraphics[scale=0.7]{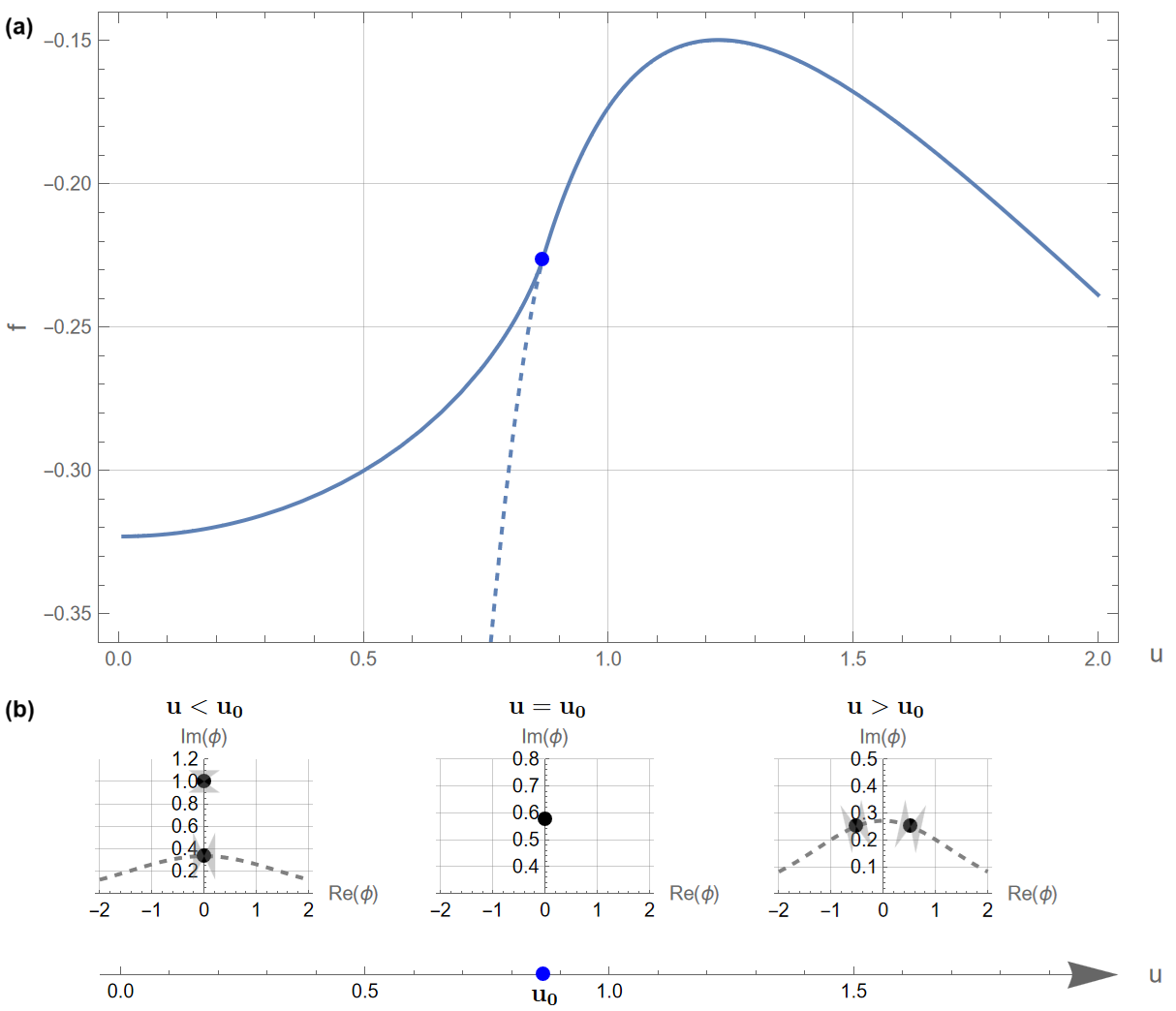}
    \caption{(a) Free energy density as a function of $u$. Two phases are separated by a phase transition occurring at the blue point $u_0 = \frac{\sqrt{3}}{2}$. For $u>u_0$, both saddles contribute, but for $u<u_0$, only one saddle contributes (solid line) while the other saddle does not (dashed line). (b) Behavior of the saddle points in the complex plane as function of $u$. For $u>u_0$, we can deform the integration contour (dashed line) to access both saddles through the allowed steepest descent region (cone). At $u=u_0$, both saddles collide. For $u<u_0$, only one of the saddles can be accessed.}
    \label{fig:free-energy-u}
\end{figure}

We find a phase transition at $u_0 = \frac{\sqrt{3}}{2}$ where two branches of saddle points collide.  For $u>u_0$, a pair of saddle points related by time-reversal symmetry both contribute, while for $u<u_0$, only one of the two saddles contributes. This is because we cannot deform the integration contour to pass through both saddles (see Appendix~\ref{appendix:coloring-details}).    
In this sense, we can say that time-reversal symmetry is spontaneously broken for $u>u_0$, and restored for $u<u_0$. It would be interesting to understand the nature of the critical theory; we leave this for future work.

Figure~\ref{fig:free-energy-u} shows a summary of these results. Some interesting limits of the free energy density to consider are: At $u\to 0$, the result approaches the exact value of two copies of the dimer model $\lim_{u \to 0} f(u) = 0.323$. At $u \to 1$, the approximate value we found for the 3-coloring model is $\lim_{u\to 1}f(u)= -0.173$,
which we can compare with Baxter's exact result $f_{\text{Baxter}}=-0.190$~\cite{baxter1970colorings}. Finally, at $u \to \infty$, the leading behavior is $\lim_{u \to \infty} f(u) \propto -\ln(u)$ coming from the coupling of the dimer term; the subleading term independent of $u$ reproduces the correct free energy for a single dimer model.

As an amusing aside, consider the following expansion of the integrand of this represenation of $Z$.
Let's call the two colors red and green.
Using the fact that 
$ \half \eta^\alpha_i \eta^\alpha_j \eta^\beta_i \eta^\beta_j 
= \eta^R_i \eta^R_j \eta^G_i \eta^G_j $, 
\bea Z & = & \nonumber
\int D\eta ~
e^{ \sum_{ij} \tilde A_{ij} \eta_i^\alpha \eta_j^\alpha }
\prod_{\vev{ij}} \( 1 - u^2 A_{ij} \eta_i^R \eta_j^R \eta_{i}^G \eta_j^G  \)
\\ \nonumber
&=& 
Z_\text{dimer}^2 - u^2 \sum_\ell Z_\text{dimer}^2(\setminus \ell)
+ u^4 \sum_{\ell, \ell'} Z_\text{dimer}^2(\setminus \ell, \ell')
+ \cdots 
\\ & =& \det(\tilde A) - u^2 \sum_\ell \det(\tilde A_\ell)
+ u^4 \sum_{\ell, \ell'} \det(\tilde A_{\ell, \ell'}) + \cdots 
\label{eq:expansion-in-u}
. \eea
Here $Z_\text{dimer}(\setminus \ell...)$ indicates the dimer model on the graph with the links $\ell...$ removed.  
In the last line we used $\Pf^2(X) = \det(X)$.  
It is interesting to compare this expansion with the formula for the characteristic polynomial of the Kasteleyn matrix
\cite{horn1990matrix}
\be \label{eq:characteristic-polynomial}
\det (1 - u \tilde A) = \sum_{k=0}^{n_s} (-u)^{k} 
\sum_{G_k} \det \tilde A_{G_k} \ee
where the sum is over all subsets $G_k$ of $k$ sites, 
and $A_{G_k}$ is the matrix of minors obtained by removing the associated rows and columns.  
Note that $\det A_{G_k}$ vanishes when $k$ is odd, 
since there are no dimer coverings of a lattice with an odd number of sites.  
But \eqref{eq:expansion-in-u} 
and 
\eqref{eq:characteristic-polynomial} 
are not the same because 
the latter is a sum over {\it all} collections of pairs of sites,
while the former only removes {\it adjacent} sites.
It is possible that the similarity between these two expressions can be used to provide a useful bound.

\subsection{An integral representation for a signed sum of colorings of an arbitrary 3-valent graph} \label{subsec:signed-3-colorings}

We come back to equation~\eqref{eq:3-coloring-partition} but now defined over $N_c=3=z$ species of grassmann variables $\eta_i^{\alpha=R,G,B}$. A nonzero contribution to this integral comes only from products of grassmann at each link $\vev{ij}$ of either $\eta_i^R \eta_j^R$, $\eta^G_i \eta_j^G$ or $\eta^B_i \eta^B_j$. We regard these as our three colors. 
That is,
\be 
Z_{N_c =z} = \int \prod_{\alpha, i} d \eta_i^\alpha \prod_{\vev{ij} }
( a + A'_{ij} \eta_i^\alpha \eta_j^\alpha )  
= \prod_{\vev{ij}} A'_{ij} \int \prod_{\alpha, i} d \eta_i^\alpha \prod_{\vev{ij} }
( \eta_i^\alpha \eta_j^\alpha )  \label{eq:Ncequalsz} \ee
is independent of $a$ when $N_c = z$, the coordination number of the graph. 
The second equality in \eqref{eq:Ncequalsz} follows because, for $N_c =z$, a nonzero contribution only happens when each link contributes exactly two $\eta$s to the integral.  
Furthermore, in this representation, 
the dependence on the $A'$ matrix simply factors out.

Note that the $A'$ matrix is antisymmetric: it involves an arbitrary choice of orientation of each link $\vev{ij}$. Thus, it defines a directed graph where $A'_{ij} = -A'_{ji} = 1$ iff the arrow goes from the site $i$ to site $j$. This is to be contrasted with Kasteleyn's matrix $\tilde A_{ij}$ which further demanded $\pi$-flux in any plaquette and only exists for planar graphs, and also with the adjacency matrix  $A_{ij}=A_{ji} = 1$ whenever there is a link between sites $i$ and $j$. 
Thus, a complete definition of \eqref{eq:Ncequalsz} requires an orientation of each link; however, different choices of orientation merely change the partition function by an overall sign.

So we are led to consider the integral 
\be \int \prod_{i,\alpha=R,G,B}d\eta_i^\alpha \prod_{\vev{ij} } \( \sum_\alpha A_{ij}' \eta_i^\alpha \eta_j^\alpha\) ~.\ee
Indeed the terms in the expansion of this integral are in one-to-one correspondence with 3-colorings of the graph, and each gives a real number of unit magnitude.  However, their contributions are not all of the same sign, as we will see.

In any 3-coloring, we can forget about one of the colors; the remaining two colors form a collection of loops that alternate between the colors.  Hence, these loops necessarily have an even number of links.
The contributions to our integral from two configurations related by flipping the colors along such a 2-color loop have the same sign, since they involve an even number of grassmann flips:
\be \eta_1^R\eta_2^R \eta_2^G \eta_3^G \cdots \eta_{2n}^G \eta_1^G = 
\eta_1^G\eta_2^G \eta_2^R \eta_3^R \cdots \eta_{2n}^R \eta_1^R . \ee
Many colorings are related to each other by such 2-color loop flips (which are used as the Monte Carlo move in simulations of the coloring model such as those in \cite{castelnovo2004dynamical}).  

However, it is known that these 2-color loop flip moves, even when including flips of non-contractible loops, are not in general quite ergodic on the configuration space of colorings \cite{2011PhRvB..84x5119F,huse1992classical,cepas2017colorings}\footnote{\cite{2011PhRvB..84x5119F} identifies an invariant (related to the Hopf invariant) on non-planar graphs. 
\cite{huse1992classical} asserts this without explanation in footnote 13.}.
For our purposes, the key observation \cite{cepas2017colorings} 
is that a certain {\it chirality} $(-1)^m$ associated with each coloring 
is conserved under the two-color loop updates. 
In a given coloring, we assign the site $i$ spin value $\sigma_i=+1$ if the colors of the adjacent links are RGB when going clockwise around the site (or a cyclic permutation thereof), and spin $\sigma_i=-1$ otherwise.  
The chirality is defined to be\footnote{Note that \cite{cepas2017colorings} and \cite{baxter1970colorings, castelnovo2004dynamical} disagree about whether the sign of the spin of a site oriented RGB should be the same or opposite on the two sublattices.  Fortunately, this difference of convention only affects the definition of $(-1)^m$ by a factor of $(-1)^{N_B}$, where $N_B$ is the number of sites in the $B$ sublattice, which is the same for every coloring.}
\be (-1)^m \equiv (-1)^{\sum_i \sigma_i/2}.\ee  

\begin{figure}
\centering
    \includegraphics[scale=0.11]{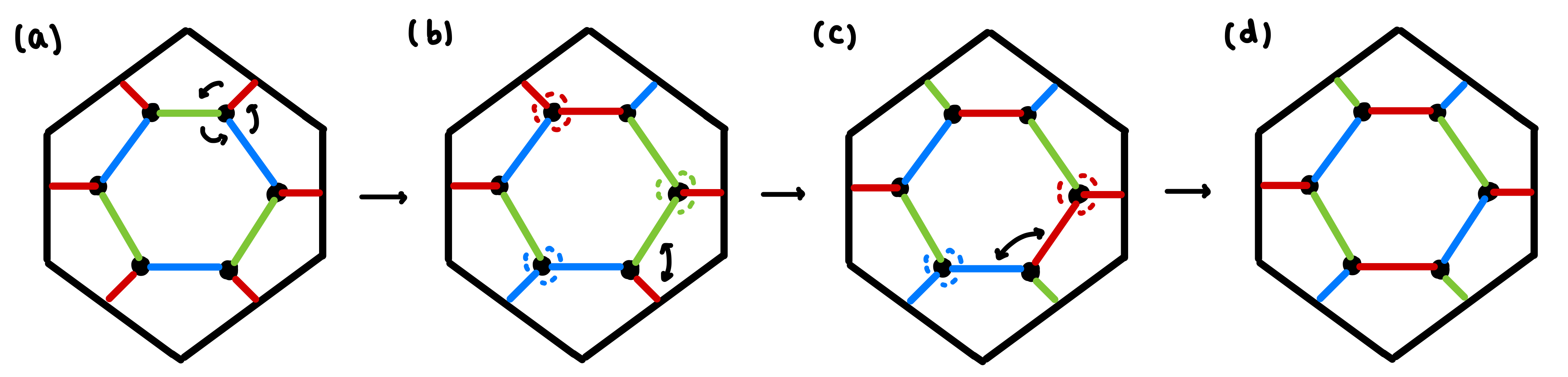}
\caption{\label{fig:cepas}A Cepas move that reverses the sign of the chirality. 
The move can be thought of as follows.  
First create three defects (sites that violate the the 3-coloring constraint) at once, by rotating the colorings around a single site (a), and then transporting a pair of them around a noncontractible loop of the torus by exchanging colors of adjacent links (b to c). The last two defects propagate (c) and annihilate (d), resulting in a change of the chirality.  
Not every such move changes the chirality.  We have checked that moves that do change the chirality also change the parity of the product of $\eta$s in the integrand of \eqref{eq:index-formula}.}
\end{figure}

We claim that the sign with which a given coloring contributes to 
\eqref{eq:Ncequalsz} is $(-1)^m$.  
To see this, we argue as follows.  The 2-color loop flips preserve both the sign of a contribution to our integral and $(-1)^m$.  
In \cite{cepas2017colorings}, a move was found that can change the sign of $(-1)^m$, see Fig.~\ref{fig:cepas}.  We conjecture that this move also reverses the sign of our integrand.  This conjecture is based on comparing the order of the $\eta$s before and after the move in many examples. Thus, we conclude that, for an arbitrary tripartite graph, not necessarily planar,
\be \label{eq:index-formula} Z(-1) 
\equiv \sum_{\text{3-colorings},~ C} (-1)^{ m(C)} 
= \int \prod_{i,\alpha=1..z=3}d\eta_i^\alpha \prod_{\vev{ij} } \( \sum_\alpha A_{ij}' \eta_i^\alpha \eta_j^\alpha\) ~.\ee

It was observed in \cite{cepas2017colorings} that the transfer matrix for this problem on the honeycomb lattice has a manifest $\gSU(3)$ symmetry, even in finite volume.  
(In contrast, the $\gSU(3)$ symmetry of the total sum of honeycomb colorings only emerges in the thermodynamic limit.)
The $\gSU(3)$ symmetry of this ``index" partition function 
$Z(-1)$
is indeed made manifest by our integral formula.  
The key point is that the honeycomb lattice is bipartite.  
So if we transform
\be \eta_{ A}^\alpha \mapsto U^{\alpha\beta} \eta_{ A}^\alpha, 
~~~\eta_{B}^\alpha \mapsto (U^{-1})^{\alpha\beta} \eta_{ B}^\alpha \ee
($A, B$ refer to the two sublattices)
then each factor in the integrand is invariant.

Let us analyze the resulting fermionic field theory associated with this signed counting problem.
The integral is expressed again in terms of the action~\eqref{eq:3-coloring-action} but with the $A'$ matrix instead of $\tilde A$, and with $N_c=3$. Since $N_c=3$ now, the difference is that $(\eta_i^\alpha \eta_j^\alpha)^3 \neq 0$ but $(\eta_i^\alpha \eta_j^\alpha)^4= 0$. Hence, the logarithmic expansion includes up to sextic terms:
\be
S[\eta] = \sum_{\vev{ij}} A'_{ij} \eta_i^\alpha \eta_j^\alpha - \frac{1}{2} A_{ij} \eta_i^\alpha \eta_j^\alpha \eta_i^\beta \eta_j^\beta + \frac{1}{3} A_{ij}' \eta_i^\alpha \eta_j^\alpha \eta_i^\beta \eta_j^\beta \eta_i^\gamma \eta_j^\gamma, \label{eq:second-action-signed-colorings}
\ee
where again we used the fact that $A'_{ij} A'_{ij} = A_{ij}$.

Using the same techniques as the subsection before, we can decouple both the quartic and sextic interactions by introducing link fields. The first link field $\varphi_{ij}$,
\be
    e^{\sum_{\vev{ij}} \frac{1}{3} A'_{ij}(\eta_i^\alpha \eta_j^\alpha)^3 -\frac{1}{2}A_{ij} (\eta_i^\alpha \eta^\alpha_j)^2} = \int \prod_{\vev{ij}} \frac{d\varphi_{ij}}{\sqrt{2\pi}}~e^{ -\frac{1}{2} \varphi_{ij}^2 - i A'_{ij} \varphi_{ij} \eta_i^\alpha \eta_j^\alpha + \frac{i}{3}A_{ij} \varphi_{ij} \eta_i^\alpha \eta_j^\alpha \eta_i^\beta \eta_j^\beta }, \label{eq:first-signed-HS}
\ee 
trades the interaction terms for a gaussian and quartic term in the $\eta$ variables, coupled with the field $\varphi_{ij}$. 
The field $\varphi$ is symmetric under exchange of indices $\varphi_{ij} = \varphi_{ji}$. This can be seen from the field's equation of motion evaluated at the saddle $\varphi_{ij}^{\text{saddle}} = i A'_{ij} \eta_i^\alpha \eta_j^\alpha - \frac{i}{3} A_{ij} \eta_i^\alpha \eta_j^\alpha \eta_i^\beta \eta_j^\beta$.

The second field $\phi_{ij}$, 
\be
    e^{\sum_{\vev{ij}}\frac{i}{3} A_{ij} \eta_i^\alpha \eta_j^\alpha \eta_i^\beta \eta_j^\beta \varphi_{ij}} = \int \prod_{\vev{ij}} \frac{d\phi_{ij}}{\sqrt{2\pi}}~e^{-\frac{1}{2}\phi_{ij}^2 - \sqrt{ 2i \varphi_{ij} \over 3} \phi_{ij} A'_{ij} \eta_i^\alpha \eta_j^\alpha }, \label{eq:second-signed-HS}
\ee
decouples the remaining grassmann quartic term. This field $\phi$ is also symmetric under index exchange $\phi_{ij} = \phi_{ji}$. Its equation of motion at the saddle is given by $\phi_{ij}^{\text{saddle}} = - \sqrt{\frac{2i \varphi_{ij}}{3}} A'_{ij} \eta^\alpha_i \eta_j^\alpha$. 

After introducing these link fields $\varphi$ and $\phi$, we find:
\be
    Z =  \int D(\varphi, \phi)~e^{\sum_{\vev{ij}} -\frac{1}{2} \phi_{ij}- \frac{1}{2} \varphi_{ij}^2 } \int \prod_{\alpha, i}  d\eta_i^\alpha~e^{\sum_{\vev{ij}} (1- i \varphi_{ij} - \phi_{ij}\sqrt{\frac{2i \varphi_{ij}}{3}} )A'_{ij} \eta_i^\alpha \eta_j^\alpha }
\ee
where we define the measure as $D(\varphi, \phi) = \prod_{\vev{ij}} \frac{d\varphi_{ij} d\phi_{ij}}{2\pi} $. 
Integrating out the $\eta$'s, we get a partition function over the link fields $Z = \int D(\varphi, \phi)~e^{-S[\varphi, \phi]}$ whose action is:
\be
S[\varphi, \phi] = \sum_{\vev{ij}}\Big(   \frac{1}{2} \varphi_{ij}^2 +\frac{1}{2} \phi_{ij}^2  \Big) - \frac{3}{2}\tr \ln \left(A'_{ij} \left( 1-i \varphi_{ij} -  \phi_{ij} \sqrt{\frac{2i}{3} \varphi_{ij}}  \right)\right).
\ee
Assuming that the mean field is homogeneous in both fields $\varphi_0$ and $\phi_0$, we find that the action for these configurations is:
\be
    S[\varphi_0, \phi_0] = n_s \left(\frac{3}{4} \varphi_0^2 + \frac{3}{4} \phi_0^2  - \frac{3}{2} \ln\left( 1- i \varphi_0 - \phi_0 \sqrt{\frac{2i \varphi_0}{3}}  \right)\right) - \frac{3}{2} \tr \ln(A'_{ij}).
\ee
We see that, at this level, all lattice dependence is encoded in the last term. 
If we consider planar lattices, we can interpret this term as the total number of coverings by three copies of the dimer model in such lattice. 
The first term has all the field dependence and it is independent of the lattice details. Hence, the saddle point equations for any 3-valent lattice will be given by:
\be
\left.\frac{\delta S}{\delta \phi_{ij}} \right|_{\phi_0,\varphi_0} \hspace{-0.4cm}= \phi_0 + \frac{\sqrt{\frac{2i \varphi_0}{3}}}{1-i \varphi_0-\phi_0 \sqrt{\frac{2i\varphi_0}{3}}}= 0, \quad  \left.\frac{\delta S}{\delta \varphi_{ij}}\right|_{\phi_0,\varphi_0} \hspace{-0.4cm} = \varphi_0 + \frac{i( 1 + \frac{1}{\sqrt{6 i \varphi_0}} \phi_0 ) }{1- i \varphi_0- \phi_0 \sqrt{\frac{2i \varphi_0}{3}}} = 0. 
\ee
There exist several solutions to these equations. Yet, not all of them can be accessed by deforming the integration contour. A detailed analysis is presented in Appendix~\ref{appendix:coloring-details} where we show that the dominant accessible saddle is located at $(\phi_0,\varphi_0)= (-0.815, -0.716 i)$ and discuss the consistency with time-reversal symmetry.  

We can check the result we get for the honeycomb lattice. The leading contributing saddle then gives a free energy density $f=-0.122$. A consistency check is that the result for $Z(-1)$ is smaller than the value of the total number of colorings $Z(1)$. 
In \cite{baxter1970colorings}, Baxter solves not only the 3-coloring problem, but also a deformation of the 3-coloring problem with a fugacity $z^{2m}$ for the chirality $m$.  His expression for the free energy in the thermodynamic limit applies for real values of this chirality.  
It should be possible to analytically continue the calculation to the value $z = \ii $ that would produce the quantity $Z(-1)$.  We leave this for future work.

\subsection{Coloring problems on non-planar graphs}  \label{subsec:non-planar-3-colorings}
The strategy we just used to study $ \tr (-1)^m$ works for an arbitrary 3-valent graph, not necessarily planar.  Combined with some tricks we learned earlier, we can now find an integral representation for the (presumably non-integrable) 3-coloring problem (without signs) on an arbitrary 3-valent graph.  

We follow the strategy described above as a ``parton construction".  
Starting from \eqref{eq:index-formula},
we replace each $\eta_i$ with a bosonic grassmann variable $\sigma_i$.
Since $\sigma_i \sigma_j = + \sigma_j \sigma_i, i \neq j$, this modification removes the dependence on the chirality. 
The nonzero contributions to the integral are in one-to-one correspondence with 3-colorings, and each contributes $+1$.  
The only problem is that bosonic grassmann integrals are hard to do.  
Then as for the dimer model, we can replace each bosonic grassmann by $ \sigma_i = \eta_i \tilde \eta_i$.
The drawback of this representation is that now the action is of degree 12 in the grassmann variables.

To be more explicit, we have
\begin{align}
Z & \equiv \sum_{\text{3-colorings}, C} 1 
= \int \prod_{i,\alpha=R,G,B}d\zeta_i^\alpha \prod_{\vev{ij} } \Big( A_{ij}\sum_\alpha \zeta_i^\alpha \zeta_j^\alpha\Big)
\\ & = \int  \prod_{i,\alpha=R,G,B}d\zeta_i^\alpha \prod_{\vev{ij} } \Big( 1 + A_{ij}  \sum_\alpha \zeta_i^\alpha \zeta_j^\alpha\Big) 
\\ & = \int \prod_{i,\alpha=R,G,B}d\zeta_i^\alpha \exp  \sum_{\vev{ij} } \log \Big( 1+ A_{ij}\sum_\alpha \zeta_i^\alpha \zeta_j^\alpha \Big) 
\\ & = \int \prod_{i,\alpha=R,G,B}d\zeta_i^\alpha \exp  \sum_{\vev{ij} }  A_{ij}\( \sum_\alpha \zeta_i^\alpha \zeta_j^\alpha  - \half  \Big( \sum_\alpha \zeta_i^\alpha \zeta_j^\alpha\Big)^2 
+ {1\over 3} \Big( \sum_\alpha \zeta_i^\alpha \zeta_j^\alpha\Big)^3 \) 
\end{align}
Now we use the same change of variables as above: for each $\alpha$, $ \zeta_i^\alpha = \eta_i^\alpha \tilde \eta_i^\alpha$ (no sum on $\alpha$).  
Thus, 
\be Z = \int D(\tilde \eta, \eta) ~e^{ S[\eta, \tilde \eta] }, \label{eq:partition-function-all-3-coloring} \ee
with
\be S[\eta, \tilde \eta] = \sum_{\vev{ij},\alpha} A_{ij} \left( \eta_i^\alpha \tilde \eta_i^\alpha \eta_j^\alpha \tilde \eta_j^\alpha 
- \half \Big( \sum_{\alpha} \eta_i^\alpha \tilde \eta_i^\alpha \eta_j^\alpha \tilde \eta_j^\alpha \Big)^2 
+ {1\over 3} \Big( \sum_\alpha \eta_i^\alpha \tilde \eta_i^\alpha \eta_j^\alpha \tilde \eta_j^\alpha \Big)^3 \right). \ee

The idea is now to introduce HS fields that decouple all these interactions.
Let us start by first focusing on the 12th degree interaction, introducing a link field $\varphi_{ij}$:
\be
    e^{ \frac{1}{3} \big(\sum_{\alpha} \eta_i^\alpha \tilde \eta_i^\alpha \eta_j^\alpha \tilde \eta_j^\alpha\big)^3 - \frac{1}{2}  \big(\sum_{\alpha} \eta_i^\alpha \tilde \eta_i^\alpha \eta_j^\alpha \tilde \eta_j^\alpha\big)^2  } = \int \frac{d\varphi_{ij}}{\sqrt{2\pi}}~ e^{- \frac{1}{2}  \varphi_{ij}^2 +i \varphi_{ij} \sum_{\alpha} \eta_i^\alpha \tilde \eta_i^\alpha \eta_j^\alpha \tilde \eta_j^\alpha -\frac{i}{3} \varphi_{ij} \big(\sum_{\alpha} \eta_i^\alpha \tilde \eta_i^\alpha \eta_j^\alpha \tilde \eta_j^\alpha\big)^2 }. \label{eq:first-non-planar-HS}
\ee 
This transformation trades the 12th degree interaction for a quartic and a 8th degree interaction. This is possible since the 4th power is equal to zero $(\sum_\alpha \eta_i^\alpha \tilde \eta_i^\alpha \eta_j^\alpha \tilde \eta_j^\alpha)^4=0$. Observe that this field is symmetric under exchange of indices $\varphi_{ij} = \varphi_{ji}$ and whose equation of motion at the saddle is $\varphi_{ij}^{\text{saddle}} = i \sum_\alpha \eta_i^\alpha \tilde \eta_i^\alpha \eta_j^\alpha \tilde \eta_j^\alpha- \frac{i}{3}(\sum_\alpha \eta_i^\alpha \tilde \eta_i^\alpha \eta_j^\alpha \tilde \eta_j^\alpha)^2$. 

We decouple the 8th degree interaction using another field $\phi_{ij}$: 
\be
    e^{- \frac{i}{3} \varphi_{ij} \big( \sum_{\alpha}  \eta_i^\alpha \tilde \eta_i^\alpha \eta_j^\alpha \tilde \eta_j^\alpha \big)^2 } = \int  \frac{d\phi_{ij}}{\sqrt{2\pi}} ~e^{ -\frac{1}{2} \phi_{ij}^2 +  \sqrt{\frac{2i}{3} \varphi_{ij}} \, i \phi_{ij} \sum_{\alpha} \eta_i^\alpha \tilde \eta_i^\alpha \eta_j^\alpha \tilde \eta_j^\alpha}. \label{eq:second-non-planar-HS}
\ee
At the saddle, the equations of motion is given by $\phi_{ij}^{\text{saddle}}= i \sqrt{\frac{2i}{3}\varphi_{ij}} \sum_{\alpha} \eta_i^\alpha \tilde \eta_i^\alpha \eta_j^\alpha \tilde \eta_j^\alpha$. This link field is also symmetric under exchange of indices $\phi_{ij} = \phi_{ji}$.

We are only left with decoupling the quartic interaction, which we do by introducing a color-dependent link field $P^{\alpha}_{ij}$:
\be
    e^{ (1+i\varphi_{ij} +i \phi_{ij}\sqrt{\frac{2i}{3}\varphi_{ij}} )  \eta_i^\alpha \tilde \eta_i^\alpha \eta_j^\alpha \tilde \eta_j^\alpha } =  \int  \frac{dP^\alpha_{ij}}{\sqrt{2\pi}}\,e^{ -\frac{1}{2}(P^{\alpha}_{{ij}})^2 +  P^{\alpha}_{ij}  \sqrt{1+i\varphi_{ij} +i \phi_{ij}\sqrt{\frac{2i}{3}\varphi_{ij}} }\big( \eta_i^\alpha \tilde \eta_i^\alpha + \eta_j^\alpha \tilde \eta_j^\alpha \big)}.  \label{eq:third-non-planar-HS}
\ee
Note that this is color-dependent link field is also symmetric under exchange of indices $P^\alpha_{ij} = P^{\alpha}_{ji}$. The equations of motion, at the saddle, is given by $P^\alpha_{ij} = (1+i\varphi_{ij} +i \phi_{ij}\sqrt{\frac{2i}{3} \varphi_{ij}})^{1/2} (\eta_i^\alpha \tilde \eta_i^\alpha + \eta_j^\alpha \tilde \eta_j^\alpha)$.

We replace back all these HS transformations into~\eqref{eq:partition-function-all-3-coloring} and integrate out the variables $\eta, \tilde \eta$. The partition function is then given by 
\be
    Z = \int D(\phi,\varphi,P)~ e^{-S[\phi, \varphi, P]},
\ee
where we define the measure as $D(\phi,\varphi,P) = \prod_{\vev{ij}} \frac{d\phi_{ij}d\varphi_{ij}}{2\pi} \prod_{\vev{i,j},\alpha} \frac{dP^\alpha_{ij}}{\sqrt{2\pi}}$, and whose resulting action is: 
\be
    S[\phi, \varphi, P] = \frac{1}{2}\sum_{\vev{i,j}} \Big( \phi_{ij}^2 + \varphi_{ij}^2 + \sum_{\alpha} (P^{\alpha}_{ij})^2 \Big) - \sum_{i, \alpha} \ln\Big( \sum_{\vev{i|j}} P^\alpha_{ij} \Big(1+i\varphi_{ij} +i \phi_{ij}\sqrt{2i\varphi_{ij}/3} \Big)^{1/2}  \Big). \label{eq:2nd-representation-action}
\ee
Assuming uniform saddles, the action describing the coloring problem on non-planar graphs is 
\be
    S[\varphi_0,\phi_0,P_0] = \frac{3}{4}n_s (\phi_0^2 + \varphi_0^2 +3 P_0^2) - 3 n_s \Big( \ln(3P_0) + \frac{1}{2} \ln(1+i\varphi_0+i\phi_0\sqrt{2i\varphi_0/3})\Big),  
\ee
where $n_s$ is the number of sites. At this level, there is no lattice dependence at all, unlike the other integral representations studied before. All 3-valent graphs will have the same contribution. The contributions of luctuations about the saddle point do depend on the lattice information encoded in the adjacency matrix. 

Taking the variation of the action with respect to the fields gives:
\begin{align}
    \left.\frac{\delta S}{\delta \varphi_{ij}} \right|_{\phi_0, \varphi_0, P_0} &= -  \varphi_0 +  i \Big(  \frac{1+\phi_0 \sqrt{\frac{i}{6\varphi_0}}}{1+i \varphi_0 +i \phi_0 \sqrt{2i \varphi_0/3 }} \Big)= 0, \\
    \left.\frac{\delta S}{\delta \phi_{ij}} \right|_{\phi_0, \varphi_0, P_0} &= -  \phi_0 + \Big( \frac{i \sqrt{\frac{2i\varphi_0}{3}}}{1+i \varphi_0 +i \phi_0 \sqrt{\frac{2i \varphi_0}{3}}} \Big) =0 , \\
    \left.\frac{\delta S}{ \delta P_{ij} } \right|_{\phi_0, \varphi_0, P_0} &= - \frac{9}{2} P_0 + 3  \Big( \frac{1}{P_0} \Big) = 0.
\end{align} 
Solutions to these coupled equations can be found. As in the previous calculations, not all of them contribute to the free energy density. The calculation is described in Appendix~\ref{appendix:coloring-details}, where the location of the dominant accessible saddle is $(\varphi_0, \phi_0, P_0) = (0.72i, -0.82,\sqrt{\frac{2}{3}})$ and the approximate value of the free energy density we find is $f=-0.825$. 

This leading-order estimate for the number of honeycomb 3-colorings is not as good as that from the simpler representation above.  
But the true value of this second integral representation of the 3-coloring problem is that it works even for non-planar graphs.  We expect that for a tripartite non-planar graph, such as a hyperhoneycomb lattice \cite{Lee_2014,Kimchi_2014}, the 3-coloring problem is not integrable.  
But we can use our integral representation to give an estimate of the number of colorings on such a lattice.

We note that the two representations of the 3-coloring problem we have given can be generalized to
coloring problems on arbitrary $z$-valent graphs, by placing $N_c = z-1$ or $N_c = z$ grassmann species at each site of the graph, respectively.

\subsection{Deformations of the honeycomb coloring model}
In \cite{baxter1970colorings,castelnovo2004dynamical}, generalizations of the 3-coloring model on the honeycomb lattice are described,
which take advantage of the correspondence between colorings of the honeycomb lattice and spins on the sites of the honeycomb lattice discussed above.  
We can weight the sum over colorings by various functionals of the spin configuration, such as the total magnetization \cite{baxter1970colorings} (for which the model is still integrable), or 
nearest-neighbor Ising interactions $  \sum_{\vev{ij}} \sigma_i \sigma_j$ \cite{castelnovo2004dynamical}.  

Here we describe a somewhat crude method to incorporate this deformation into our grassmann integral representation.  We describe the method for the first, planar representation for simplicity.  
The first step is to generalize the weights to include a separate fugacity for each link and each color: 
\be Z(w) = \sum_\text{colorings, $C$} \prod_{\text{links}, \ell} w_\ell^{\alpha_\ell(C)} \ee
where $\alpha_\ell(C)$ is the color of link $\ell$ in the coloring $C$.  
By the same logic as above, $Z(w)$ has the following integral representation 
in terms of $N_c =z-1=2$ grassmann variables on each site 
\be Z(w) = \int D\eta  \prod_{\vev{ij} } \( w_{ij}^0 + w_{ij}^1 \eta_i^1 \eta_j^1 + w_{ij}^2 \eta_i^2 \eta_j^2 \) . \ee
Taking the log as above, this is a path integral for a field theory with quartic interactions, albeit now without translation symmetry.  

With these 3 independent couplings for each link, we keep track of all information about each coloring, in particular the spin configuration.  
The following formula then produces an integral representation for
\be 
Z(y) \equiv \sum_\text{colorings, $C$} y^{\sum_i \sigma_i(C)} 
= \prod_{\text{sites}, i} \sum_{\alpha\beta\gamma} |\eps_{\alpha\beta\gamma}|
y^{\eps_{\alpha\beta\gamma}} {\partial \over \partial  \log w_{mi}^\alpha }
 {\partial \over \partial  \log w_{mj}^\beta} {\partial \over \partial  \log w_{mk}^\gamma } 
 Z(w)|_{w=1} \ee
 where $i,j,k$ denote the three neighboring sites of site $m$.  
 A related, though even more unwieldy, formula can be written for 
 \be 
 Z(q) \equiv  \sum_\text{colorings, $C$} q^{\sum_{\vev{ij}} \sigma_i(C)\sigma_j(C)} . \ee

\section{Quantum dimer piling models}\label{sec:QDM}

The quantum dimer model associated to a graph $\Gamma$ is a quantum system whose Hilbert space is spanned by a collection of orthonormal states, each associated to a dimer covering of $\Gamma$~(e.g.~\cite{Rokhsar:1988zz,moessner2001resonating}).  It can be regarded as a toy model for a system of spin-half particles on $\Gamma$ in the subspace of states spanned by local singlets, as favored by local antiferromagnetic interactions.  It is a toy model because different local singlet configurations are in fact not orthogonal (for a review, see~\cite{moessner2010quantum}).  

In the same spirit, we can construct a quantum dimer piling model on $\Gamma$.  The Hilbert space is spanned by orthonormal states labelled by dimer piling configurations on $\Gamma$.  This is a toy model for an antiferromagnet with spin-$N/2$ degrees of freedom on the sites of $\Gamma$ in the same sense as above.  

First recall the RK model \cite{Rokhsar:1988zz}. 
The general Hamiltonian considered by RK is 
\be H= - \sum_p  \( t T_p - v V_p \) ,~ T_p =  \ketbra{||}{=}+ \ketbra{=}{||} ,~ 
V_p = 
\ketbra{||}{||}+ \ketbra{=}{=} .  \ee
(For simplicity we draw the dimer configurations as if they are on the square lattice, but nothing we say here is special to that case.)
The operators $T_p$ and $V_p$ are supported on a plaquette $p$; the arguments of the states are caricatures of local dimer configurations around that plaquette.
At the special point $t=v$, the groundstate is exactly known, in the following manner.  
The following Hamiltonian is frustration free and agrees with $H$ when $t=v=1$: 
\be H_\text{RK} = \sum_{p}\sum_C Q_{p,C}\ee
where the sum is over plaquettes and configurations $C$ where the plaquette $p$ is flippable,
$C_p$ is the result of flipping the plaquette $p$ in the configuration $C$, and 
\be Q_{p,C} \equiv \ketbra{C}{C} - \ketbra{C_p}{C} ~.\ee
Note that $H_\text{RK}$ is Hermitian because of the sum over $C$.  

To see that this model is frustration-free, note that $Q_{p,C} \geq 0$, and 
\be Q_{p,C}^2 = Q_{p,C} . \ee
Thus its eigenvalues are $0, 1$, and the groundstate of $H$ is the simultaneous eigenstate of $Q_{p,C}$ with eigenvalue zero for each, if it exists.  
But the condition $Q_{p,C} \ket{\Psi_\text{RK}}= 0 $ is the condition that the wavefunctions for $C$ and $C_p$ are equal, 
so the groundstate is 
\be \ket{\Psi_\text{RK}} = {1\over \sqrt{Z}} \sum_{C} \ket{C} . \ee

Following \cite{castelnovo2005quantum2}, this construction can be generalized to include nontrivial relative weights as follows.  
Take instead
\be Q_{p,C} \equiv \sqrt{ p(C_p) \over p(C)} \ketbra{C}{C} - \ketbra{C_p}{C} ~.\ee
Then 
\be Q_{p,C}^2 = \sqrt{p(C_p) \over p(C)} Q_{p,C} , \ee
and the spectrum of $Q_{p,C}$ is instead $\{0, \sqrt{p(C_p) \over p(C)} \}$.  
The simultaneous zero eigenstate of all the $Q_{p,C}$s now requires the weight of $C_p$
to be $\sqrt{p(C_p) \over p(C)}$ times that of $C$, and so 
\be \ket{\Psi_\text{RK}} = \sum_{C}  \sqrt{ p(C) } \ket{C} . \ee

Now we construct a quantum dimer piling model associated with model A on a lattice $\Gamma$.
The quantity $p(C)$, up to normalization, is 
\be p(C) = \prod_{\ell} {1\over N^{n_\ell} n_\ell !} . \ee
An orthonormal basis of the Hilbert space is labelled by $N$-dimer pilings on the lattice.
To be explicit, this Hilbert space can be realized by placing a harmonic oscillator on each link of the lattice, and restricting to the linear subspace of states satisfying $ \sum_{\ell \in v(i)} n_\ell = N$ for all sites $i$, where $v(i)$ indicates the links incident on site $i$.  

What is the kinetic term?  Recall that an $N$-dimer piling can be regarded (not canonically) as $N$ independent dimer coverings of the lattice; this is a coloring of the dimer piling (see for example Fig.~\ref{fig:sample-piling}).  
To update a dimer piling, we can simply choose such a representative coloring, identify a flippable plaquette of one of the colors, and flip it.  This produces a new dimer piling.  
Thus we have a notion of ``flippable plaquette in dimer piling $C$".
As in the $N=1$ case, the question of how ergodic is this update in the space of dimer pilings is an interesting one that we leave for future work.

We observe that with model A weights for two configurations related by such an update on plaquette $p$, 
\be \label{eq:p-over-pc} \sqrt{p(C_p) \over p(C)}  = \sqrt{  \prod_{\ell \in p_\up} n_\ell(C) \over \prod_{\ell \in p_\down} (n_\ell(C) +1)} \ee
where $ p_\up$ denotes the set of links in the boundary of $p$ that gain a dimer, while $p_\down$ denotes the complementary set of links in the boundary of $p$. 
The equation \eqref{eq:p-over-pc} uses the fact that a flippable plaquette must have an even number of links, so the factors of $N$ cancel.

Now define 
\be Q_{p,C} \equiv \sqrt{ p(C_p) \over p(C)} \ketbra{C}{C} - \ketbra{C_p}{C} ~\ee
where $p(C)$ is the model-A probability, and $p$ is a plaquette which is flippable in configuration $C$.  
In this expression, $ p(C_p)/p(C)$ is a c-number
determined by the occupation numbers in configuration $C$.
Therefore, 
\be Q_{p,C}^2 = \sqrt{p(C_p) \over p(C)}  Q_{p,C}. \ee

The Hamiltonian 
\be H_\text{$N$RK} = \sum_{p} \sum_{C} Q_p \ee
in which $C$ is a sum over configurations in which the plaquette $p$ is flippable, 
is frustration free, and its groundstate is the simultaneous zero eigenstate of all the $Q_{p,C}$s, which 
is 
\be \ket{\psi_{\text{RK}}} = \sum_C \sqrt{p(C)} \ket{C}  ~.\ee
In this state, the equal-time correlations functions of operators diagonal in the dimer basis, are computed by model A.  

The rough features of the phase diagram for this model, as we vary $t/v$, is much the same as in the $N=1$ case.
Large positive $v/t$ favors a minimal number of flippable plaquettes, 
and thus a staggered dimer solid,
while large negative $v/t$ wants to maximize this number.  
For small $v/t$, the details of the phase diagram depend on the nature of the lattice.  We expect that, as in the $N=1$ case, frustrated lattices will produce a regime with topological order.  Based on the form of the RK wavefunction, we expect that it is still $\IZ_2$ topological order for any $N$, since winding numbers around the torus are still only well-defined mod two.

Ideally, we would like to give an example where we know the RK point 
of the quantum dimer piling model represents a gapped phase.  In the case of $N=1$, such an example is provided by the kagome lattice dimer model \cite{misguich2002quantum}, which has a simple commuting parent Hamiltonian.
This possibility relies on the following arrow representation of kagome lattice dimers introduced in \cite{elser1993kagome}.
Each kagome dimer covering determines a configuration of arrows on the kagome site: the arrow points towards the triangle containing the adjacent dimer.  
However, not every arrow configuration is obtained this way -- each triangle must have an even number of arrows pointing in (since each dimer has two arrows pointing at it).  

One can try to generalize this arrow representation to $N$-pilings on kagome.
An $N$-piling can be obtained by placing $N$ dimer coverings on top of each other, which we regard as different colors.  Therefore it determines a configuration of $N$ arrows on each site by the rule described above.
However, the dimer piling forgets the colors of the dimers, so the arrows must be regarded as indistinguishable.  Thus, each site has $N+1$ possible states, determined by the number of arrows, out of $N$, which point in each direction. Yet, already for $N=2$, there are configurations of such arrows, with an even number pointing in to each triangle, which do not correspond to any $N$-piling
(for example, consider a triangle whose corners have 2 in, 0 in, 0 in respectively).   Thus, there is not a simple rule for allowed arrow configurations, and one can check that the operator on a hexagon that acts as a kinetic term for dimer pilings cannot be written in a simple way in terms of an operator that flips the arrows.  

We must admit that the particular form of the weights computed by the grassmann integral did not play a crucial role in this construction.  
A natural microscopic construction of such a model begins from bosons living on the links of the graph.  
Then consider the following Hamiltonian
\be H_\text{microscopic} = \sum_{\vev{\ell \ell'}} J b_{\ell}^\dagger b_{\ell'}^\nd + h.c.
+ \Gamma \sum_i \( \sum_{\vev{i|j}} b_{ij}^\dagger b_{ij}^\nd  - N \)^2 . 
\ee
Here $\vev{\ell \ell'}$ indicates links that share a site, and $\sum_{\vev{i|j}}$ indicates a sum over the links ending at site $i$.
We interpret $b_\ell^\dagger b_\ell^\nd$ as the number of dimers on link $\ell$.
Here $\Gamma$ is a large energetic penalty for violating the constraint that $N$ dimers end at every site.  
If $\Gamma \gg J$, degenerate perturbation theory in $J$ produces as its leading term a plaquette flip operator of the form above, with $t = -J^2/\Gamma$.  
However, with the simplest boson hopping term, its coefficient is not the same as the weight function for the models defined above.

\section{Discussion}\label{sec:discussion}

The integral representations of dimer problems we have discussed raise a number of interesting questions.  

{\bf Monte Carlo method.}  Under certain symmetry assumptions, the Hubbard-Stratonovich integral in \eqref{eq:HS_B} has positive measure, and therefore may be efficiently evaluated by sampling\footnote{We are grateful to Tarun Grover for sharing his expertise on this subject.}.  
This would be a zero-dimensional analog of auxiliary-field Monte Carlo commonly used to study certain problems of itinerant fermions \cite{assaad2008world}.

For example, consider model B on a bipartite lattice with even $N$. Arrange the colors into two groups labelled $\up a$ and $\down a$, $a=1..N/2$.   The operation $U$ that takes
\be \eta_{A\up a} \to \tilde\eta_{A\down a}, ~~
\eta_{B\up a} \to -\tilde\eta_{B\down a}, ~~
\eta_{A\down a} \to - \tilde\eta_{A\up a},~~ 
\eta_{B\down a} \to \tilde\eta_{B\up a},~~ \ii \to - \ii
\ee
is an antiunitary symmetry of 
the action  of model B after the Hubbard-Stratonovich transformation : 
\begin{align}  
S_B &= |\phi_{AB}|^2 + \eta_{A\up a}\eta_{B\up a} \phi_{AB}^\star 
- \tilde \eta_{A\up a}\tilde\eta_{B\up a} \phi_{AB}
+ \eta_{A\down a} \eta_{B\down a} \phi_{AB}^\star
- \tilde \eta_{A\down a} \tilde \eta_{B\down a} \phi_{AB}
\nonumber 
 \\  
 & = |\phi|^2 - 
\psi^T M \psi, ~~~
\text{with } 
\psi \equiv 
 \begin{pmatrix} 
\tilde \eta_{A\up a} \\ 
 \tilde \eta_{B\down a} \\ 
 \eta_{A\down a} \\
 \eta_{B \up a}
 \end{pmatrix}
.\end{align}
The result of the grassmann integral is $\int D\psi e^{ \psi^T M \psi} = \det M$.
A theorem commonly used in the context of quantum Monte carlo is the following: 
if $ U^\dagger M U = M$ with $U$ antiunitary and $U^2 = - \Ione$, then the eigenvalues of $M$ come in complex-conjugate pairs.  Hence $ \det M\geq 0$.  
Therefore, at least in this class of examples, the integral over $\phi$ may be done efficiently by Monte Carlo sampling.  We leave its evaluation to future work.  

{\bf Related models.}
The recent paper 
\cite{douglas2022dimers}
considers ensembles of dimer pilings (called `multiwebs') with a different set of weights than we study.  
Their weights seem to be chosen so that $ Z_N = Z_1^N$.
For their $n$-piling problem, they include in their input data a $\mathsf{SL}(n,\IR)$ matrix $\Phi_{ij}^{ab}$ on each link of the graph (called a `local $\mathsf{SL}_n$ system').  Such data can also be incorporated into our models by replacing the quartic term by 
(\eg~generalizing model A)
\be \Phi_{ij}^{ab} \eta_i^a \tilde \eta_i^a \eta_j^b \tilde \eta_j^b .\ee

The models studied in \cite{kenyon2021multinomial} are closer to our model A, but we believe they count dimer pilings with a different weight (in particular, without a factor of $1/n_\ell!$ that makes our dimers indistinguishable).  They obtain results on the large-$N$ asymptotics of their models by a different set of methods, and it would be interesting to find some overlapping regime of validity.  
\cite{kenyon2023higherrank,kenyon2023planar} also study related questions.

{\bf More general tiling problems.}
We can consider tiling a graph by other shapes.  For example, consider tiling a region of the square lattice by {\it wedges} -- collections of three neighboring sites that form an `L'.  An integral representation for the number of such tilings can be obtained by the methods above.  
Now the input is not just the adjacency matrix of the graph, but a collection of possible tiling objects.   
We can represent these as a tensor $A_{ijk}$ with a number of indices equal to the number of sites covered by each object (three, in the case of wedges), which is nonzero if $ijk$ represents a possible tile, and zero otherwise.  
In terms of even grassmanns, the partition sum is 
\be Z_1(A) = \int D\zeta~ e^{ {1\over 3!} \sum_{ijk} A_{ijk} \zeta_i \zeta_j \zeta_k } . \ee
In terms of ordinary grassmann variables, this can be written as 
\be Z_1(A) = \int D(\tilde \eta, \eta)~ e^{ {1\over 3!} \sum_{ijk} A_{ijk} \eta_i\tilde \eta_i \eta_j\tilde \eta_j \eta_k\tilde \eta_k } .\ee
A monomer or dimer fugacity can be introduced in the form of quadratic and quartic terms respectively, again preserving the 
symplectic local gauge invariance.

A similar strategy works to provide a Grassmann integral representation of the square ice model \cite{lieb1967residual}.
The degrees of freedom of the square ice model are arrows on the links; 
the (zero temperature version of the) model counts configurations of the arrows satisfying the 2-in 2-out rule at each vertex.
This model is equivalent to 3-colorings of the faces of the square lattice (for an explanation see page 342 of \cite{lieb1972two}),
which has the following representation along the lines we've been discussing:
\be Z \equiv \sum_{\text{3-colorings~of~faces}} 1 
= \int \prod_{\alpha=1}^3\prod_{\text{links}, l} d\sigma_l^\alpha \prod_{\text{faces}, f} \( 
\sum_{\alpha=1}^3 \prod_{l \in \partial f} \sigma_l^\alpha  \)\ee
where $\sigma_l^\alpha$ are 3 bosonic grassmanns on each link of the square lattice.  
Now we can replace each bosonic grassmann by $ \sigma_l^\alpha = \eta_l^\alpha \tilde \eta_l^\alpha$ as before.  
There are some interesting open questions about such integrals.
If we instead replace each bosonic grassmann in this integral with a single fermionic grassmann in this case, does it instead count 
$\sum_\text{3-colorings~of~faces} (-1)^{m} $
where $m$ is some invariant of 
3-colorings of the faces of the square lattice? What invariant is it?
As for the honeycomb 3-coloring case,
we can write the integrand of $Z$ above in the form $e^{ S}$ where $S$ comes from the Taylor expansion of the log, which truncates after the cubic term in this case.  
What happens if we vary the coefficients of this series away from the special values from the Taylor expansion of the log?  

In fact there is simpler, more direct, representation of the square ice model.
Put a single bosonic grassmann on each link of the square lattice, and consider the following: 
\be 
\int \prod_l d\sigma_l \prod_{sites~i} \( \sum_{ {jk}, ~\text{pairs of neighbors of $i$}}  \sigma_{ij} \sigma_{ik} \) ~.
\ee
The nonzero contributions are in one-to-one correspondence with configurations of arrows satisfying the ice rules. 
The idea is that the six terms in each factor of the integrand correspond to possible configurations of the vertex, where we interpret the presence of $\sigma_l \sigma_m$ as saying that links $l$ and $m$ are pointing toward the vertex in question.  
Then the integrand only gets a nonzero contribution if every vertex contributes two grassmanns in a way that is compatible with its neighbors -- the one going out from site $i$ on link $ij$ (so that site $i$ does not contribute a grassmann $\sigma_{ij}$) must be the one going in to site $j$ (so that site $j$ does contribute that grassmann).  
However, in this case 
we cannot usefully write the integrand as $e^{-S}$.
If, like we did in the 3-coloring integral,
we replace the integrand factors with 
$ \(1 +  \sum_{ {jk}, ~\text{pairs of neighbors of $i$}}  \sigma_{ij} \sigma_{ik} \) $
the result would include configurations where all the links were pointing out of site $i$.

The square ice model is exactly solvable.  One can ask how that exact solvability reflected in the form of the action for the grassmanns.
Finally, as in the 3-coloring problems, higher dimensional versions of the ice model are not exactly solvable, while a grassmann representation still exists and may be useful.

{\bf Littlewood-Richardson coefficients.} 
Littlewood-Richardson recoupling coefficients, $n_{\lambda \nu}^\mu$, count the number of representations 
of $\mathsf{GL}(n)$ of each kind in the tensor product of two other representations, $ \Phi_\lambda \otimes \Phi_\nu = \sum_\mu n_{\lambda\nu}^\mu \Phi_\mu$.  
There is a representation of these numbers as a tiling problem on a section of triangular lattice with certain boundary conditions, called {\it Knutson-Tao puzzles} \cite{1998math......7160K}.
By our methods, a grassmann integral representation of these numbers can be given, involving at most octic terms in the action (because the tiling involves some 4-sided objects).  
It involves four grassmann variables on each link $\eta^{a=0,1}, \tilde \eta^{a=0,1}$.
Perhaps a mean-field analysis of such an integral representation can provide simple asymptotic formula for the Littlewood-Richardson coefficients.  Alternatively, this same set of numbers compute the intersection product on Grassmannians \cite{2001math.....12150K}, \ie~the symmetric spaces of embedded planes.  So this would be a happy union of two very different objects named after Grassmann.

Actually, a few new ingredients worth mentioning are required to realize this integral representation.  
We will explain them using the simpler representation in terms of bosonic grassmann variables $\sigma$.
To make the integral representation with ordinary grassmanns, 
we use the relation as above $\sigma = \eta \tilde \eta$.  
The first new ingredient is that the grassmann integrals we've introduced so far behave as $\mathsf{AND}$ operations, in the sense that they give nonzero contributions only when a collection of conditions (the presence of a tile) are all met, most basically 
\be \int d\sigma_1 d\sigma_2 e^{ a \sigma_1 + b \sigma_2} 
= ab .\ee
For the above application, we need the analog of an $\mathsf{XOR}$ operation, which produces $a+b$ instead.  
A simple solution is to modify the measure as follows:
\be \int d\sigma_1 d\sigma_2 ( \sigma_1 + \sigma_2)  e^{ a \sigma_1 + b \sigma_2}  = a+b . \ee
A second ingredient is a notion of `interlocking' of tiles.  That is, in the Knutson-Tao puzzles, the edges of the tiles are labelled ($a=0, 1$), and two tiles may share an edge only if they carry the same label.  
This can be accomplished by associating a {\it pair} of grassmans for each species and each edge $ \sigma_l^{La}, \sigma_l^{Ra}, a=0,1$, and using the measure
\be \int  \prod_{a=0,1} d\sigma^{La} d\sigma^{Ra} ( \sigma^{L0}\sigma^{R0} + \sigma^{L1}\sigma^{R1})  ... \ee
The generalization to more than two labels is instead 
\be \int  \prod_{a=0...n-1} d\sigma^{La} d\sigma^{Ra} 
\sum_{a} \( \prod_{b\neq a} \sigma^{Lb}\sigma^{Rb}\) ...  \ee

This construction suggests a different method of giving integral representations for tiling problems on hypergraphs (by a hypergraph, we mean a graph with some notion of faces, \ie~a 2-skeleton), where the grassmanns live on the links rather than the sites.  For each link, we associate one (bosonic) grassmann variable for each face incident on the link.  
(For a planar graph, we can simply choose an orientation of the links and call the two variables $\sigma^L_\ell$ and $\sigma^R_\ell$ as above.)
Then the following integral counts tilings of a general hypergraph by a collection of objects $O$:
\be Z(O) = \int \prod_\ell \prod_{f, \partial f \ni \ell} d \sigma_{\ell f} 
e^{ S_O[\sigma]}, \ee
where $S_O$ is a sum of terms associated with the objects in $O$, as follows.
If an object covers a collection of faces $ f_1, .. f_n$, 
the associated term is 
\be \prod_{f \in O, \ell \in  \partial f}  \sigma_{\ell f} .\ee

{\bf Height representations.} 
In the case of the ordinary dimer model, there is a representation in terms of a periodic height variable.  
In the case where $\Gamma$ is a 2d lattice this provides a sort of bosonization of the grassmann integral given above.  

For general tilings, 
the analog of the height variable is more complicated: it takes values in Conway's tiling group
\cite{thurston1990conway}.  
For example, in the case of linear trimers, the analogous variable lives in the 2-torus \cite{ghosh2007random}.
In most of the literature on this subject (\eg~\cite{kenyon1992tiling}), the tiling group is used only to ask whether tilings {\it exist}, 
not to count them.  How is this bosonic representation related to that in terms of the Hubbard-Stratonovich variable $\phi$?

{\bf Generalized Heilmann-Lieb argument and transitions at finite monomer fugacity.} Heilmann and Lieb demonstrated \cite{heilmann1972theory} that in the $N=1$ monomer-dimer problem on any lattice, there is no phase transition at finite monomer fugacity $ x \in (0, \infty)$. Using~\eqref{eq:SD} as their main tool, they showed that all zeroes of the partition function lie on the pure-imaginary axis of the complex $x$ plane. Thus the only real value of $x$ that can be pinched by the zeros is $x=0$.

All the models we have studied reduce to this problem when $N \to 1$.  
It is interesting to ask if such an argument exists for the models we have studied for arbitrary $N$. 
For $N=2,3$ and small system size on the periodic chain and square lattice, we have found that the zeroes of the partition function of model $A$ are still on the imaginary axis. 
On the other hand, for model $B$ and $C$ under the same conditions, the zeroes are in the complex plane. 
Therefore, the possibility remains open that in models B and C for $N>1$, these complex roots can pinch the real $x$ axis at some nonzero value.  
Indeed, our large-$N$ free energies do have singularities at $x=1$.  

We can see how the Heilmann-Lieb argument breaks down for $N>1$, because the Schwinger-Dyson equation is quite different.   
In order to write the equation, we introduce the notation
$ Z(\Gamma, \{N_k\}_k) $ for the dimer piling problem with $N_k$ colors at site $k$.  
Can it be written as 
\be Z(\Gamma, N) \buildrel{?}\over{=} x Z(\Gamma - i, N-1) + \sum_{j} A_{ij} Z(\Gamma- i - j, N-1)? \ee 
No -- because only at the indicated sites are there fewer grassmanns!  
The simplest one takes the form
\be Z(\Gamma, \{N_k\}_k) =  x Z(\Gamma,  \{... N_i-1 \}) + \sum_{j} A_{ij} Z(\Gamma, \{...N_i-1, N_j-1\}),\ee
where $N_k$ is the number of colors at site $k$.  Thus, for $N>1$, the Schwinger-Dyson equations only close on a much larger collection of inhomogeneous functions.  

As is often the case, our models exhibit a tension between tractibility and excitement.   
We showed above that on a bipartite graph, models B and C are the same, and for even $N$ have an integral representation that can be efficiently evaluated by sampling.  However, non-bipartite graphs, where models B and C differ, have the new ingredient of geometrical frustration.   We anticipate that many interesting phenomena are hidden there.  

{\bf Matrix models. }  In the case where we choose the complete graph, $A_{ij} = 1, \forall i,j$, 
these dimer piling models become matrix models 
like those of \cite{Hartnoll:2019pwe}.
The actions for colorings A, B and C take the form
\be S_A  = \(\tr (S^\dagger S) \)^2 , ~~~~
S_B = \tr (SS^\dagger SS^\dagger), ~~~~
S_C = \tr ( S^T S S^\dagger S^{\dagger T}) . 
\ee
in terms of the (grassmann-valued) matrix
\be (S)_{\alpha i } \equiv \eta_i^\alpha, 
(S^\dagger)_{i\alpha} \equiv \tilde \eta_i^\alpha. \ee
These models have an additional symmetry permuting the sites.  
Such models can have phases that spontaneously break this permutation symmetry (analogous to the translation-symmetry-breaking states found above).  
For a grassmann-valued matrix, the analog of the eigenvalue distribution (and its topology) is mysterious to us.

{\bf Consequences for computational complexity?}
The dimer model on a generic non-planar graph is a Hard Problem (as described in \cite{moore2011nature}, \S13.3).  Specifically, computing the permanent of a binary matrix is $\mathsf{\#P}$-complete.  
Can we leverage that guaranteed hardness to say something about the difficulty of solving interacting field theories?
In this paper we described a formula for the dimer problem on an arbitrary graph in terms of a quartic grassmann integral.
Solving this field theory must be at least as hard as the complexity class that contains the non-planar dimer model.

One can also ask about the difficulty of {\it approximating} the path integral.  The difficulty of approximating the permanent is discussed in \cite{moore2011nature}, \S13.4.  And the conclusion there is that approximating the permanent is actually in P.  
Does this mean that we should expect a Monte Carlo algorithm for our quartic integral on any graph?  
It would be interesting to compare the polynomial-time algorithm described in \cite{moore2011nature}, \S13.5,
for approximating the number of perfect matchings (no monomers) on bipartite graphs \cite{jerrum2004polynomial}
with the auxiliary-field Monte Carlo algorithm discussed above.  
For graphs whose perfect matchings can be efficiently counted in polynomial time (such as planar graphs), \cite{alimohammadi2023fractionally} and 
\cite{barvinok2018approximating} give a 
polynomial-time algorithm for the monomer-dimer problem.

\vskip.4in

{\bf Acknowledgements.} 
Thanks to Greg Huber for inspiration about tiling problems.
We are grateful to Tarun Grover, Isaac McGreevy and Yi-Zhuang You for helpful discussions and comments.
We thank Tyler Helmuth for telling us about references \cite{kenyon2021multinomial,alimohammadi2023fractionally,barvinok2018approximating}.
This work was supported in part by
funds provided by the U.S. Department of Energy
(D.O.E.) under cooperative research agreement 
DE-SC0009919, 
and by the Simons Collaboration on Ultra-Quantum Matter, which is a grant from the Simons Foundation (652264).  JM received travel reimbursement from the Simons Foundation;
the terms of this arrangement have been reviewed and approved by the University of California, San Diego in accordance with its conflict of interest policies. 

\vfill\eject

\appendix
\renewcommand{\theequation}{\Alph{section}.\arabic{equation}}

\section{Combinatorial problems on dimer pilings} \label{appendix:combinatorial-problems}

The different colorings, defined in section~\S\ref{sec:dimer-piling-problems}, lead to distinct combinatorial problems when $N>1$. We will be interested in understanding their different interpretations in terms of monomers and dimers fully covering a lattice. Before proceeding, though, we will first need to introduce some notation and definitions.

Let us define a monomer-dimer piling, corresponding to a graph $\Gamma$, as a fully-packed configuration of colorless monomers/dimers on $\Gamma$ such that, at each vertex, the total number of incident monomers and dimers must be $N$. We shall also define the monomer piling as the monomer-only packed configuration, and a dimer piling as a dimer-only packed configuration, both satisfying the same constraint as the monomer-dimer piling. For the sake of simplicity, we will interchangeably use piling when referring to the monomer-dimer piling.

Graphically, we will represent such pilings using $x$'s on the sites as monomers, and bonds on the links as dimers. 
As an example, we present below the periodic two-site chain for $N=2$ and their corresponding pilings. The graph is presented on the left side of the equality, while the pilings are on the right side. Numbers next to the pilings correspond to the number of equivalent pilings one can get. We identify the first piling as the monomer piling, and the last two as the dimer pilings.

\begin{figure}[h!]
    \centering
    \includegraphics[width=\textwidth]{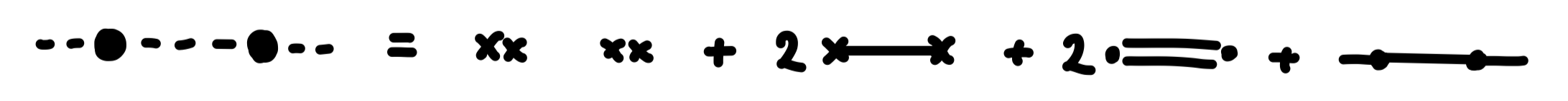}
    \caption{Two-site periodic chain and its corresponding monomer-dimer pilings.}
    \label{fig:two-site-pilings}
\end{figure}

Each of these pilings will come with a weight corresponding to its structure, i.e. the monomers at the sites and the dimers at the links of such piling.  We weight them with $1/(N^{n_l}n_l!)$ associated to the number of dimers $n_l$ at link $l$, times $x^{n_{m_i}}/n_{m_i}!$ associated to the number of monomers $n_{m_i}$ at site $i$. As shall we see, these pilings determine the weights in the combinatorial problem each model defines.

Having defined what a piling is, let us move on to the first coloring, type $A$. The combinatorial problem this coloring describes is one where we sum over all monomer-dimer pilings of a given graph $\Gamma$. All of these pilings are multiplied by an overall factor $N!$ to the total number of sites $n_s$. The partition function is thus, 
\begin{equation}
    Z_{N}^A(\Gamma,x) = (N!)^{n_s} \sum_{\substack{ \text{pilings } \mathcal{C} \text{ of } \Gamma }} \Bigg( \prod_{\text{sites }i \, \in \, \mathcal{C}} \frac{x^{n_{m_i}}}{n_{m_i}!} \Bigg) \Bigg( \prod_{\text{links }l\,\in\,\mathcal{C}}\frac{1}{N^{n_l\,} n_l!}\Bigg). \, \label{eq:coloring_A}
\end{equation}

The overall factor of model $A$ can be understood as follows. Recall that dimers of this model comes from the quartic term $V_A = A_{ij} \eta_i^\alpha \tilde \eta_i^\alpha \eta_j^\beta \tilde \eta_j^\beta$. We can think of the type-$A$ dimer as arising from two independent monomer terms: one at site $i$, and the other at site $j$. Equations~\eqref{eq:single-site} and~\eqref{eq:dimer-piling-condition} then say that every site of the graph will be filled with $N$ pairs of grassmann $\eta_i^\alpha \tilde \eta_i^\alpha$ of different colors (either coming from monomers or dimers), the $N!$ in~\eqref{eq:single-site} being the number of ways of arranging the $N$ colors at a single site. Hence, the factor $(N!)^{n_s}$ in the partition function $Z^A_N$ is the total number of arrangements at all sites $n_s$ of the graph $\Gamma$. 

In contrast, the two remaining models define a much different combinatorial problem. We can no longer understand the dimer as two independent monomers, since model $B$ and $C$ interactions couple the color indices of the grassmanns at site $i$ with those at site $j$. Each of the models define a unique way of placing the colorful dimers in the graph $\Gamma$.
And its weight will correspond to the number of ways of arranging such colorful dimers in the copies $\mathcal{C}_\eta$ and $\mathcal{C}_{\tilde \eta}$, as previously mentioned in \S\ref{subsec:combinatorial-integral}, following the rule that at each site of both $\mathcal{C}_\eta$ and $\mathcal{C}_{\tilde \eta}$ there must be $N$ different colorful dimers. 
In the following, we will make sense of this coloring problem and its combinatorial weight by means of a simple example: the last dimer piling in Figure~\ref{fig:two-site-pilings}. 

Let's first consider model $B$, whose interaction $V_B=A_{ij} \eta_i^\alpha \tilde \eta_i^\beta \eta_j^\alpha \tilde \eta_j^\beta$ decouples colorwise the $\eta$'s from the $\tilde \eta$'s. This type-$B$ dimer paints the link $\vev{ij}$ from $\mathcal{C}_\eta$ with color $\alpha$, and from $\mathcal{C}_{\tilde\eta}$ with color $\beta$. That is, each copy of $\mathcal{C}$ gets to be painted independently, always respecting that, at each site of $\mathcal{C}_\eta$ and $\mathcal{C}_{\tilde \eta}$, there must be $N$ different colorful dimers. So for the simple example, this gives four colorings:
\begin{figure}[h!]
    \centering
    \includegraphics[width=.8\textwidth]{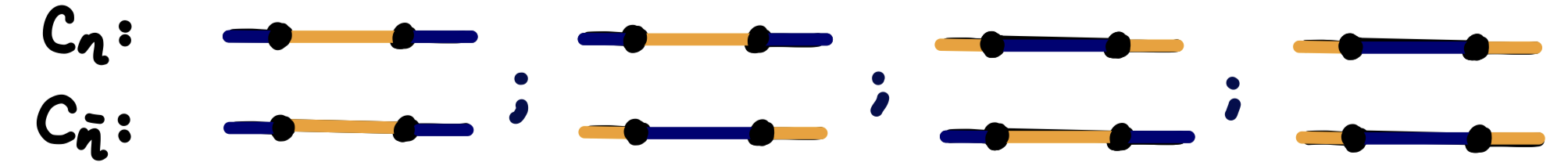}
    \caption{All four possibilities of the $N=2$ two-site periodic chain using coloring type-$B$.}
    \label{fig:two-site-coloring-B}
\end{figure}

On the other hand, model $C$ paint the copies of such piling in another manner. Instead of fully painting the dimer in $\mathcal{C}_\eta$ or $\mathcal{C}_{\tilde \eta}$ as in $B$, the interaction $V_C = A_{ij} \eta_i^\alpha \tilde \eta_i^\beta \eta_j^\beta \tilde \eta_j^\alpha$ paints with the same color half of the dimer in $\mathcal{C}_\eta$ and the other half in $\mathcal{C}_{\tilde \eta}$.
\begin{figure}[h!]
    \centering
    \includegraphics[width=.7\textwidth]{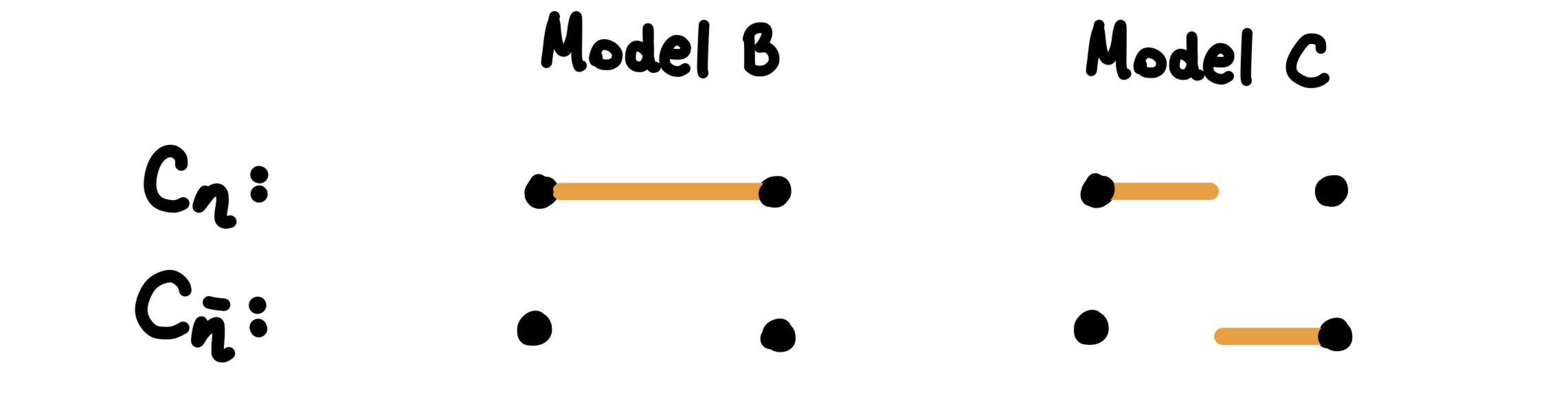}
    \caption{Model $B$ paints the link only in $\mathcal{C}_\eta$, while model $C$ paints half link in $\mathcal{C}_\eta$ and half in $\mathcal{C}_{\tilde \eta}$}
    \label{fig:B-vs-C}
\end{figure}

\begin{figure}[h!]
    \centering
    \includegraphics[width=\textwidth]{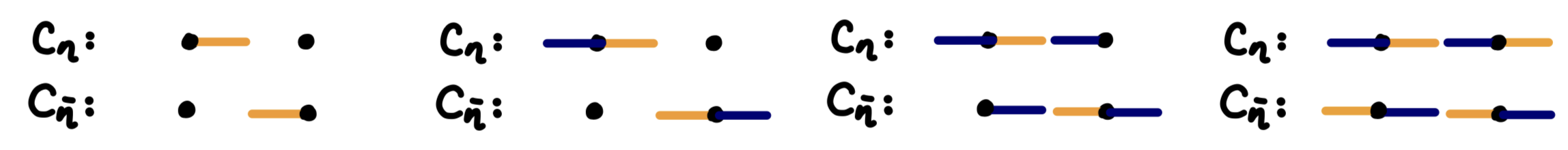}
    \caption{Step-by-step painting of one of the four possible colorings using type-$C$ dimers. Each site of either copy of $\mathcal{C}$, satisfies there are $N=2$ different colorful dimers. 
    }
    \label{fig:two-site-model-C-painting}
\end{figure}

One can then paint the example using this type-$C$ coloring rule and find that the number of colorings is again equal to four. One such coloring is presented figure~\ref{fig:two-site-model-C-painting}. 
From this example, we can understand why $B$ and $C$ are equivalent on bipartite graphs. They both define the coloring counting problem over disjoint sets of sites: model $B$ over the disjoint sites $\mathcal{C}_\eta$ and $\mathcal{C}_{\tilde \eta}$, and model $C$ over the two disjoint sublattices. Furthermore, we can see the action of the transformation in one of the two sublattices of the bipartite graph $(\eta_i^\alpha, \tilde \eta_i^\alpha) \to (\tilde \eta_i^\alpha, - \eta_i^\alpha)$ relating model $B$ with $C$: it maps the disjoint sets from one model to the other.

However, if one considers non-bipartite graphs, the models do differ. There exist piling terms that do not contribute to the partition function of model $B$, but do for model $C$. As an example, consider the following piling of the three-site chain: 
\begin{figure}[h!]
    \centering
    \includegraphics[width=.8\textwidth]{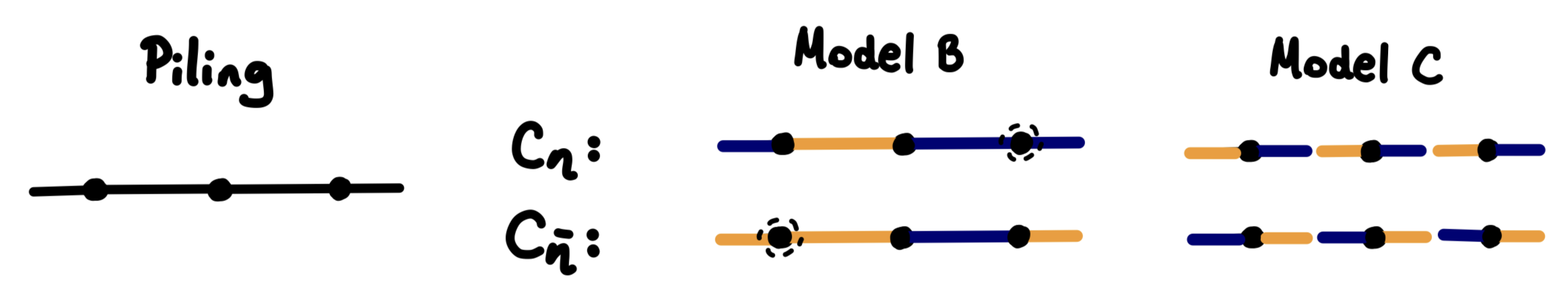}
    \caption{Three-site piling, painted using model $B$ and $C$. Using type-$B$ dimers, it's not possible. Circles show where the color rule is violated. Yet, using type-$C$ dimers, two colorings are possible.}
    \label{fig:coloring-B-vs-C}
\end{figure}

Moreover, this piling comes with an overall negative sign while none of the model $B$ pilings are negative. This begs the question: How can we tell whether a piling coloring comes with negative sign?  

Let us set up the ground by considering what colorings we are sure do come with a positive sign. And let us follow another dimer piling example, in this case of the $N=4$ two-site periodic chain and one of its colorings:
\begin{figure}[h!]
    \centering
    \includegraphics[width=.6\textwidth]{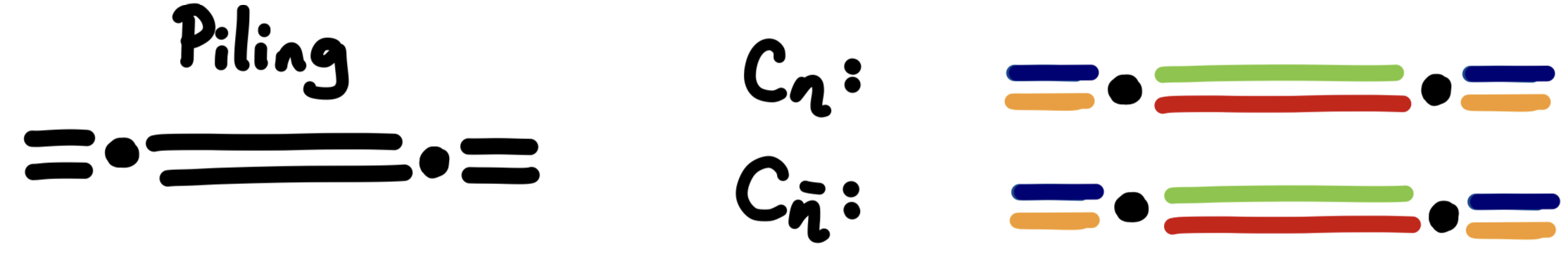}
    \caption{Two-site piling for $N=4$ and one of its homogeneous colorings, using either type-$B$ or $C$ dimers. The color structure is the same on $\mathcal{C}_\eta$ and $\mathcal{C}_{\tilde \eta}$. }
    \label{fig:N=4-two-site-piling}
\end{figure}
\\ We see that this coloring has the same color structure in both $\mathcal{C}_\eta$ and $\mathcal{C}_{\tilde \eta}$. It is an exact copy. We will denote these type of colorings as homogeneous colorings. Since they are the same colorwise, then the contributions associated to this coloring is associated with terms of the form $\eta_i^a \tilde \eta_i^a \eta_j^a \tilde \eta_j^a$, where the superscript $a$ is any of the colors. The integral of such terms does not produce a minus sign since we just need to move pairs of grassmanns ($\eta^a_i \tilde \eta^a_i$ or $\eta^a_j \tilde \eta_j^a$) to integrate them out. Hence, the homogeneous colorings will always come with a positive sign.  

What about the signs of colorings where $\mathcal{C}_\eta$ and $\mathcal{C}_{\tilde \eta}$ are different? Consider another coloring of the mentioned piling using type-$B$ dimers without loss of generality:
\begin{figure}[h!]
    \centering
    \includegraphics[width=.66\textwidth]{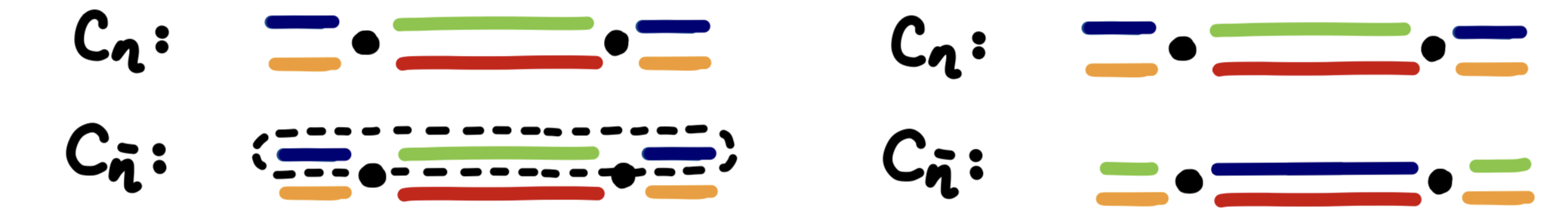}
    \caption{On the left, the homogeneous coloring. On the right, a contribution with different colorings in $\mathcal{C}_{\eta}/ \mathcal{C}_{\tilde \eta}$. We can arrive from the left to the right configuration by moving the circled dimers. This change carries a factor $(-1)^{l}$, where $l=2$ is the length of the loop.}
    \label{fig:N=4-two-site-piling-2}
\end{figure}
\\ As seen in the figure, this coloring can be interpreted as the result of starting from the homogeneous coloring and moving the dimers across the loop shown. Therefore, the sign factor will come from the integral associated to the loop that connects it to any homogeneous coloring. It is not hard to show that this integral carries an overall sign of $(-1)^{l}$ where $l$ is the length of the loop as measured in units of dimers.  

With all this in mind, we can see why there are no negative terms in the partition function of model $B$: there do not exist odd length loops which alternate colors, even in non-bipartite graphs. On the contrary, model $C$ can have these odd length loops and thus, negative terms in its partition function.  
Even though this raises interesting questions about these models in non-bipartite graphs, we will focus on bipartite lattices and postpone the study on non-bipartite graphs for future work. 
\begin{figure}[h!]
    \centering
    \includegraphics[width=.66\textwidth]{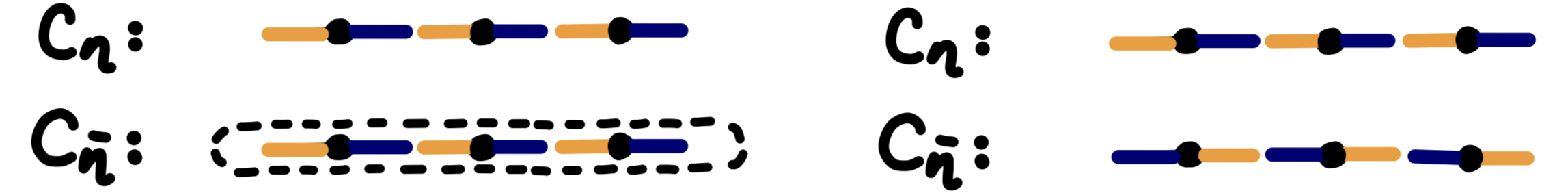}
    \caption{We can now understand the negative factor coming from this three-site piling. Starting from the homogeneous configuration on the left and moving the dimers along the loop gives a factor of $(-1)^{l}$, where $l=3$ is the length of this loop.}
    \label{fig:two-site-model-C}
\end{figure}

Finally, we can add monomers to the mix to get the weight corresponding to these colorings. These will paint both copies of $\mathcal{C}$ with the same color. The piling weight for model $B$ is then the number of ways of coloring $\mathcal{C}_\eta$ and $\mathcal{C}_{\tilde \eta}$ following the mentioned rules. All together, the partition function is 
\begin{equation}
    Z_{N}^B(\Gamma,x) = \sum_{\substack{ \text{pilings } \mathcal{C} \text{ of } \Gamma }} \Bigg( \prod_{\text{sites }i \, \in \, \mathcal{C}} \frac{x^{n_{m_i}}}{n_{m_i}!} \Bigg) \Bigg( \prod_{\text{links }l\,\in\,\mathcal{C}}\frac{1}{N^{n_l\,} n_l!}\Bigg) \bigg( \substack{ \text{\# ways coloring} \\ \text{ $\mathcal{C}_\eta$ and $\mathcal{C}_{\tilde \eta}$}  }  \bigg). \, \label{eq:coloring_B}
\end{equation}
To be more explicit about the right-most factor in \eqref{eq:coloring_B} consider the $N$-piling model on the graph which is two sites connected by a single link.  In that case, a dimer piling is specified by $k$, the number of dimers on the edge.  The contribution from $k$ dimers is then:
\be  \( {x^{2N-{2k}}\over (N-k)!^2 } \) \( {1\over N^k k! }\)  \(  C^N_k (N-k)!^2 ( k!)^2 \) . \label{eq:two-sites-coloring}\ee
In the last factor $C^N_k \equiv {N! \over (N-k)!k!}$ is the number of ways of choosing which colors to use for the monomers, $k!$ is the number of ways of coloring each set of dimers, and $(N-k)!^2$ comes from the choices of colorings of the monomers at each of the two sites.  

The description above of the sign of a contribution also applies to monomer-dimer pilings in model B (or in model C on a bipartite graph).

\section{Transfer matrix} \label{appendix:transfer-matrix}

In this appendix, we explicitly construct the transfer matrix for the $1D$ chain for all the dimer piling models $A$ to $C$.

{\bf Model A.} 
The partition function for model $A$ is
\be
Z^A_{N}(\Gamma_0) = \int D(\tilde \eta, \eta)^\alpha~e^{\sum_i x \eta_i^\alpha \tilde \eta_i^\alpha + \frac{1}{N} \sum_i \eta_i^\alpha \tilde \eta_i^\alpha \eta_{i+1}^\beta \tilde \eta_{i+1}^\beta} \label{eq:1D-chain-A},
\ee
where we denote the $1D$ chain graph as $\Gamma_0$ with adjacency matrix $A_{ij} = \delta_{i,j+1} + \delta_{i,j-1}$.

We can find the structure of the transfer matrix by expanding \eqref{eq:1D-chain-A} over the sites $i$ of the chain, 
\begin{align}
Z^A_{N}(\Gamma_0) &= \hspace{-0.1cm}\int \hspace{-0.1cm} D(\tilde \eta, \eta)^\alpha \prod_{i} \bigg(\sum_{n_{m_i}}\frac{x^{n_{m_i}}}{n_{m_i}!} (\eta_i^\alpha \tilde \eta_i^\alpha)^{n_{m_i}}  \sum_{n_{d_i}} \frac{1}{N^{n_{d_i}} n_{d_i}!} (\eta_i^\alpha \tilde \eta_i^\alpha \eta_{i+1}^\beta \tilde \eta_{i+1}^\beta )^{n_{d_i}} \bigg) \\
&= \hspace{-0.1cm} \int \hspace{-0.1cm}  D(\tilde \eta, \eta)^\alpha  \hspace{-0.25cm}\sum_{\{n_m\} \{ n_d\}} \prod_i  \(  (\eta_{i}^\alpha \tilde \eta_{i}^\alpha)^{n_{d_{i-1}}} \frac{x^{n_{m_i}}}{n_{m_i}!} (\eta_{i}^\alpha \tilde \eta_{i}^\alpha)^{n_{m_{i}}} \frac{1}{N^{n_{d_i}} n_{d_i}!} (\eta_{i}^\alpha \tilde \eta_{i}^\alpha)^{n_{d_{i}}}  \),
\end{align}
where, in the last step, we rearrange the terms to get all products of grassmann at site $i$ more explicit. We proceed by integrating over the grassmanns. Using the identity \eqref{eq:single-site}, we get
\begin{align}
    Z^A_N(\Gamma_0) &= \sum_{\{n_m\} \{ n_d\}} \prod_i \(  (N!)~ \frac{x^{n_{m_i}}}{N^{n_{d_i}} (n_{d_i})! (n_{m_{i}})! } \delta_{n_{m_i}+ n_{d_i}+ n_{d_{i-1}},N} \) \\
    &= \sum_{\{n_d\}} \prod_i \bigg( (N!)~ \frac{x^{N-n_{d_i}- n_{d_{i-1}}  }}{N^{n_{d_i}} (n_{d_i})! (N-n_{d_i}- n_{d_{i-1}})!} ~\theta(N- n_{d_i} - n_{d_{i-1}}) \bigg) \label{eq:transfer-matrix-def-A}
\end{align}
where $\theta$ is the Heaviside function which is equal to $1$ if $N-n_{d_i}-n_{d_{i-1}} \geq 0$, and $0$ otherwise. Equation~\eqref{eq:transfer-matrix-def-A} defines the transfer matrix connecting link $\vev{i-1,i}$ to $\vev{i,i+1}$, summed over the number of dimers on such links. 

As a first example, let's look at $N=1$. Here the possible values of the dimers on the links are $n_d = 0, 1$. The transfer matrix is hence a $2\times 2$ matrix whose rows correspond to the dimer states on the link $\vev{i-1,i}$, and whose columns correspond to the dimer state on the link $\vev{i,i+1}$. That is
\be\label{eq:transfer matrix N=1}
T_A = ~\parfig{.2}{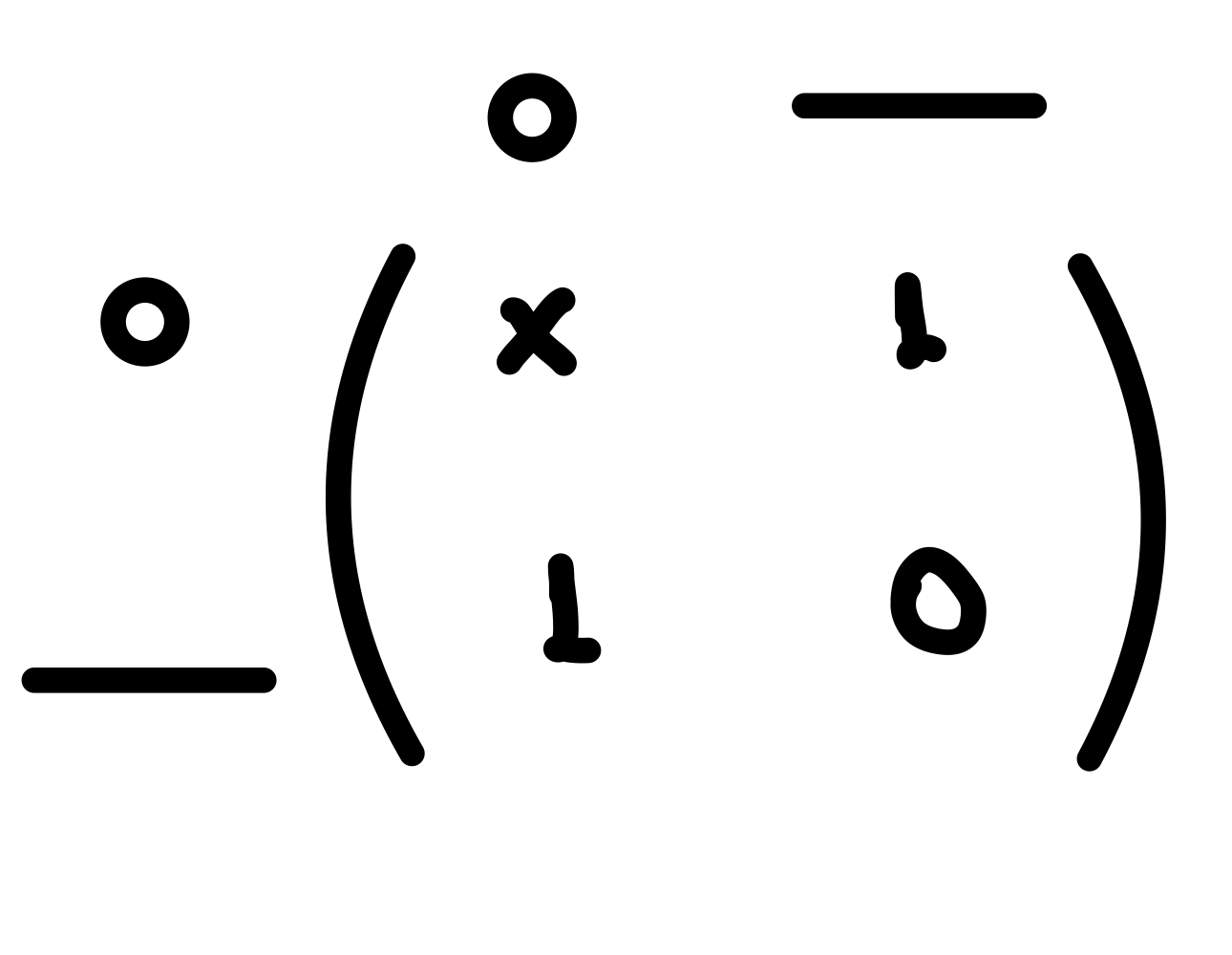}
\ee
and the partition function for a periodic chain is $Z = \tr (T_A^{n_s})$. This gives the free energy density $f=-\lim_{n_s\to \infty} \frac{1}{n_s} \ln Z$, the same as the known result $f = \ln (\frac{1}{2}(x+ \sqrt{x^2 +4}))$. 

For $N=2$ the transfer matrix is
\be
T_A =~\parfig{.4}{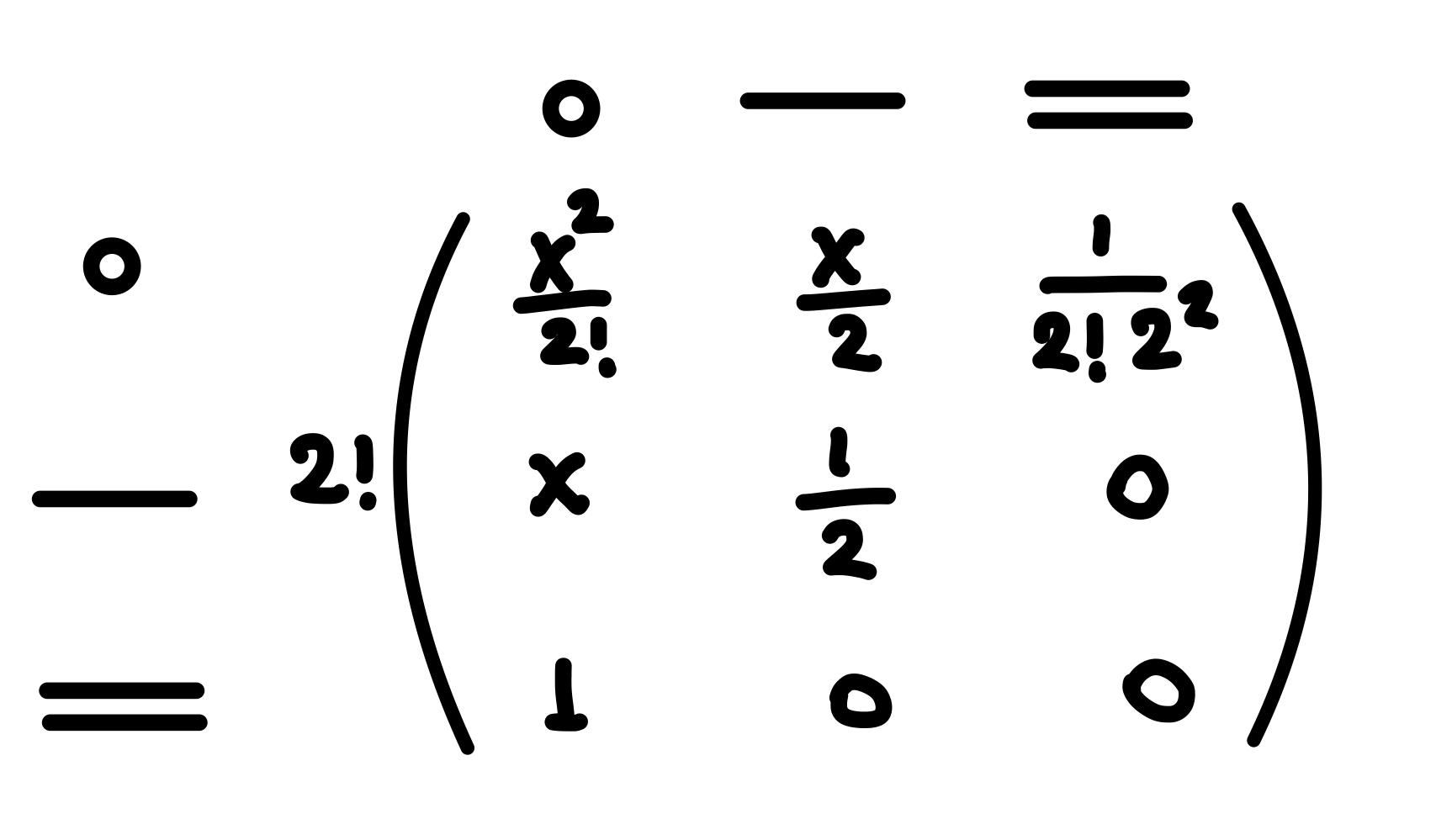}
\ee
where the vector space is now generated by no dimers, one dimer and two dimers on the links. 
Quite generally, for arbitrary $N$, the transfer matrix is given by 
\be
T^A_{ij} = N! \( {x^{N-i-j} \over (N-i-j)!} \) \( {1 \over N^j j!}\) \theta(N-i-j).
\ee
Free energy densities can then be obtained for finite $N$. Some examples are plotted in figure~\ref{fig:coloringA}.

It's difficult to find a general expression for the largest contribution to the partition function for arbitrary $N$. However, we can find this for two limits. 
For $x\to \infty$, the dominating term comes from the configuration with no dimers at any link, fully covered by monomers. Its free energy density is just $f = - N \ln(x)$. 

For $x \to 0$, the matrix is anti-diagonal and dominated by the homogeneous configuration with $N/2$ dimers on each link of the chain. In this limit, the free energy density is $f = - \frac{N}{2} \ln(\frac{2}{e}) - \frac{1}{2} \ln(2)$. This coincides with the result~\eqref{eq:1D-free-energy-density} we got using saddle point.

{\bf Models B and C.}  We can do the same procedure for the remaining models, and find their corresponding transfer matrix. It's important to point out that the transfer matrix for models $B$ and $C$ is now defined over the links of the two copies $\mathcal{C}_\eta$ and $\mathcal{C}_{\tilde \eta}$ (as explained in the appendix~\ref{appendix:combinatorial-problems}). The rows are the state on the link before $\vev{i-1,i}$ and the columns are the state on the link after $\vev{i,i+1}$. The entries of the matrix follow that for each $n_d$ dimer on the next link, we get a factor of $n_d!/N^{n_d}$; and for $n_m$ monomers, we get $x^{n_m}/n_m!$. Further more there is the constraint that all monomers and dimers must be ($N$) different colors at each site. 

To be more clear, let's do the simplest non-trivial example $N=2$ for model $B$. The transfer matrix is 
\be\label{eq:transfer-matrix-B-N=2}
T_B =~ \parfig{.55}{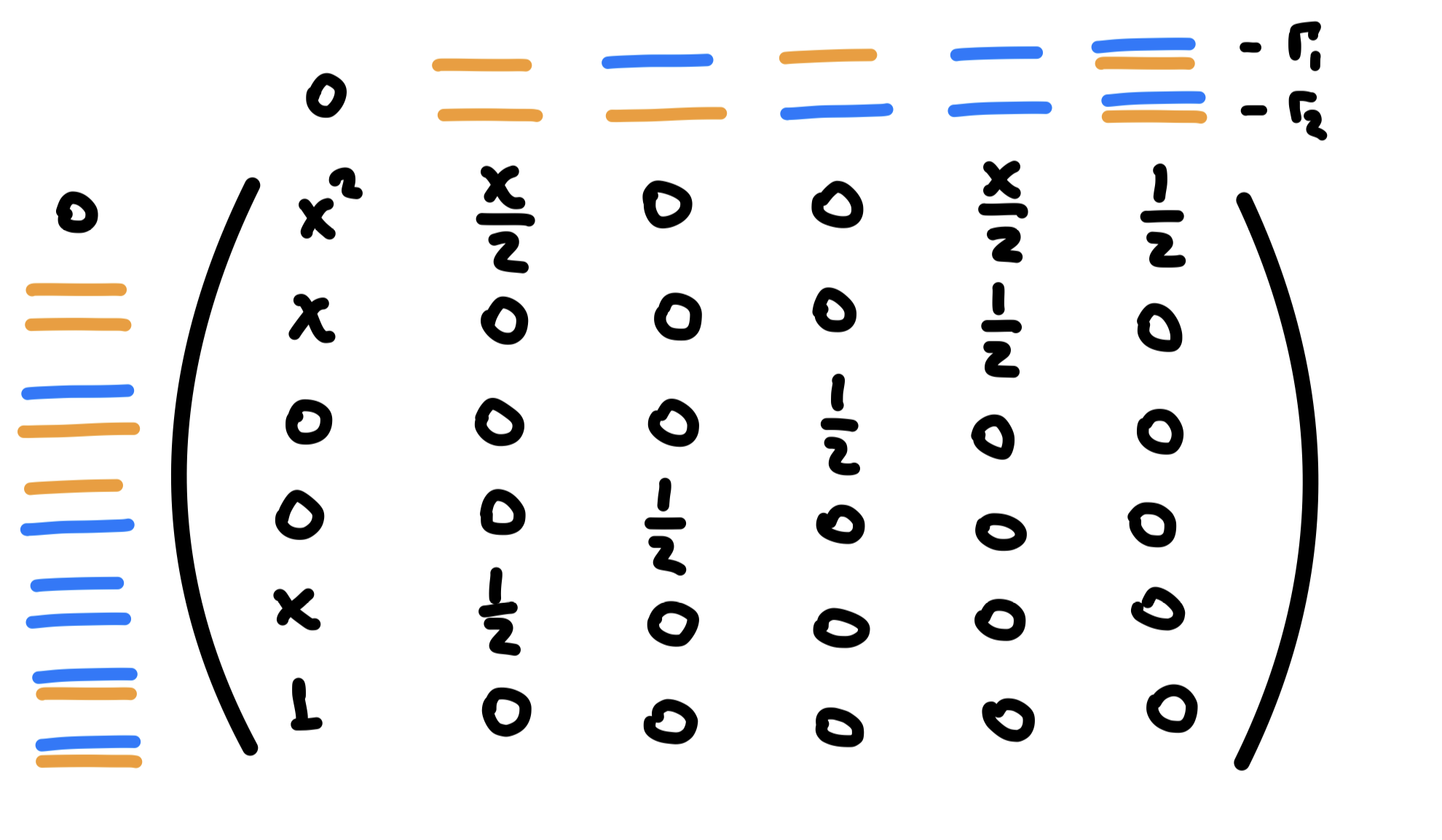}
\ee
where we have chosen the two colors to be blue and orange. As indicated in \eqref{eq:transfer-matrix-B-N=2}, the basis vectors are labelled as follows: no dimers, one orange dimer in both $\mathcal{C}_\eta$ and $\mathcal{C}_{\tilde \eta}$, one orange in $\mathcal{C}_\eta$ and one blue in $\mathcal{C}_{\tilde \eta}$, one blue in both $\mathcal{C}_\eta$ and $\mathcal{C}_{\tilde \eta}$, and two colorful dimers blue and orange in both $\mathcal{C}_\eta$ and $\mathcal{C}_{\tilde \eta}$. This transfer matrix is then used to plot figure~\ref{fig:1d_type_B_free_energy}~(a). It is not difficult to construct the matrices for the other $N$. 

We can again look at the dominant term in the transfer matrix in the two regimes mentioned for model $A$. For $x\to \infty$, the sum is dominated by fully covering the chain with monomers, with free energy density $-N \log x$. But for $x\to 0$, now we find that the dominant term is the chain with alternating structure: $N$ dimers on one link and $0$ on the next one. 
We find that the free energy density is given by $f=\frac{1}{2}N -\frac{1}{4} \ln(2\pi N)$. This agrees with the alternating mean field we used in the saddle point analysis for model $B$ up to order $\ln(N)$.  

Lastly, we can do $N=2$ for model $C$. The transfer matrix is
\be
T_C = ~\parfig{.55}{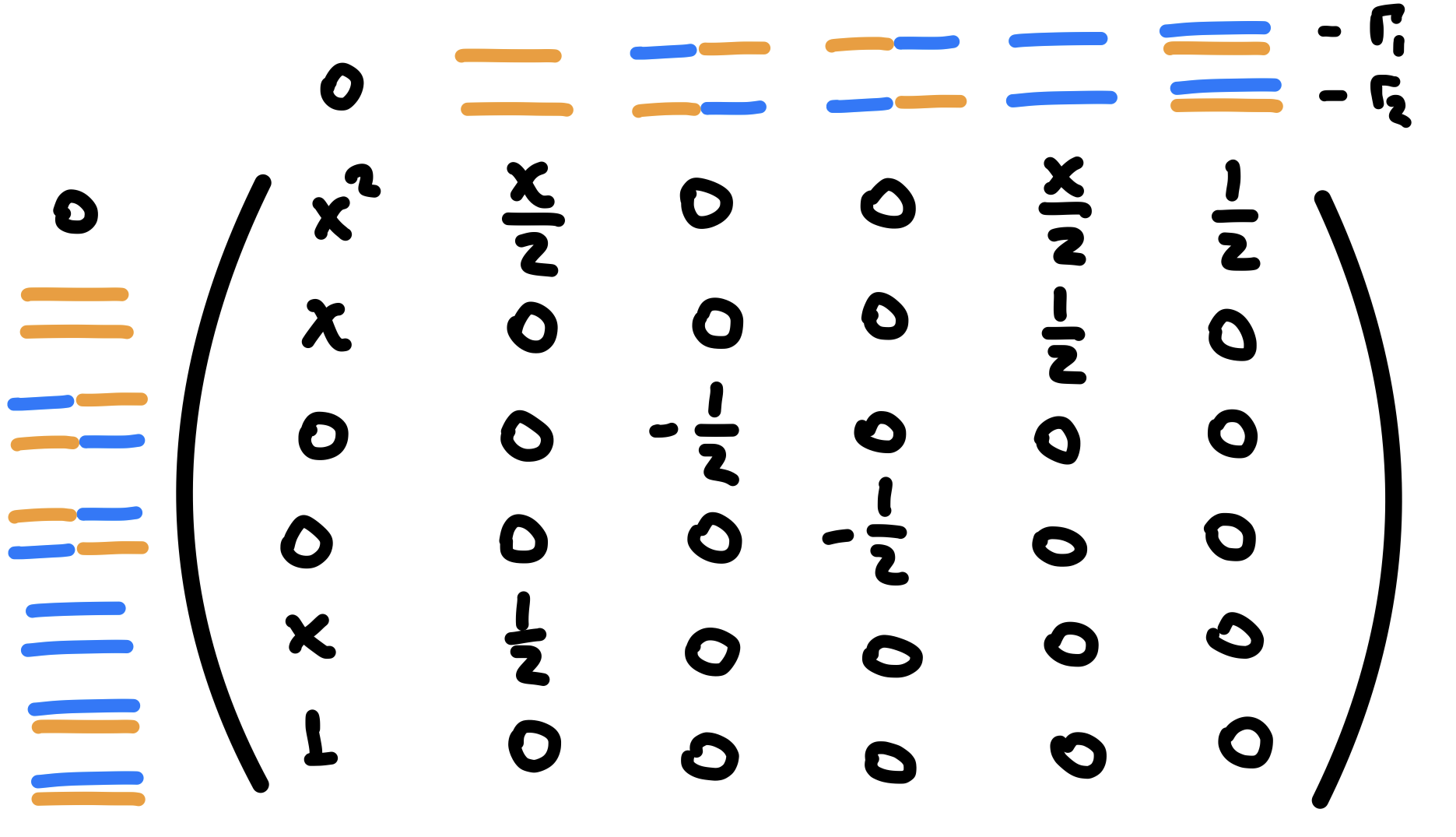}
\ee
where the difference now is that the basis vectors are labelled with half of the dimer painted with one color and the other half with another, and the entries are diagonal within that subspace.  

This transfer matrix also shows the negative terms that were discussed in appendix~\ref{appendix:combinatorial-problems}. We can see this by taking the partition function $Z = \tr(T^{n_s})$ of any odd number of sites. For example, for $n_s=1$, the partition function is $Z= x^2 -1$. These terms come from the fully covered configuration by monomers $(x^2)$, and the dimers-only configuration $(-1)$ respectively.

\section{Details of saddle point calculation for dimer piling models} \label{appendix:saddle-point-details}

{\bf Uniform saddle point for each link direction in model A.} In the body of the paper we studied model $A$ on a homogeneous lattice, where each link had the same weight, and found a homogeneous mean field solution, independent of the link direction. This was possible since there was no preferred direction, as the action~\eqref{eq:effective-action-A} only depends on the symmetric $A_{ij}$ matrix. We can ask how this picture would change if we instead consider a dimer fugacity that depends on the direction of the link.  This is included in the integral representation by generalizing $A_{ij}$ to include a direction-dependent weight.

Let's consider the action of model $A$~\eqref{eq:effective-action-A} on a Bravais lattice with weight $\omega_i$ in the direction $\hat e_i$:
\be
S_A[\phi,\phi^*]= \sum_{\vec r, i} |\phi_{\vec r, \vec r + \hat e_i}^{(\hat e_i)}|^2- \sum_{\vec r} \ln\left(x + \sum_{i} \omega_i (\phi_{\vec r, \vec r + \hat e_i}^{(\hat e_i)}+ \phi_{\vec r-\hat e_i, \vec r}^{(\hat e_i)*})\right)~.
\ee
Taking the first variation of the action, 
\begin{align}
    \frac{\delta S}{ \delta \phi^{(\hat e_i)*}_{\vec r, \vec r + \hat e_i}} &= \phi^{(\hat e_i)}_{\vec r, \vec r + \hat e_i} - \frac{\omega_i}{x+\sum_j \omega_j(\phi^{(\hat e_j)}_{\vec r + \hat e_i, \vec r +\hat e_i + \hat e_j} + \phi^{(\hat e_j) *}_{\vec r + \hat e_i - \hat e_j, \vec r + \hat e_i})} \\[0.2cm]
    \frac{\delta S}{ \delta \phi^{(\hat e_i)}_{\vec r, \vec r + \hat e_i}} &= \phi_{\vec r, \vec r + \hat e_i}^{(\hat e_i)*} - \frac{\omega_i}{x+ \sum_j \omega_j (\phi^{(\hat e_j)}_{\vec r, \vec r + \hat e_j} + \phi^{(\hat e_j)*}_{\vec r -\hat e_j, \vec r})}. 
\end{align}
We see that the mean field is still real ($\phi_0^{(\hat e_i)}= \phi_0^{(\hat e_i)*}$) with solution given by 
\be
\phi_0^{(\hat e_i)} = \omega_i \left( \frac{-x\pm \sqrt{x^2 +8\omega^2} }{4 \omega^2} \right),
\ee
where $\omega^2 = \sum_i \omega_i^2$. This reduces to the case with homogeneous mean field in all directions~\eqref{eq:meanfield_type_A} when $\omega_i =1$ as expected. 

We can take the second variation with respect to the field $\phi$ to find the second-order fluctuations~\eqref{eq:second-order-variation} around the mean field $\phi_0^{(\hat e_i)}$,
\be
S^{(2)}_A[\tilde \phi,\tilde \phi^*] = \sum_{\vec r, i} |\tilde \phi_{\vec r, \vec r + \hat e_i}^{(\hat e_i)}|^2 + \sum_{\vec r,i,j} \phi_0^{(\hat e_i)} \phi_0^{(\hat e_j)} \( \tilde \phi_{\vec r, \vec r + \hat e_i}^{(\hat e_i)} \tilde \phi_{\vec r- \hat e_j, \vec r }^{(\hat e_j)*} + \frac{1}{2} \tilde \phi_{\vec r, \vec r + \hat e_i}^{(\hat e_i)} \tilde \phi_{\vec r, \vec r + \hat e_j}^{(\hat e_j)} +\frac{1}{2} \tilde \phi_{\vec r, \vec r + \hat e_i}^{(\hat e_i)*} \tilde \phi_{\vec r+\hat e_i - \hat e_j, \vec r + \hat e_i}^{(\hat e_j)*}  \),
\ee
which we can simplify by going to Fourier space
\be
    S^{(2)}_A[\tilde \phi, \tilde \phi^*] = \sum_{\vec p, \hat e_i} |\tilde \phi_{\vec p}^{(\hat e_i)} |^2 + \sum_{\vec p, \hat e_i, \hat e_j}  \phi_0^{(\hat e_i)} \phi_0^{(\hat e_j)}  \left( \tilde \phi_{\vec p}^{(\hat e_i)} \tilde \phi_{\vec p}^{*(\hat e_j)} e^{i \vec p \, \hat e_j} + \frac{1}{2} \tilde \phi_{\vec p}^{(\hat e_i)} \tilde \phi_{-\vec p}^{(\hat e_j)} + \frac{1}{2} \tilde \phi_{\vec p}^{* (\hat e_i)} \tilde \phi_{-\vec p}^{*(\hat e_j)} e^{i\vec p\,(\hat e_i - \hat e_j)} \right). \label{eq:second_order_var_A_momentum_space}
\ee
We can find the spectrum of the Gaussian fluctuations to get the effective free energy density. We get 
\be
    f_A = N \bigg( \sum_{i} |\phi_0^{(\hat e_i)}|^2 - \ln\big|x+\sum_i 2\omega_i \phi_0^{(\hat e_i)} \big| \bigg) +  \frac{\nu_0}{2} \int_{BZ} \frac{d^dp}{(2\pi)^d} \ln\Big(1+ \sum_{i}2 |\phi_0^{(\hat e_i)}|^2 \cos(\vec p \cdot \hat e_i) \Big), \label{eq:weighted-free-energy-density-A}
\ee
where $\nu_0$ is the unit cell volume, $z$ is the coordination number, and $\hat e_i$ are the primitive vectors. Again, this simplifies to $\eqref{eq:free-energy-density-A}$ when taking the limit $\omega_i =1$.

Singularities arise from the $1/N$ corrections at $x=0$ where the argument of the log function becomes zero. For example, for the triangular lattice, this can only happen when one of the weights $\omega_1=0$ along one of the directions. 
In this limit, the triangular lattice model degenerates into the square lattice dimer model.
This is in accordance with the fact that the square-lattice dimer model is indeed critical when $N \to 1$, with algebraic decay of correlators~\cite{kasteleyn1963dimer,fisher1963statistical,fendley2002classical}.

{\bf Uniform saddle point in model B.} As we saw in figure~\ref{fig:action-solutions}, for model B, the homogeneous solution $\(\phi_0\)_{ij} = \phi_0$ is not the energetically superior mean field. 
Here we present a detailed analysis of the homogeneous ansatz to illustrate the instability.  It is conceivable that there is some observable for which the contribution of the textured saddle vanishes and the homogeneous one therefore dominates.

Using the homogeneous ansatz $\phi_0$ in the $1D$ periodic chain, we find that the action is 
\be
S_B[\phi,\phi^*] = n_s |\phi_0|^2- \frac{1}{2} \tr \ln  \begin{pmatrix}
        -B(\phi^*_0) & -x\mathbb{1} \\ 
        x\mathbb{1} & B(\phi_0)
    \end{pmatrix},
\ee
where the block matrix $B$ is, in momentum space, 
\be
B(\phi_0) = -2i\phi_0 \sum_p \sin(p) \ket{p}\hspace{-0.1cm}\bra{p}~.
\ee
We can simplify the action to get 
\be
S_B[\phi_0,\phi_0^*] = n_s |\phi_0|^2-\frac{1}{2} \sum_p \ln(x^2 + 4 |\phi_0|^2 \sin^2(p)),
\ee
and minimize the action with respect to $\phi_0$ 
\be
\frac{\delta S_B}{\delta\phi_0^*} = n_s \phi_0 - \frac{1}{2} \sum_p \frac{4\phi_0 \sin^2(p)}{x^2 + 4 |\phi_0|^2 \sin^2(p)} = 0. 
\ee
The solution to this gap equation is presented in figure~\ref{fig:action-solutions}.

We can do the same procedure to find the gap equation for different lattices. If we consider Bravais lattices, we find that the gap equation is 
\be
\phi_0 = \frac{\nu_0}{z} \int \frac{d^dp}{(2\pi)^d} \(\frac{4\phi_0 \(\sum_{\hat e_i} \sin(\vec p \cdot \hat e_i)\)^2}{x^2 + 4 |\phi_0|^2\(\sum_{\hat e_i} \sin(\vec p \cdot \hat e_i)\)^2} \), \label{eq:gap-equation-bravais}
\ee
where $\nu_0$ is the unit cell volume, $z$ is the coordination number, and $\hat e_i$ are the primitive vectors. 

\begin{figure}[t!]
    \centering
    \includegraphics[width=.5\textwidth]{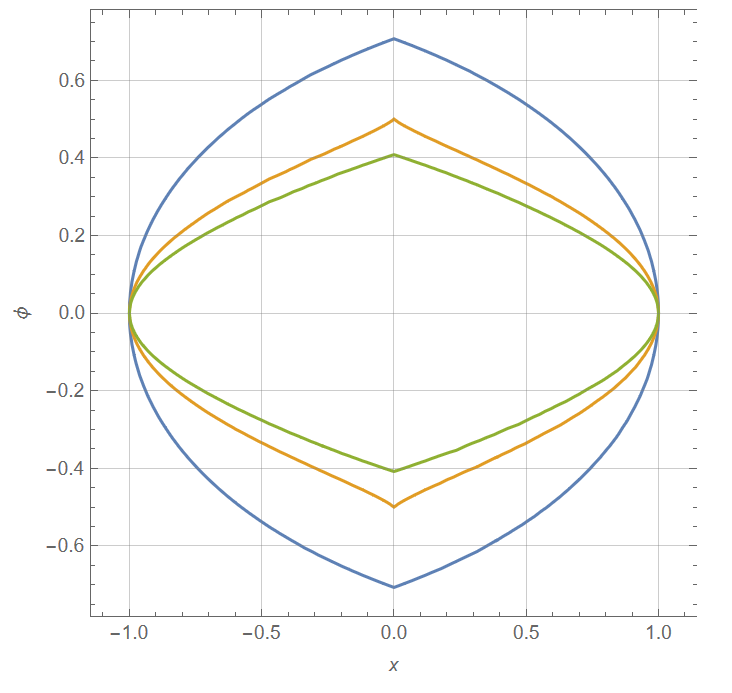}
    \caption{Order parameter $\phi$ as a function of $x$ for different dimensions $d=1,2,3$ where the blue curve is $d=1$, the orange curve is $d=2$ and the green curve is $d=3$. Two critical points can be seen at $x=0, x=\pm1$. }
    \label{fig:order_parameter_B}
\end{figure}

Figure~\ref{fig:order_parameter_B} shows the solution of~\eqref{eq:gap-equation-bravais} for the $d$-dimensional cube. The different curves represent $\phi_0$ as a function of the monomer fugacity $x$, for $d=1,2,3$. For all $d$, two special points can be spotted in the plot $x=0$ and $x=\pm 1$.

We proceed to calculate the behavior around those points. Let's first start with the points $x_c=\pm 1$ where the gap becomes zero. Expanding~\eqref{eq:gap-equation-bravais} about $x=x_c-\delta x$ and $\phi = \delta \phi$, we get 
\be
|\delta \phi| = \frac{1}{\sqrt{\alpha x_c}} \(\delta x\)^{\frac{1}{2}},
\ee
where the value of the constant $\alpha$ is given by 
\be
\alpha = \frac{2\nu_0}{z x_c^4} \int \frac{d^dp}{(2\pi)^d} \Big(\sum_{\hat e_i} \sin(\vec p \cdot \hat e_i )\Big)^4. 
\ee
We see that the scaling near $x_c$ is of the square-root form $\delta \phi \sim (x_c-x)^{\frac{1}{2}}$, independent of the lattice details.

Next, let's consider the behavior of the gap $\phi_c$ at $x=0$. Once again, we can find the scaling by expanding~\eqref{eq:gap-equation-bravais} around $\phi = \phi_c + \delta \phi$ and $x = \delta x$ to find
\be
\delta \phi = - \delta x^2 \( \frac{\nu_0}{2 z \phi_c} \int \frac{d^dp}{(2\pi)^d} \frac{1}{\delta x^2 + 4\phi_c^2 \(\sum_{\hat e_i} \sin(\vec p \cdot \hat e_i)\)^2}\).
\ee
We see that there will be factors of $\delta x$ coming from the integral. Low energy degrees of freedom will play an important role in evaluating the scaling around $x=0$. Assuming that there's a Fermi surface, then the scaling is $\delta \phi \sim |\delta x|$. In contrast, if we have Dirac points, then it will depend on the number of dimensions. For example, for the honeycomb lattice, the scaling near the gap is $\delta \phi \sim \delta x^2 \ln \delta x$.

{\bf Alternative decoupling channel in model B.} Previously, we have studied the partition function for model $B$ in a charged $\gU(1)$ decoupling channel \eqref{eq:HS_B}. We concluded that the uniform saddle in model B is energetically inferior to a textured one. Here, we show that decoupling in a different channel (the \gU(1) preserving channel) is also subleading.

We introduce the auxiliary Hubbard-Stratonovich fields $P^{\alpha \beta}_{ij}$ to decouple the type-$B$ interaction in this other channel: 
\be
    e^{\frac{1}{N} \sum_{\vev{i,j}} A_{ij} \eta_i^\alpha \tilde \eta_i^\beta \eta_j^\alpha \tilde \eta_j^\beta } = \int DP~ e^{\sum_{\vev{i,j}}  -\frac{1}{2} (P^{\alpha \beta}_{ij})^2 + \sqrt{\frac{A_{ij}}{N}} P^{\alpha \beta}_{ij} (\eta^\alpha_i \tilde \eta^\beta_j + \eta_j^\alpha \tilde \eta_j^\beta ) }.
\ee
Integrating out the $\eta, \tilde \eta$ variables we get 
\be
    Z^B_N(\Gamma,x)= \int DP~e^{-S[P]}, 
\ee
where the measure is $DP = \prod_{\vev{i,j}}\prod_{\alpha, \beta} \frac{dP_{ij}^{\alpha \beta}}{\sqrt{2\pi}}$, and the action $S[P]$ is given by 
\be
    S[P] = \sum_{\vev{i,j}} \frac{1}{2} (P_{ij}^{\alpha \beta})^2 - \sum_i \frac{1}{2} \tr \ln \begin{pmatrix}
    0 & M_i \\
    - M_i^T & 0 
\end{pmatrix} 
\ee
and the block matrix $M_i$ is 
\be
    M_i = \sum_{\alpha, \beta} \bigg( x \delta_{\alpha \beta} + \sum_{\vev{i|j}} \sqrt{\frac{A_{ij}}{N}} (P_{ij}^{\alpha \beta} + P_{ji}^{\alpha \beta} ) \bigg).
\ee

Taking the $\gSO(N)$-invariant and translation-invariant solution $P^{\alpha \beta}_{ij} = P_0 \, \delta_{\alpha \beta}$, the action becomes 
\be
    S_0 = \frac{1}{2} n_s N P_0^2 - n_s N \ln \Big| x+ \frac{z P_0}{\sqrt{N}} \Big|. 
\ee
Taking the first variation of the action, we find that the value of $P_0$ is given by 
\be
    P_{0,\pm} = \frac{1}{4}\left(- \sqrt{N} x \pm \sqrt{N x^2 +16} \right),
\ee
where $P_{0,+}$ dominates when $x>0$ and $P_{0,-}$ when $x<0$. 

Leading order behavior in powers of $N$ is plotted in Figure~\ref{fig:action-solutions}. We see that, for all values of $x$, the textured ansatz is energetically superior to this new solution. Hence, although these two variables are in different decoupling channels, they compete with each other and we find that the charged $\gU(1)$ channel is dominant.

{\bf Gauge fixing for model B.} We have seen that the spectrum of~\eqref{eq:singular-matrix-1D-chain-B} has zero eigenvalues, making the matrix singular. We have argued that this is a result of the gauge redundancy $U: \phi_{ij} \to e^{i(\theta_i+\theta_{j})} \phi_{ij}$ of the model. We will show how to lift the redundancy using the procedure known as gauge fixing, introduced by Faddeev and Popov~\cite{FADDEEV196729}. 

Let us denote the gauge condition that we will impose by $g(\phi,\phi^*)=0$, where we take 
\be
g(\phi,\phi^*) = \phi_{i,i+1}- \phi^*_{i,i+1},
\ee
and consider the expression defined by the group integral 
\be
\Delta_g^{-1}[\phi,\phi^*] = \int DU~\delta(g(\phi^U, \phi^{*U})), \label{eq:faddeev-popov-determinant}
\ee
where $\phi^U,\phi^{*U}$ are configurations related by the gauge transformation $U$ to $\phi,\phi^*$ respectively. 

The quantity $\Delta_g[\phi,\phi^*]$ is known as the Faddeev-Popov determinant. This quantity is gauge invariant by construction. This can be shown to be 
\be
\Delta_g[\phi,\phi^*] = \det \left|\frac{\delta g}{\delta \theta}\right|_{g=0} = \prod_{i=1}^{n_s} \left| 2 \phi_{i,i+1}\right|.
\ee
Since the mean field we are expanding around is 
\be
\(\phi_0\)_{i,i+1} = \phi_\pi \( 1+e^{i \pi x_i}\), \label{eq:alternating-mean-field}
\ee
we just need to impose the gauge-fixing condition on the even links $\vev{2i,2i+1}$. Else, the gauge would be singular.

Let's insert the identity in the form \eqref{eq:faddeev-popov-determinant} in the path integral for the $1D$ periodic chain we have been discussing, 
\begin{align}
    Z^N_B &= \int D(\phi,\phi^*) \int DU~ \delta(g(\phi,\phi^*))\Delta_g [\phi,\phi^*]~e^{-N S_B[\phi,\phi^*]} \\
    &= \big({2\pi}\big)^{\frac{n_s}{2}}\int D(\phi,\phi^*)~\delta(g(\phi,\phi^*))\Delta_g [\phi,\phi^*]~e^{-N S_B[\phi,\phi^*]}, \label{eq:result-gauge-fixing}
\end{align}
where, in the last line, we changed the order of integration and get a factor of the volume of the gauge group $(2\pi)^{n_s/2}$. 
The result after gauge fixing~\eqref{eq:result-gauge-fixing} is a mixture of real and complex fields: real fields on the even links $\vev{2i,2i+1}$ and complex fields on the odd links $\vev{2i-1,2i}$. We will denote the real fields as $\varphi$ and the complex still as $\phi$.

We can proceed with the saddle point analysis. Expanding around the alternating mean field $(\phi_0)_{2i-1,2i}= 2\phi_\pi$ and $\varphi_{2i,2i+1}=0$, we get that the action governing the second-order fluctuations~\eqref{eq:second-order-variation} is
\be
S^{(2)}[\varphi, \phi, \phi^*] = \sum_{i} (1-x^2) |\tilde \phi_{2i-1,2i}|^2+\sum_i 2 (1-x^2) ( \tilde \varphi_{2i,2i+1} )^2,
\ee
leading to the effective partition function~\eqref{eq:effect-partition} to be 
\begin{align}
Z_B^{N}(\Gamma_0) &= (8 \pi \phi_\pi )^{\frac{n_s}{2}} e^{-N S_0} \int D(\phi,\phi^*)~e^{-N (1-x^2) \sum_i  |\tilde \phi_{2i-1,2i}|^2 } \int D\varphi~e^{-2N(1-x^2) \sum_i \tilde \varphi^2_{2i,2i+1}} \nonumber \\
&=  (8 \pi \phi_\pi )^{\frac{n_s}{2}} e^{-N S_0} \( (1-x^2)^{-\frac{n_s}{2}} \) \bigg( \(\frac{N}{\pi}\)^{\frac{n_s}{4}} \( 2(1-x^2)\)^{-\frac{n_s}{4}} \bigg),
\end{align}
where $\phi_\pi^2 = \frac{1}{4}(1-x^2)$. We see that we get terms proportional to $N^{\frac{n_s}{4}}$ due to the measure of the integral $D\varphi = \prod_i \(\frac{N}{\pi} d\tilde\phi_{2i,2i+1}\)$. 

We finally find that the free energy density is 
\be
f_{1D} = \frac{N}{2}(1-x^2) - \frac{1}{4} \ln(8\pi N) + \frac{1}{2}\ln(1-x^2), \quad \left|x\right|<1.
\ee

{\bf Singularity at $x=1$.}
Where does this singularity come from? If we do a study of the contributions at large $x$ ($x>1$) we see that corrections come from a gas of dilute dimers.
Doing the expansion, we will see that it matches perfectly with the values of the partition function but only in the $N \to \infty$ limit. 

Let's be more precise about these statements. Consider the free energy density we obtained in~\eqref{eq:free-energy-density-general-B}. Its corresponding partition function is given by 
\be
    Z_B \equiv e^{-n_s f_B} = \frac{x^{n_s N}}{(1-1/x^2 )^{\frac{n_s z}{2}}}, \quad \quad \left|x\right|>1. 
\ee
Expanding and collecting terms in $x$, we get 
\be
    Z_B = x^{n_sN} + \frac{1}{2} n_s z\, x^{n_sN -2} + \frac{1}{2!} \left(\frac{n_s z}{2}\right) \left(\frac{n_s z}{2}+1 \right)x^{n_s N-4} + \mathcal{O}({x^{n_sN-6}}). \label{eq:general-lattice-expansion-large-N}
\ee
Each of these terms can be traced to contributions coming from dilute dimers: the $k$th term comes from $2k$ dilute dimers in a lattice of coordination number $z$ in the limit $N \to \infty$. 
As an example, let's explicitly look where the weight from the two-dimer contribution $x^{n_sN-4}$ comes from. At this order we can have two independent dimers, two connected dimers at a single site, or two dimers on top of each other at the same link. Each of these comes with the corresponding weight:
\begin{align}
    &= x^{n_sN-4}\left( \frac{1}{2!} \left(\frac{n_s z}{2} \right) \left(\frac{n_s z}{2} -(2z-1) \right) + \frac{n_s z(z-1) }{2} \left( \frac{N-1}{N}\right)  + \frac{ n_s z}{2} \left( \frac{N-1}{N} \right) \right) \nonumber \\
    &\buildrel{N\to\infty}\over{=} x^{n_s N-4} \left( \frac{1}{2!} \left( \frac{n_s z}{2} \right) \left( \frac{n_s z}{2} +1 \right) \right)
\end{align}
which matches exactly with~\eqref{eq:general-lattice-expansion-large-N} only in the limit $N \to \infty$. 
A similar match can be made at each order in $k$, the number of dimers.
This also agrees with the result for any graph of uniform coordination number that we found using the large-$N$ diagrammatic expansion, \eqref{eq:free-energy-density-general-B}. 

\section{Details of calculations for coloring models} \label{appendix:coloring-details}

{\bf Time-reversal in the coloring models.} The models defined in section~\ref{sec:graph-coloring} are symmetric under time-reversal: there exists an antiunitary operator $\mathcal{T}$ satisfying $\mathcal{T}^2 = -1$ and $i \rightarrow \mathcal{T}^{-1} i \,\mathcal{T} = -i$. In the following, we will see the explicit realizations of such symmetry and the transformation properties of the HS fields under time-reversal in each of our integral representations.
This provides a useful check on the saddle point calculations, and guarantees that the resulting estimate for the free energy is real.

Let's start with the integral representation of the planar 3-coloring model described in subsection~\ref{subsec:planar-three-coloring}. Time-reversal is implemented via 
\be
    \eta_A^\alpha \rightarrow \mathcal{T}^{-1} \eta_A^\alpha \mathcal{T} = \eta_B^\alpha, \quad \quad \eta_B^\alpha \rightarrow \mathcal{T}^{-1} \eta_B^\alpha \mathcal{T} =- \eta_A^\alpha.
\ee
which indeed leaves the action invariant~\eqref{eq:deformed-first-action}. On the other hand, the HS field we introduced is odd under time-reversal,
\be
\phi_{ij} \rightarrow \mathcal{T}^{-1} \phi_{ij} \mathcal{T} = - \phi_{ij},
\ee 
as can be seen from \eqref{eq:HS-planar-3-coloring}, and therefore so are the $\phi$ saddle points. 

Next, let's consider time-reversal on the representation for a signed sum of colorings in~\ref{subsec:signed-3-colorings}. Assuming the 3-valent graph is bipartite, time reversal acts as
\be
    \eta^\alpha_A \rightarrow \mathcal{T}^{-1} \eta_A^\alpha \mathcal{T} = \eta^\alpha_B, \quad \quad \eta_B^\alpha \rightarrow \mathcal{T}^{-1} \eta_B^\alpha \mathcal{T} = - \eta_A^\alpha.
\ee
Note that this leaves all terms in the action~\eqref{eq:second-action-signed-colorings} invariant. In particular, the Gaussian term is invariant since the matrix $A'_{ij}$ is antisymmetric.  Equation~\eqref{eq:first-signed-HS} shows that $\varphi_{ij}$ is odd under time-reversal,
\be\label{eq:signed-T1}
    \varphi_{ij} \to \mathcal{T}^{-1} \varphi_{ij} \mathcal{T} = -\varphi_{ij}.
\ee
On the other hand, equation~\eqref{eq:second-signed-HS} says that $\phi_{ij}$ is even, 
\be\label{eq:signed-T2}
    \phi_{ij} \to \mathcal{T}^{-1} \phi_{ij} \mathcal{T} = \phi_{ij},
\ee
as long as $i \varphi_{ij}>0$, or, otherwise, it is odd 
\be\label{eq:signed-T3}
    \phi_{ij} \to \mathcal{T}^{-1} \phi_{ij} \mathcal{T} = -\phi_{ij}.
\ee

Lastly, we consider the integral representation for the 3-coloring problem on an arbitrary 3-valent graph studied in subsection~\ref{subsec:non-planar-3-colorings}. In this model, time-reversal is realized by 
\be
    \eta_i^\alpha \rightarrow \mathcal{T}^{-1} \eta_i^\alpha \mathcal{T} = \tilde \eta_i^\alpha, \quad \quad \tilde \eta_i^\alpha \rightarrow \mathcal{T}^{-1} \tilde \eta_i^\alpha \mathcal{T} = - \eta_i^\alpha.
\ee
The first field, $\varphi$, introduced in~\eqref{eq:first-non-planar-HS} is odd under time-reversal, 
\be\label{eq:nonplanar-T1}
    \varphi_{ij} \to \mathcal{T}^{-1} \varphi_{ij} \mathcal{T} = - \varphi_{ij} 
\ee 
The second HS field $\phi$ in equation~\eqref{eq:second-non-planar-HS} is odd if $i\varphi >0$,
\be\label{eq:nonplanar-T2}
    \phi_{ij} \to \mathcal{T}^{-1} \phi_{ij} \mathcal{T} = - \phi_{ij}
\ee
or otherwise even
\be\label{eq:nonplanar-T3}
    \phi_{ij} \to \mathcal{T}^{-1} \phi_{ij} \mathcal{T} = \phi_{ij}.
\ee 
Finally, the last link field $P$ in~\eqref{eq:third-non-planar-HS} is time-reversal odd
\be\label{eq:nonplanar-T4}
    P_{ij}^\alpha \to \mathcal{T}^{-1} P_{ij}^\alpha \mathcal{T} = - P_{ij}^\alpha.
\ee

{\bf Saddle point analysis for planar 3-coloring model.}
Here we give the details of the saddle point calculations presented in the main text for the planar 3-coloring model. That is, we will show which saddles contribute to the free energy density and which do not. This will explain figure~\ref{fig:free-energy-u} and the results therein.   

Let us start by considering the planar $3$-coloring model, described in \ref{subsec:planar-three-coloring}. By introducing a HS link field $\phi$, we found a partition function with action~\eqref{eq:action-deformed-u-model} and whose saddles satisfy~\eqref{eq:mean-field-equations-deformed-u-model}. The solutions to the mean field equations are: 
\be
    \phi_0^{\pm} = \pm \frac{1}{6u} \sqrt{12u^2-9} + \frac{1}{2u}i.
\ee
A critical point exists at $u_0=\sqrt{3}/2$, separating two phases where time reversal is respectively broken $(u>u_0)$ and unbroken $(u<u_0)$. 

The idea is now to evaluate the Hessian at these points and check whether we can deform the integration contour to pass through them. To that end, let's take the second variation of the action with respect to $\phi_0$: 
\be
    \left.\frac{\delta^2 S}{\delta  \phi_{0}^2} \right|_{\phi_0^\pm} = \frac{3}{2} n_s + \frac{n_s}{2} \left(\frac{ iu}{1+iu\phi^\pm_0} \right)^2 = \frac{3}{2} n_s \left( 1 +  3 \phi_0^{\pm\, 2} \right),
\ee
where we have used the mean field equation in the last equality. We see that depending on the value of $u$, the second variation is either complex $(u>u_0)$ or real $(u<u_0)$. 

Given the second variation of the action evaluated at the saddle points, the angle in the complex plane at which the path of steepest descent passes through the saddle is given by the phase:
\be
    \alpha = -\frac{1}{2} \arg  \left.\frac{\delta^2 S}{\delta  \phi_{0}^2} \right|_{\phi_0^\pm}.
\ee
 For $u>u_0$, the phases at $\phi_0^{\pm}$ are complex, and depend on the value of $u$. Yet, we can always deform the original contour (the real line) to pass through both of these saddles. On the other hand, for $u<u_0$, the phase $\alpha$ does not depend on $u$ but only on the saddle $\phi_0^{\pm}$. For $\phi_0^+$, the second order variation is always negative, and its associated phase $\alpha = - \pi/2, \pi/2$. In other words, the steepest descent path is in the imaginary axis. Therefore, it is not possible to deform the contour in the real axis to pass through this saddle. For $\phi_0^-$, the second order variation is instead always positive, and the phase $\alpha = 0, \pi$. Hence, we can access this saddle and it will be the only one contributing to the free energy density in this regime. All these results are shown in figure~\ref{fig:free-energy-u}.

Alternatively, we can check which saddles contribute to the free energy density of the planar 3-coloring model by computing the zeromode integral. That is, consider the integral
\be
    I = \int_{-\infty}^\infty \frac{d \phi_0}{\sqrt{2\pi}}~e^{- n_s S_0[\phi_0]},
\ee
where
\be
    S_0[\phi_0] = \frac{3}{4} \phi_0^2 + \frac{1}{2} \ln(1+iu\phi_0),
\ee
is the $\phi_0$-dependent part of the action~\eqref{eq:action-tree-level-planar-3-colorings}. In the thermodynamic limit $n_s\to \infty$, using steepest descent method, the integral can be approximated by 
\be
    I \simeq \sum_{\text{saddles } \alpha = \pm } \sqrt{\frac{2\pi}{-n_s |S_0''(\phi_0^\alpha)|}}~e^{-n_s S_0[\phi_0^\alpha]}.
\ee
Comparing this with the numerical integration, we can identify the dominant saddle. One can show that indeed the integral is well-approximated by the two contributions $\phi_0^+$ and $\phi_0^-$ when $u>u_0$, and by $\phi_0^-$ when $u<u_0$. At $u=u_0$, the Hessian is zero at the saddle point; a more detailed analysis is needed to take care of this case.  

\begin{figure}[b!]
    \centering
    \includegraphics[scale=0.55]{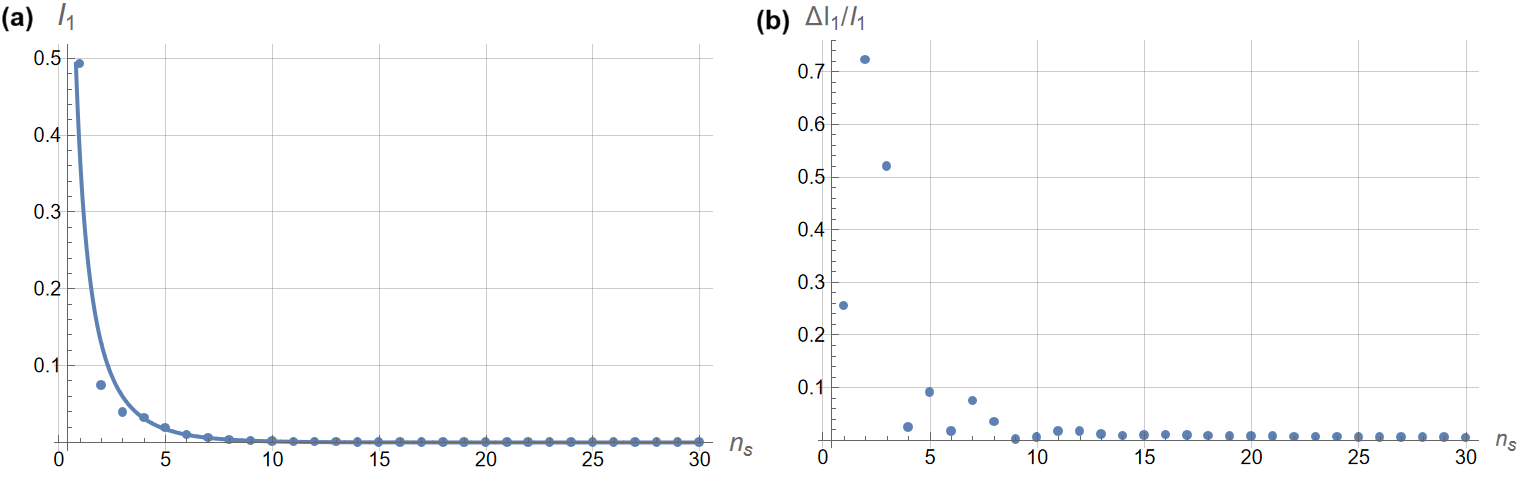}
    \caption{(a) Asymptotic result of the $I_1$ integral (solid line), and its numerical results (dots) as a function of the number of sites $n_s$. (b) Relative error between these values.   }
    \label{fig:error-saddle-vs-numerics-signed-3-colorings}
\end{figure}

{\bf Zeromode analysis for the other coloring models.}
For the other models, it becomes increasingly difficult to do the saddle point analysis. However, we can still do the zeromode integral to find the correct saddles. Let's consider the associated integral for the signed 3-coloring model described in~\eqref{subsec:signed-3-colorings}:
\be
    I_1 = \int_{-\infty}^{\infty} \frac{d\varphi_0\, d\phi_0}{2\pi}~e^{-n_s S_0[\varphi_0,\phi_0]}, 
\ee
where 
\be
    S_0[\varphi_0, \phi_0] = \frac{3}{4} \varphi_0^2 + \frac{3}{4}\phi_0^2 - \frac{3}{2} \ln(1-i\varphi_0 -\phi_0 \sqrt{2i\varphi_0/3}).
\ee
Taking the first variation, we found that the mean-field equations are: 
\be
\left.\frac{\delta S_0}{\delta \phi_{0}} \right|_{\phi_0,\varphi_0} \hspace{-0.4cm}= \phi_0 + \frac{\sqrt{\frac{2i \varphi_0}{3}}}{1-i \varphi_0-\phi_0 \sqrt{\frac{2i\varphi_0}{3}}}= 0, \quad  \left.\frac{\delta S_0}{\delta \varphi_{0}}\right|_{\phi_0,\varphi_0} \hspace{-0.4cm} = \varphi_0 + \frac{i( 1 + \frac{1}{\sqrt{6 i \varphi_0}} \phi_0 ) }{1- i \varphi_0- \phi_0 \sqrt{\frac{2i \varphi_0}{3}}} = 0,
\ee
the solutions of which are: 
\begin{align}
    (\phi_0,\varphi_0) = \{& (0.607 + 0.68i, -1.10-0.288i); (0.607 -0.68i, 1.10 - 0.288i); \nonumber \\  &(-2.08i,  0.54 i); (-0.815, -0.716i) \}.
\end{align}
We see that all of these saddle points are orbits of the time-reversal operation \eqref{eq:signed-T1},\eqref{eq:signed-T2},\eqref{eq:signed-T3}. The idea is to check which saddle dominates in the asymptotic expansion. In this case, the dominant saddle is given by $(\phi_0, \varphi_0) = (-0.815,-0.716i)$. 
Figure~\ref{fig:error-saddle-vs-numerics-signed-3-colorings} shows the comparison between the asymptotic expansion via saddle point of the dominant saddle (solid line), and the numerical integration results (dots). It also shows the relative error between these values. We see that, as $n_s$ increases, the asymptotic value comes closer and closer to the numerical exact value, and the relative error approaches zero.   

Finally, consider 
\be
    I_2 = \int_{-\infty}^{\infty}  \frac{d\phi_0 \,d\varphi_0\, dP_0}{ (2\pi)^{3/2}}~e^{-n_s S_0[\phi_0,\varphi_0,P_0]}~,
\ee 
the associated integral to the non-planar 3-coloring problem, where:
\be
    S_0[\phi_0, \varphi_0, P_0] = \frac{3}{4}\phi_0^2 + \frac{3}{4} \varphi_0^2 + \frac{9}{4} P_0^2 - 3 \ln P_0 - \frac{3}{2} \ln(1+i\varphi_0 + i\phi_0 \sqrt{2i\varphi_0/3} ).
\ee

\begin{figure}[t!]
    \centering
    \includegraphics[scale=0.55]{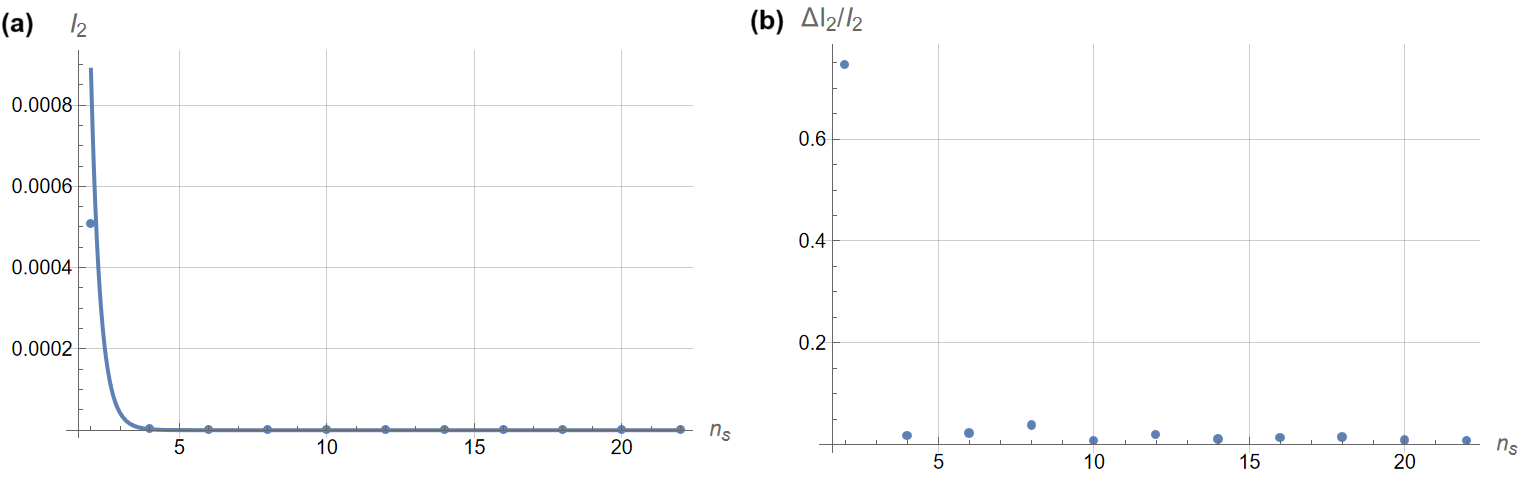}
    \caption{(a) Asymptotic result of the $I_2$ integral (solid line), and its numerical results (dots) as a function of the number of sites $n_s$. (b) Relative error between these values.}
    \label{fig:error-saddle-vs-numerics-nonplanar-3-colorings}
\end{figure}

The mean-field equations are given by:
\begin{align}
    \left.\frac{\delta S_0}{\delta \varphi_{0}} \right|_{\phi_0, \varphi_0, P_0} &= -  \varphi_0 +  i \Big(  \frac{1+\phi_0 \sqrt{\frac{i}{6\varphi_0}}}{1+i \varphi_0 +i \phi_0 \sqrt{2i \varphi_0/3 }} \Big)= 0, \\
    \left.\frac{\delta S_0}{\delta \phi_{0}} \right|_{\phi_0, \varphi_0, P_0} &= -  \phi_0 + \Big( \frac{i \sqrt{\frac{2i\varphi_0}{3}}}{1+i \varphi_0 +i \phi_0 \sqrt{\frac{2i \varphi_0}{3}}} \Big) =0 , \\
    \left.\frac{\delta S_0}{ \delta P_{0} } \right|_{\phi_0, \varphi_0, P_0} &= - \frac{9}{2} P_0 + 3  \Big( \frac{1}{P_0} \Big) = 0.
\end{align}
Solutions to the last equation are $P_0 = \pm \sqrt{2/3}$. For the other fields, we get:
\begin{align}
    (\phi_0, \varphi_0) = \{& (0.607+0.68i,1.10+0.288i); (-0.607+0.68i, -1.10+0.288i); \nonumber \\
    &(2.08i, -0.54i);(-0.815, 0.716i)\}.
\end{align}
Again, we see that saddles $(\phi_0, \varphi_0, P_0)$ are orbits of the time-reversal operation \eqref{eq:nonplanar-T1},\eqref{eq:nonplanar-T2},\eqref{eq:nonplanar-T3},\eqref{eq:nonplanar-T4}, the dominant ones being $(-0.815,0.716i,\pm \sqrt{2/3})$.
The comparison between numerical integration and the contribution of the dominant saddle is given in the plot~\ref{fig:error-saddle-vs-numerics-nonplanar-3-colorings}. 

\newpage

\phantomsection\addcontentsline{toc}{section}{References}
\bibliographystyle{ucsd}
\bibliography{Bibliography}

\end{document}